\documentclass[structabstract]{aa}  

\usepackage{graphicx}
\usepackage{txfonts}
\usepackage{longtable}
\usepackage{natbib}
\bibpunct{(}{)}{;}{a}{}{,}

\def\figw{5cm}
\def\LL{\left(L\right)}
\def\SS{\left(S\right)}

\begin{document}

\title{The HARPS search for southern extra-solar planets\thanks{Based on observations made with the HARPS instrument on the ESO 3.6-m telescope at La Silla Observatory (Chile), under program IDs 072.C-0488 and 183.C-0972.}}

\subtitle{XXVII. Up to seven planets orbiting HD 10180: \\ probing the architecture of low-mass planetary systems}

\author{C. Lovis\inst{1}
\and D. S\'egransan\inst{1}
\and M. Mayor\inst{1}
\and S. Udry\inst{1}
\and W. Benz\inst{2}
\and J.-L. Bertaux\inst{3}
\and F. Bouchy\inst{4,5}
\and A. C. M. Correia\inst{6}
\and \\ J. Laskar\inst{7}
\and G. Lo Curto\inst{8}
\and C. Mordasini\inst{9,2}
\and F. Pepe\inst{1}
\and D. Queloz\inst{1}
\and N. C. Santos\inst{10,1}
}

\institute{Observatoire de Gen\`eve, Universit\'e de Gen\`eve, 51 ch. des Maillettes, CH-1290 Versoix, Switzerland \\
\email{christophe.lovis@unige.ch}
\and Physikalisches Institut, Universit\"at Bern, Sidlerstrasse 5, CH-3012 Bern, Switzerland
\and Universit\'e Versailles Saint-Quentin, LATMOS-IPSL, 11 Boulevard d'Alembert, F-78280 Guyancourt, France
\and Institut d'Astrophysique de Paris, UMR7095 CNRS, Universit\'e Pierre \& Marie Curie, 98bis Bd Arago, F-75014 Paris, France
\and Observatoire de Haute-Provence, CNRS/OAMP, F-04870 St. Michel l'Observatoire, France
\and Department of Physics, I3N, University of Aveiro, Campus Universit\'ario de Santiago, 3810-193 Aveiro, Portugal
\and ASD, IMCCE-CNRS UMR8028, Observatoire de Paris, UPMC, 77 Av. Denfert-Rochereau, F-75014 Paris, France
\and European Southern Observatory, Karl-Schwarzschild-Str. 2, D-85748 Garching bei M\"unchen, Germany
\and Max-Planck-Institut f\"ur Astronomie, K\"onigstuhl 17, D-69117 Heidelberg, Germany
\and Departamento de F\'isica e Astronomia, Faculdade de Ci\^encias, Universidade do Porto, 4150-762 Porto, Portugal
}

\date{Received 12 August 2010 / Accepted 18 November 2010}

\abstract
{Low-mass extrasolar planets are presently being discovered at an increased pace by radial velocity and transit surveys, opening a new window on planetary systems.}
{We are conducting a high-precision radial velocity survey with the HARPS spectrograph which aims at characterizing the population of ice giants and super-Earths around nearby solar-type stars. This will lead to a better understanding of their formation and evolution, and yield a global picture of planetary systems from gas giants down to telluric planets.}
{Progress has been possible in this field thanks in particular to the sub-m\,s$^{-1}$ radial velocity precision achieved by HARPS. We present here new high-quality measurements from this instrument.}
{We report the discovery of a planetary system comprising at least five Neptune-like planets with minimum masses ranging from 12 to 25\,$M_\oplus$, orbiting the solar-type star HD 10180 at separations between 0.06 and 1.4 AU. A sixth radial velocity signal is present at a longer period, probably due to a 65-$M_\oplus$ object. Moreover, another body with a minimum mass as low as 1.4\,$M_\oplus$ may be present at 0.02 AU from the star. This is the most populated exoplanetary system known to date. The planets are in a dense but still well-separated configuration, with significant secular interactions. Some of the orbital period ratios are fairly close to integer or half-integer values, but the system does not exhibit any mean-motion resonances. General relativity effects and tidal dissipation play an important role to stabilize the innermost planet and the system as a whole. Numerical integrations show long-term dynamical stability provided true masses are within a factor $\sim$3 from minimum masses. We further note that several low-mass planetary systems exhibit a rather "packed" orbital architecture with little or no space left for additional planets. In several cases, semi-major axes are fairly regularly spaced on a logarithmic scale, giving rise to approximate Titius-Bode-like (i.e. exponential) laws. These dynamical architectures can be interpreted as the signature of formation scenarios where type I migration and interactions between protoplanets play a major role. However, it remains challenging to explain the presence of so many Neptunes and super-Earths on non-resonant, well-ordered orbits within $\sim$1-2 AU of the central star. Finally, we also confirm the marked dependence of planet formation on both metallicity and stellar mass. Very massive systems are all found around metal-rich stars more massive than the Sun, while low-mass systems are only found around metal-deficient stars less massive than the Sun.}
{}


\maketitle

\section{Introduction}

Over the past 15 years, the field of extrasolar planets has been witnessing uninterrupted developments and several major milestones. Among these one can mention: the initial proof of existence of extrasolar gas giants \citep{mayor95}, the first detection of a transiting planet, providing a precise mass and radius \citep{charbonneau00,henry00}, the detection of a large sample of gas giants with a variety of masses and orbital properties, the characterization of bulk properties and atmospheres of transiting gas giants, and the detection of objects in the Neptune-mass and super-Earth range. Recently, two transiting planets with masses and radii close to those of Earth have been discovered: CoRoT-7b \citep{leger09, queloz09} and GJ~1214~b \citep{charbonneau09}. High-precision radial velocity surveys are now able to find planets with minimum masses as low as 1.9 $M_{\oplus}$ \citep{mayor09a}. Preliminary results from the HARPS survey are hinting at a large population of Neptune-like objects and super-Earths within $\sim$0.5 AU of solar-type stars \citep{lovis09}. Moreover, hundreds of small-radius candidate planets have been announced by the Kepler Team \citep{borucki10}. Clearly, the exploration of the low-mass planet population has now fully started, and will become the main focus of the field in the coming years. It is expected that the characterization of planetary system architectures, taking into account all objects from gas giants to Earth-like planets, will greatly improve our understanding of their formation and evolution. It will also allow us to eventually put our Solar System into a broader context and determine how typical it is in the vastly diverse world of planetary systems. The characterization of a significant sample of low-mass objects, through their mean density and some basic atmospheric properties, is also at hand and will bring much desired insights into their composition and the physical processes at play during planet formation.

As part of this broad effort to explore the low-mass exoplanet population, we are conducting a high-precision radial velocity survey of about 400 bright FGK stars in the solar neighbourhood using the HARPS instrument \citep{mayor03}. Observations of this sample were obtained during HARPS GTO time from 2004 to 2009 (PI: M. Mayor), and then were continued as an ESO Large Program (PI: S. Udry) until today. Several stars from this survey have already revealed orbiting low-mass objects: HD~160691 \citep{santos04, pepe07}, HD~69830 \citep{lovis06a}, HD~4308 \citep{udry06}, HD~40307 \citep{mayor09b}, HD~47186, HD~181433 \citep{bouchy09}, and HD~90156 \citep{mordasini10}. More and more candidates are detected as measurements accumulate, and many new systems are about to be published (Queloz et al., Udry et al., S\'egransan et al., Benz et al., Dumusque et al., Pepe et al., in prep.). Following 400 stars to search for radial velocity signals at the m\,s$^{-1}$ level requires a lot of telescope time, and this survey is by construction a long-term project. Over the years, we chose to focus on a smaller sample of stars showing a low level of chromospheric activity to minimize the impact of stellar noise on our planet detection limits. Based on measured Ca\,II activity levels $\log{R'_{\mathrm{HK}}}$, we kept about 300 stars which we are monitoring regularly. Once a sufficient number of observations has been gathered for each star, we will be able to derive important statistical properties of the low-mass planet population (Mayor et al., in prep.).

In this paper we present the discovery of a new low-mass planetary system comprising at least 5 Neptune-mass planets and, probably, a longer-period object and a close-in Earth-mass planet. The parent star is the G1V dwarf HD~10180, located 39 pc away from the Sun towards the southern constellation Hydrus.

\section{Observations and data reduction}

\addtocounter{table}{1}

The data presented in this paper have been obtained with the HARPS spectrograph at the ESO 3.6-m telescope at La Silla Observatory (Chile). This instrument has demonstrated a sub-m\,s$^{-1}$ radial velocity precision over more than 6 years \citep[e.g.][]{lovis06b, mayor09b} and has led to the detection of the majority of the low-mass planets known to date.

We have obtained a total of 190 data points on HD~10180, spread over more than 6 years. This star is part of the high-precision planet-search sample of about 400 stars that we have been following closely since 2004. Exposure times were set to 15 min to average out stellar oscillations. The achieved SNR at 550 nm ranges from 120 to 270, depending on weather conditions. The estimated photon noise limit to the radial velocity precision ranges from 80 to 30\,cm\,s$^{-1}$, respectively. Including other measurable instrumental errors (wavelength calibration, noise on instrumental drift measurement), we obtain error bars between 1.3 and 0.4\,m\,s$^{-1}$. This does not include other instrumental systematics like telescope guiding (light injection) errors, which are expected to be small but difficult to estimate. Data reduction was performed with the latest version of the HARPS pipeline (see Lovis et al. 2011, in prep. for a more detailed description).

The end products of the reduction are barycentric radial velocities with internal error bars, bisector span measurements, parameters of the cross-correlation functions (FWHM and contrast), and Ca\,II activity indices $S$ and $\log{R'_{\mathrm{HK}}}$. Being a stabilized, well-controlled instrument, HARPS provides not only precise radial velocities, but also precise spectroscopic indicators in general, which is very useful to better understand the stars under consideration \citep[see e.g. the case of the active star CoRoT-7,][]{queloz09}. The whole set of radial velocities and spectroscopic measurements of HD~10180 can be found in Table~\ref{TableData} and in electronic form at CDS.

\section{Stellar properties}

\begin{table}
\caption{Fundamental properties of HD 10180.}
\label{TableHD10180}
\centering
\begin{tabular}{l l c}
\hline\hline
Parameter & & HD 10180 \\
\hline
Spectral type & & G1V \\
$V$ & [mag] & 7.33 \\
$B - V$ & [mag] & 0.629 \\
$\pi$ & [mas] & 25.39 $\pm$ 0.62 \\
$M_V$ & [mag] & 4.35 \\
$T_{\mathrm{eff}}$ & [K] & 5911 $\pm$ 19 \\
$\log{g}$ & [cgs] & 4.39 $\pm$ 0.03 \\
$\mathrm{[Fe/H]}$ & [dex] & 0.08 $\pm$ 0.01 \\
$L$ & [$L_{\odot}$] & 1.49 $\pm$ 0.02 \\
$M_*$ & [$M_{\odot}$] & 1.06 $\pm$ 0.05 \\
$v \sin{i}$ & [km\,s$^{-1}$] & $<$ 3 \\
$\log{R'_{\mathrm{HK}}}$ & & -5.00 \\
$P_{\mathrm{rot}} (\log{R'_{\mathrm{HK}}})$ & [days] & 24 $\pm$ 3 \\
Age $(\log{R'_{\mathrm{HK}}})$ & [Gyr] & 4.3 $\pm$ 0.5 \\
\hline
\end{tabular}
\end{table}

The fundamental properties of HD~10180 (G1V, $V$=7.33) are taken from the Hipparcos catalogue \citep{esa97} and the spectroscopic analysis by \citet{sousa08}. They are listed in Table~\ref{TableHD10180}. HD~10180 is a solar-type star with a mass $M$ = 1.06 $\pm$ 0.05 $M_\odot$ and metallicity [Fe/H] = 0.08 $\pm$ 0.01 dex. With a mean activity index $\log{R'_{\mathrm{HK}}}$ = -5.00, measured on the HARPS spectra presented here, it is clearly an inactive star. Furthermore, it does not show any well-defined activity cycle such as the solar one (the rms dispersion of the $\log{R'_{\mathrm{HK}}}$ measurements is only 0.012 dex). Given this low activity level and the early-G spectral type, we expect a stellar RV jitter at the level of $\sim$1\,m\,s$^{-1}$ based on comparisons with similar stars in the HARPS sample (see Dumusque et al. 2011, in prep., Lovis et al. 2011, in prep.). Among inactive stars, early-G dwarfs appear to have slightly more jitter than late-G and early-K dwarfs, possibly due to more vigorous convection and thus increased granulation noise. We thus adopt a value of 1.0\,m\,s$^{-1}$ for the stellar jitter in this paper and add this number quadratically to the instrumental error bars. The main purpose for doing this is to avoid large, unjustified differences in the individual weights ($w_i$= 1/$\sigma_i^2$) used in the $\chi^2$-minimization process.

\section{Analysis of the radial velocity data}
\label{SectSignals}

\subsection{Detection of 5 strong signals}

\begin{figure}
\centering
\includegraphics[width=85mm]{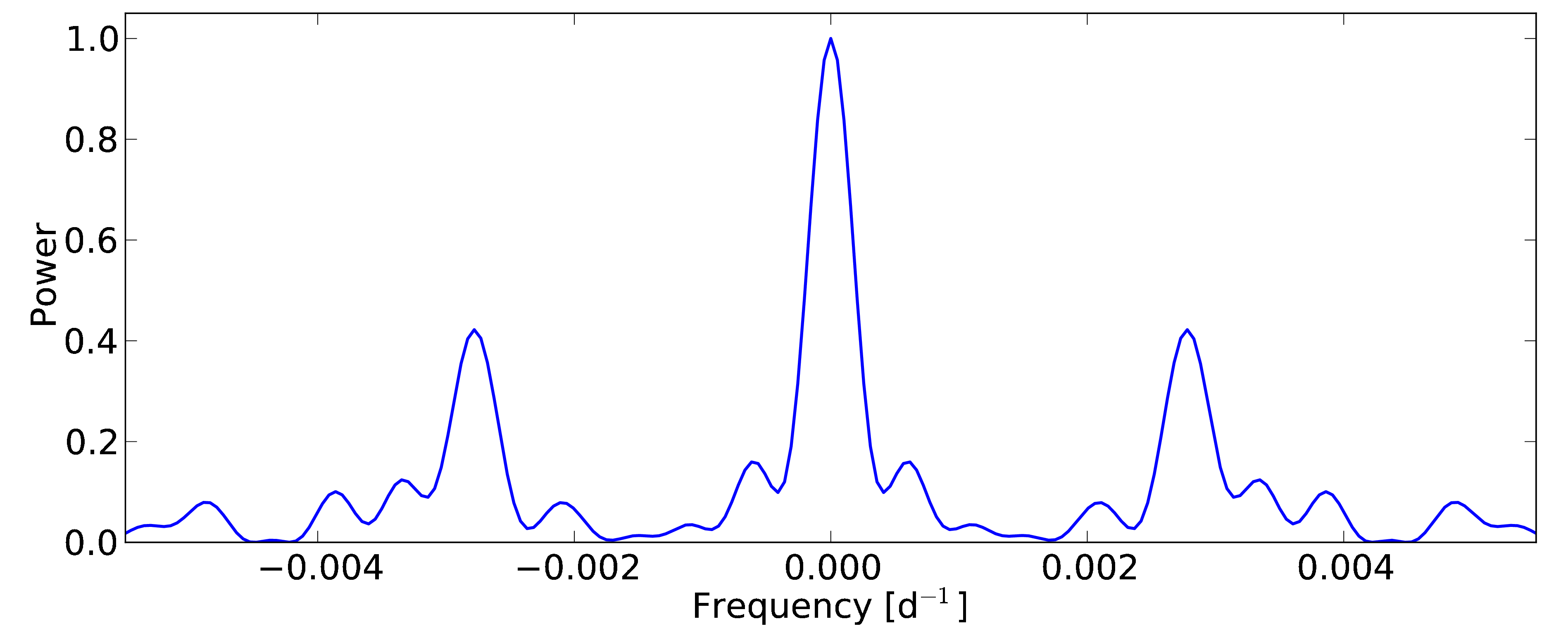}
\includegraphics[width=85mm]{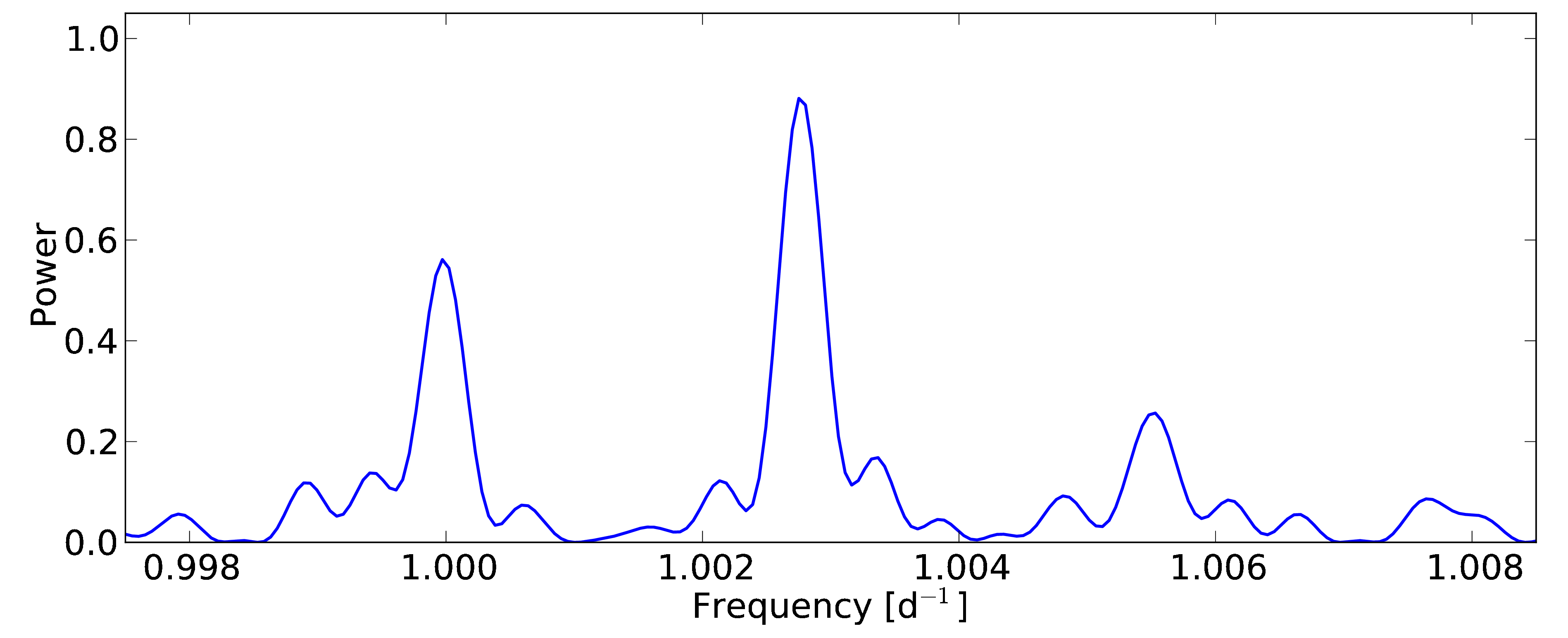}
\includegraphics[width=85mm]{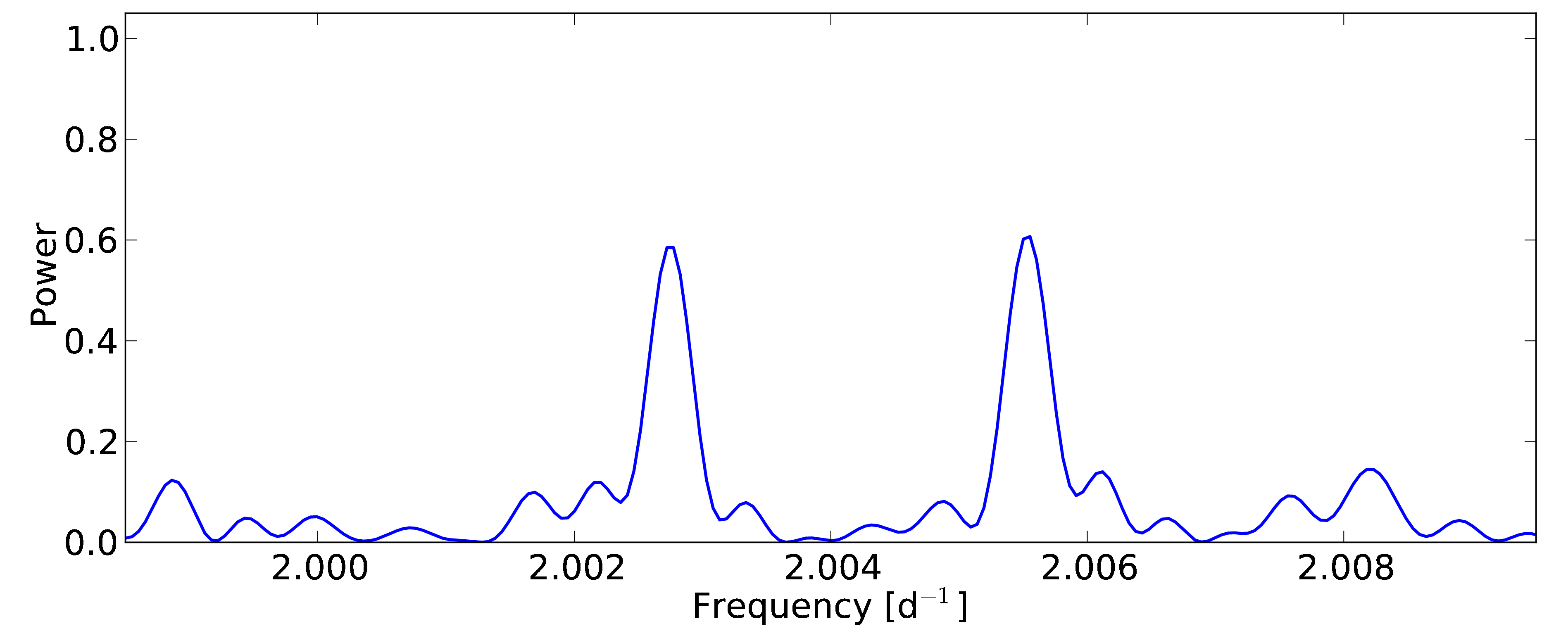}
\caption{Close-up views of the window function of the radial velocity measurements, centered on the regions that have significant peaks and that may induce aliases in the RV data.}
\label{FigWF}
\end{figure}

\begin{figure}
\centering
\includegraphics[bb=0 47 595 255,width=85mm,clip]{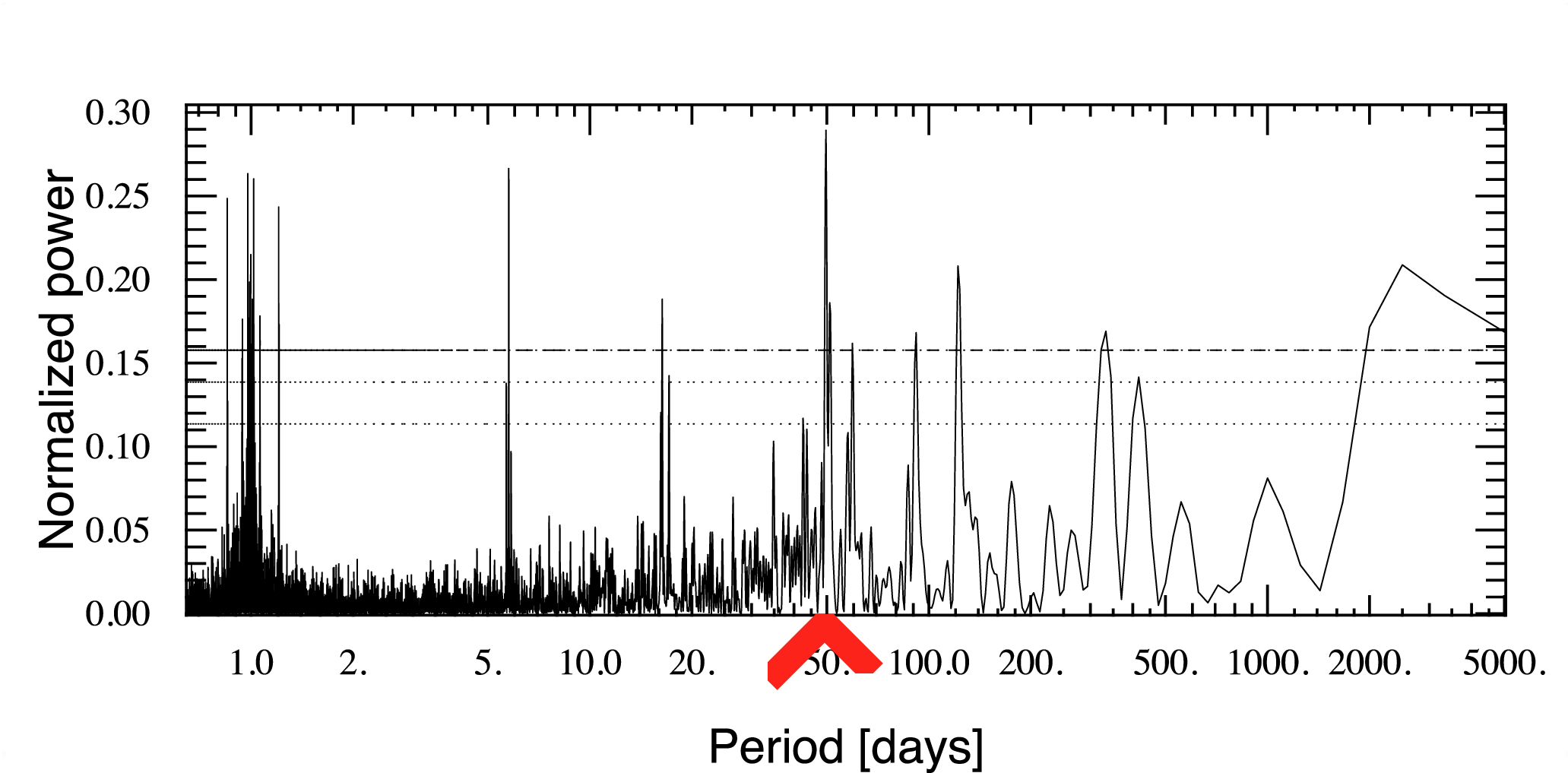}
\includegraphics[bb=0 47 595 255,width=85mm,clip]{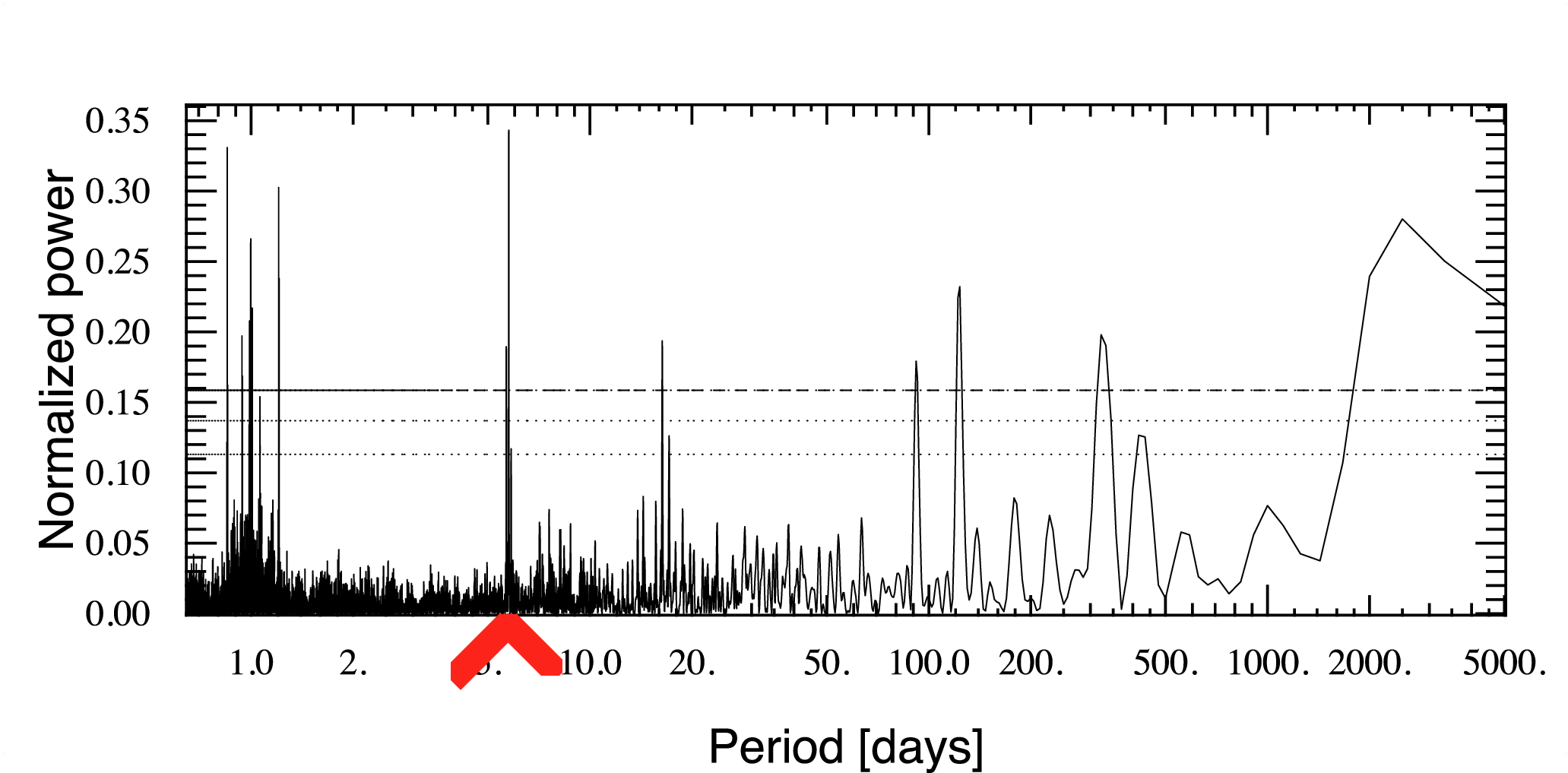}
\includegraphics[bb=0 47 595 255,width=85mm,clip]{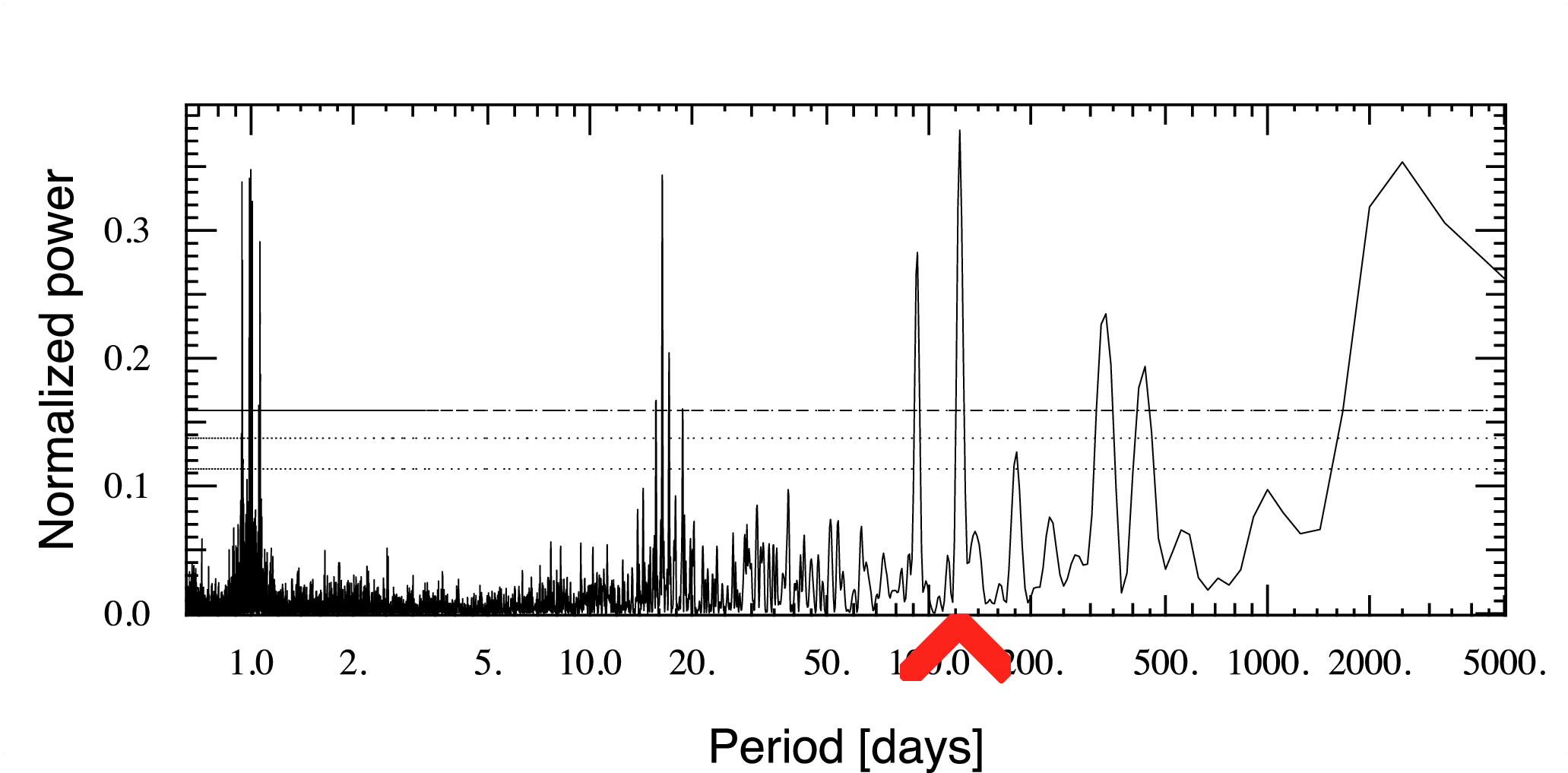}
\includegraphics[bb=0 47 595 255,width=85mm,clip]{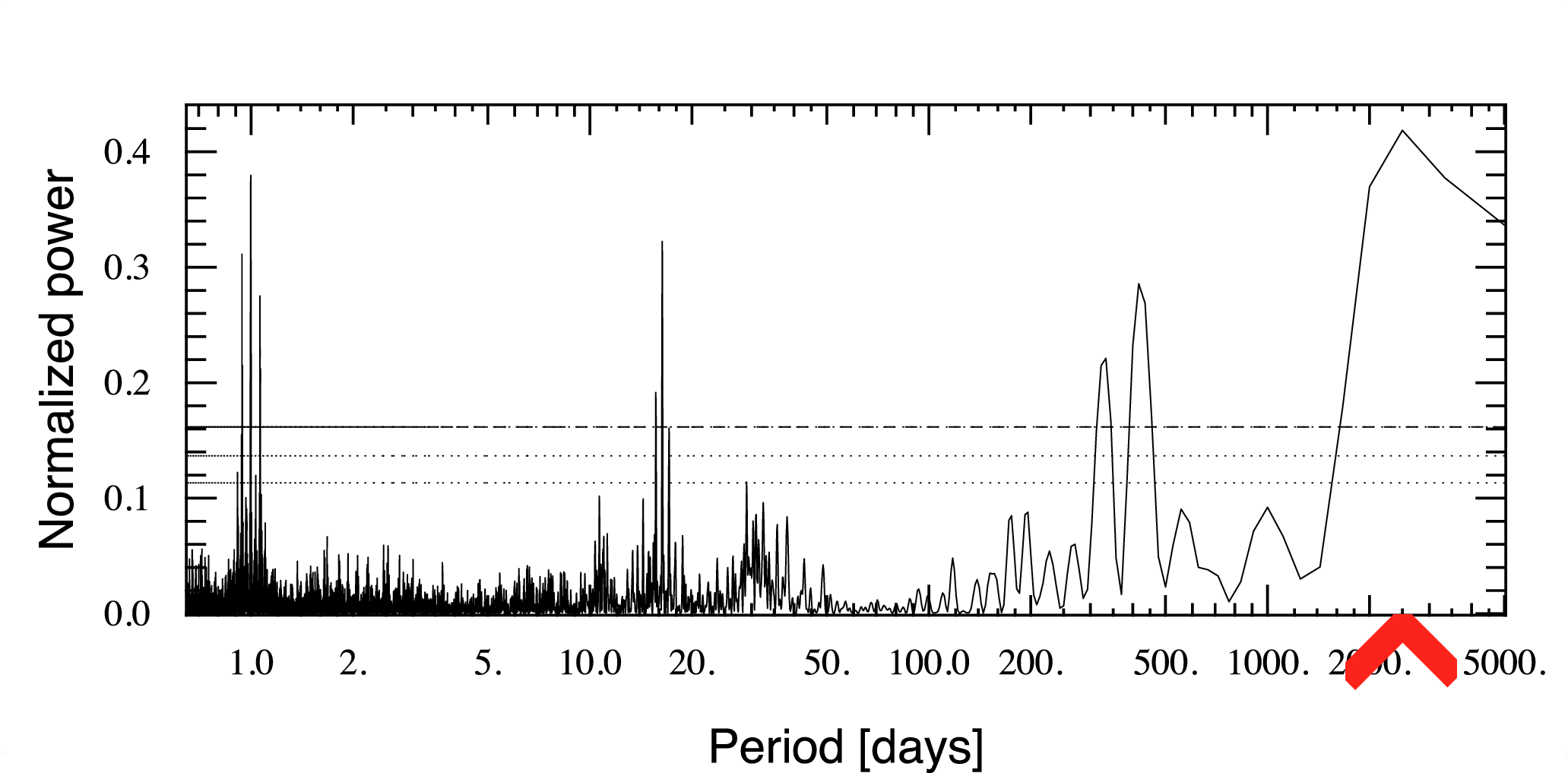}
\includegraphics[bb=0 47 595 255,width=85mm,clip]{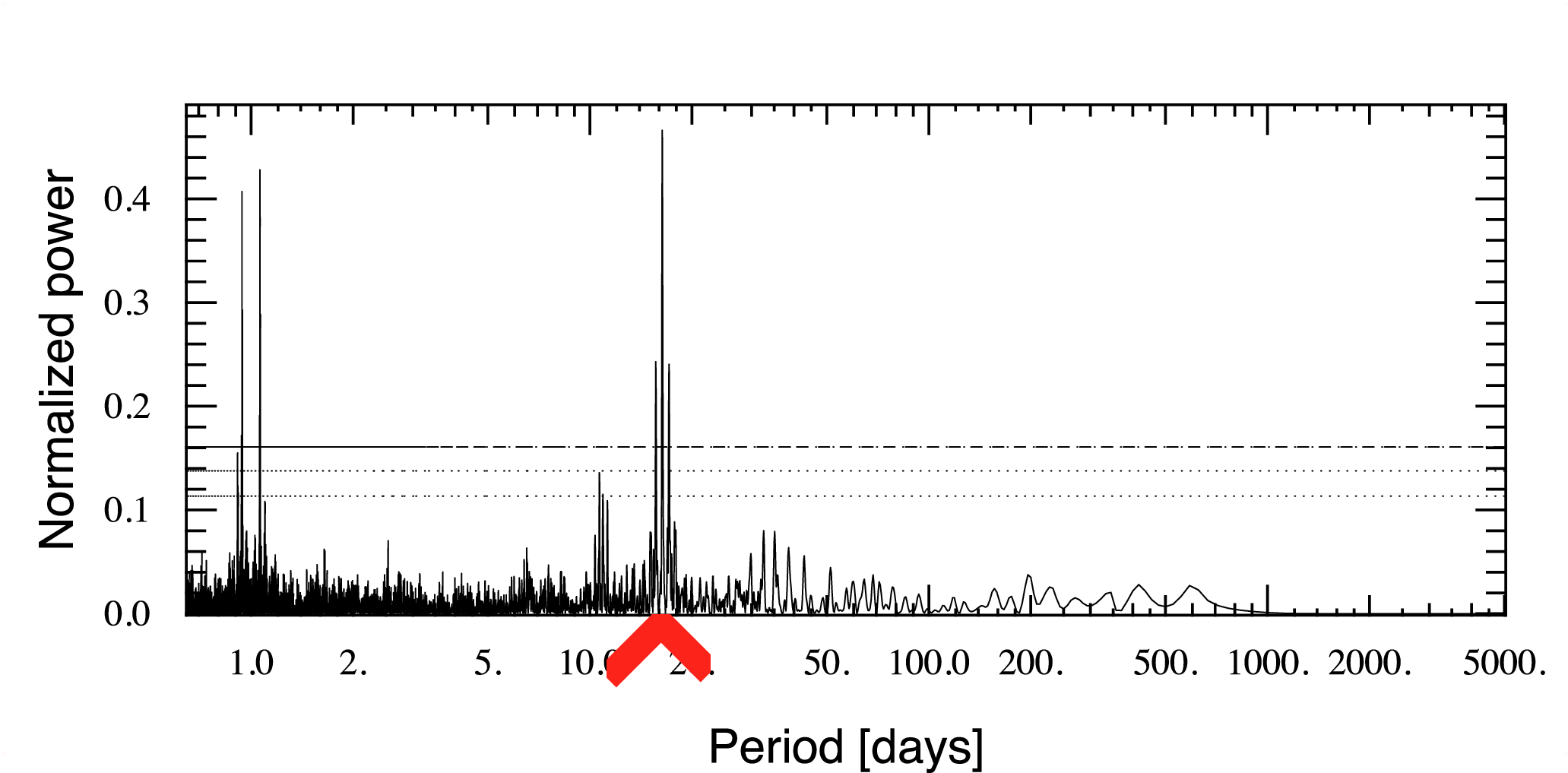}
\includegraphics[bb=0 47 595 255,width=85mm,clip]{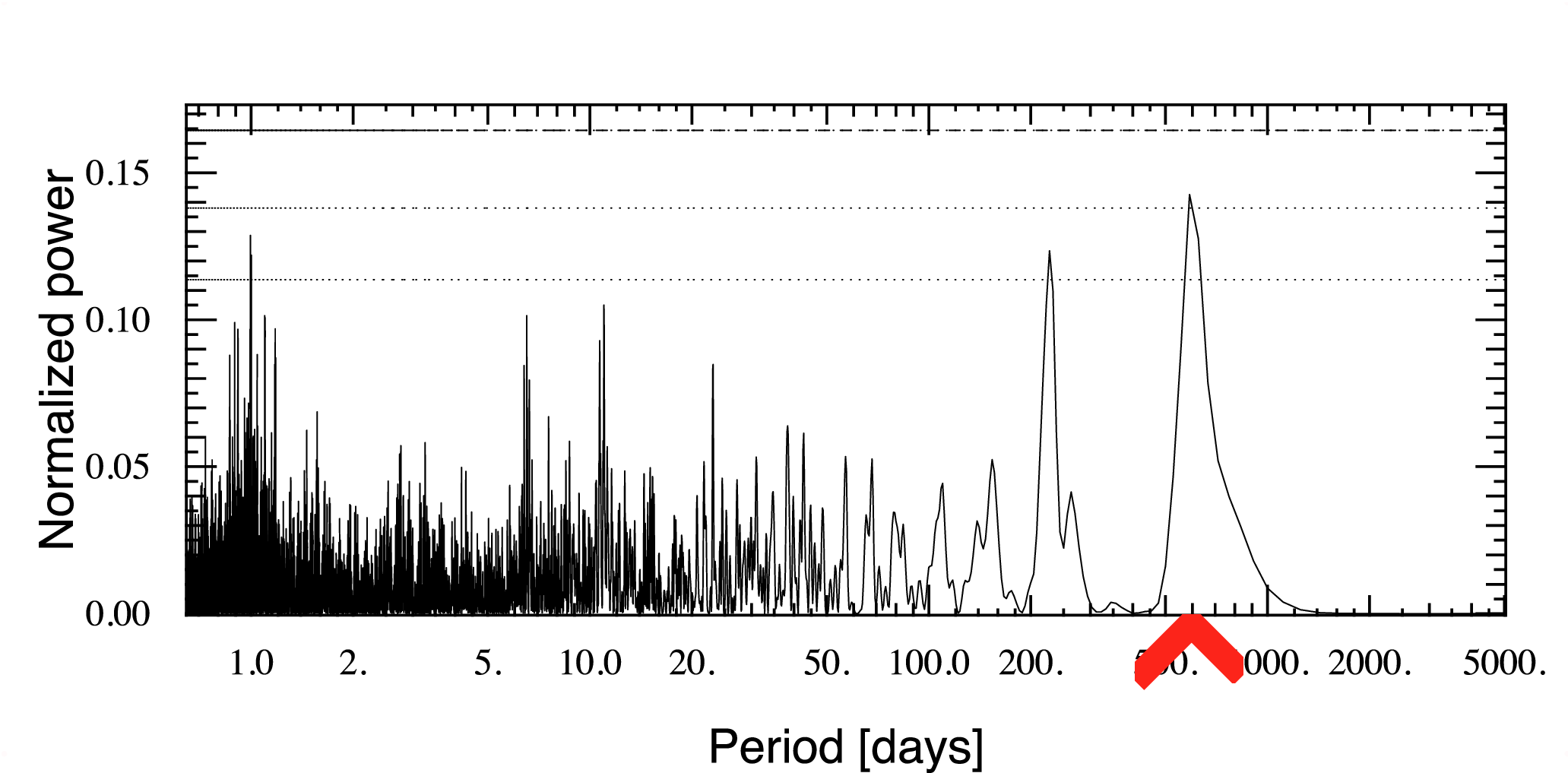}
\includegraphics[bb=0 47 595 255,width=85mm,clip]{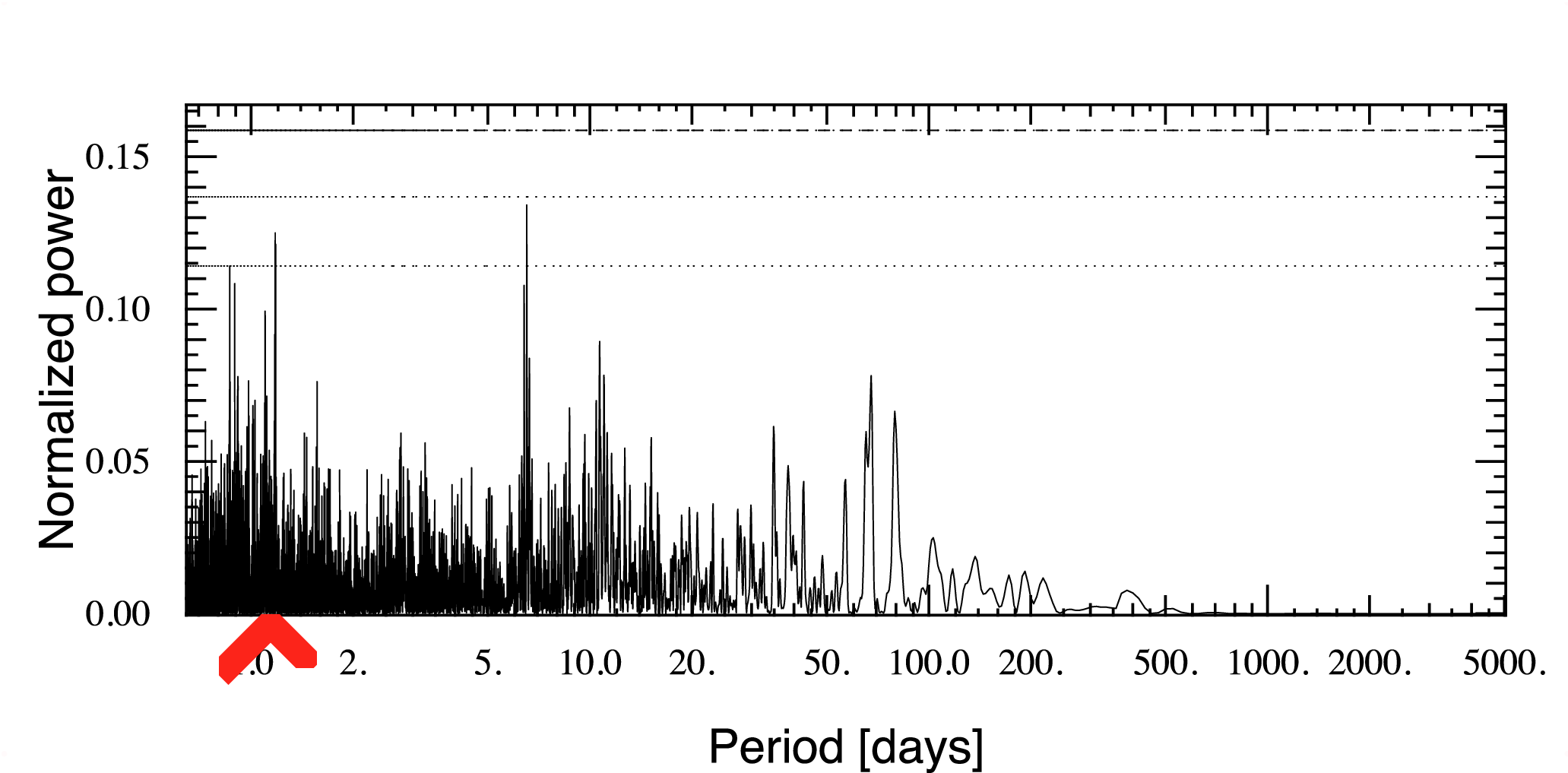}
\includegraphics[bb=0 0 595 255,width=85mm,clip]{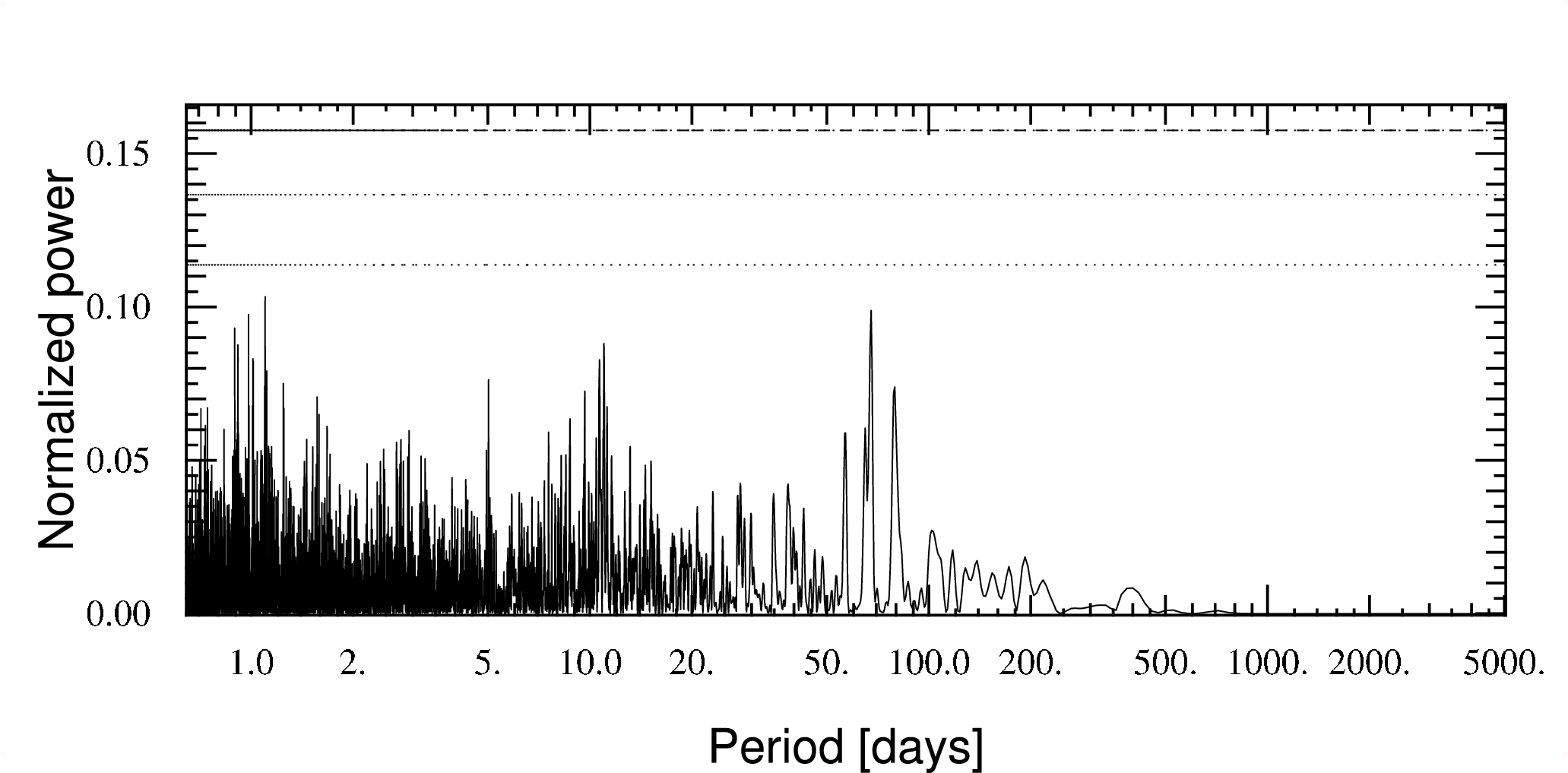}
\caption{Successive GLS periodograms of the HD 10180 RV time series, where the main signal (indicated by an arrow) is subtracted at each step by adding one more planet in the multi-Keplerian model. FAP thresholds of 10\%, 1\% and 0.1\% are indicated as dashed lines.}
\label{FigGLS}
\end{figure}

\begin{figure*}
\centering
\includegraphics[bb=30 20 395 400,width=40mm,clip]{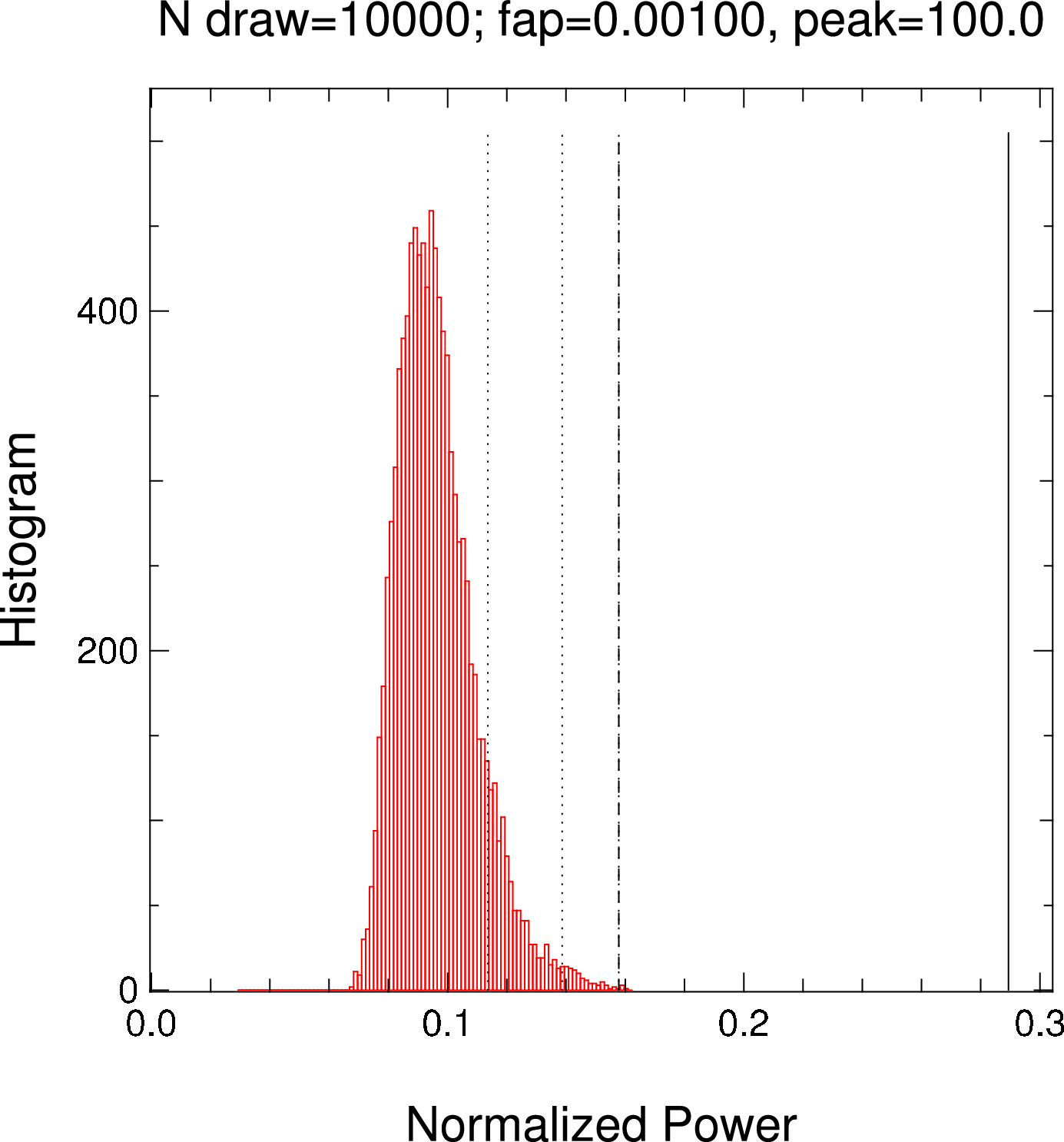}
\includegraphics[bb=30 20 395 400,width=40mm,clip]{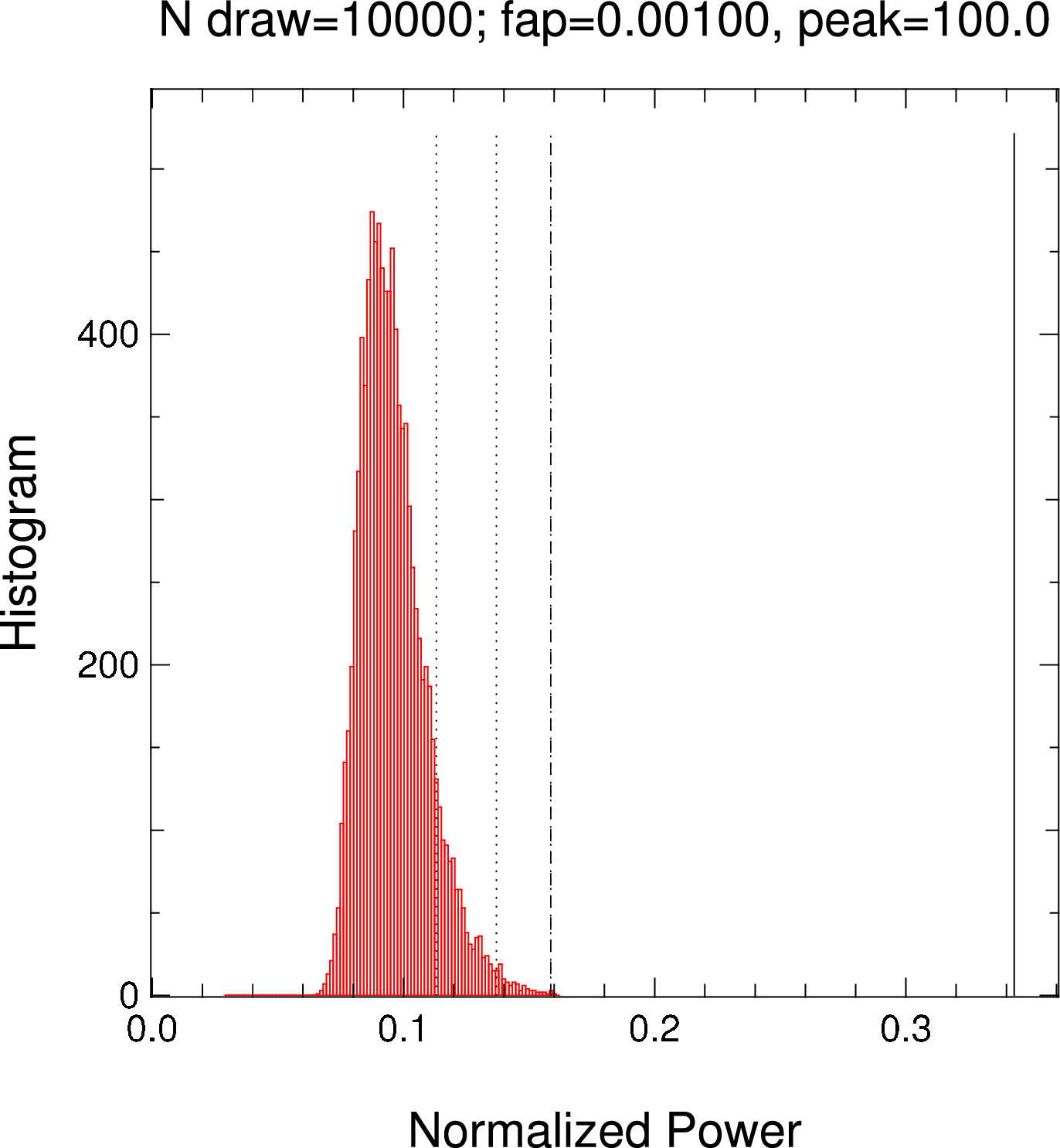}
\includegraphics[bb=30 20 395 400,width=40mm,clip]{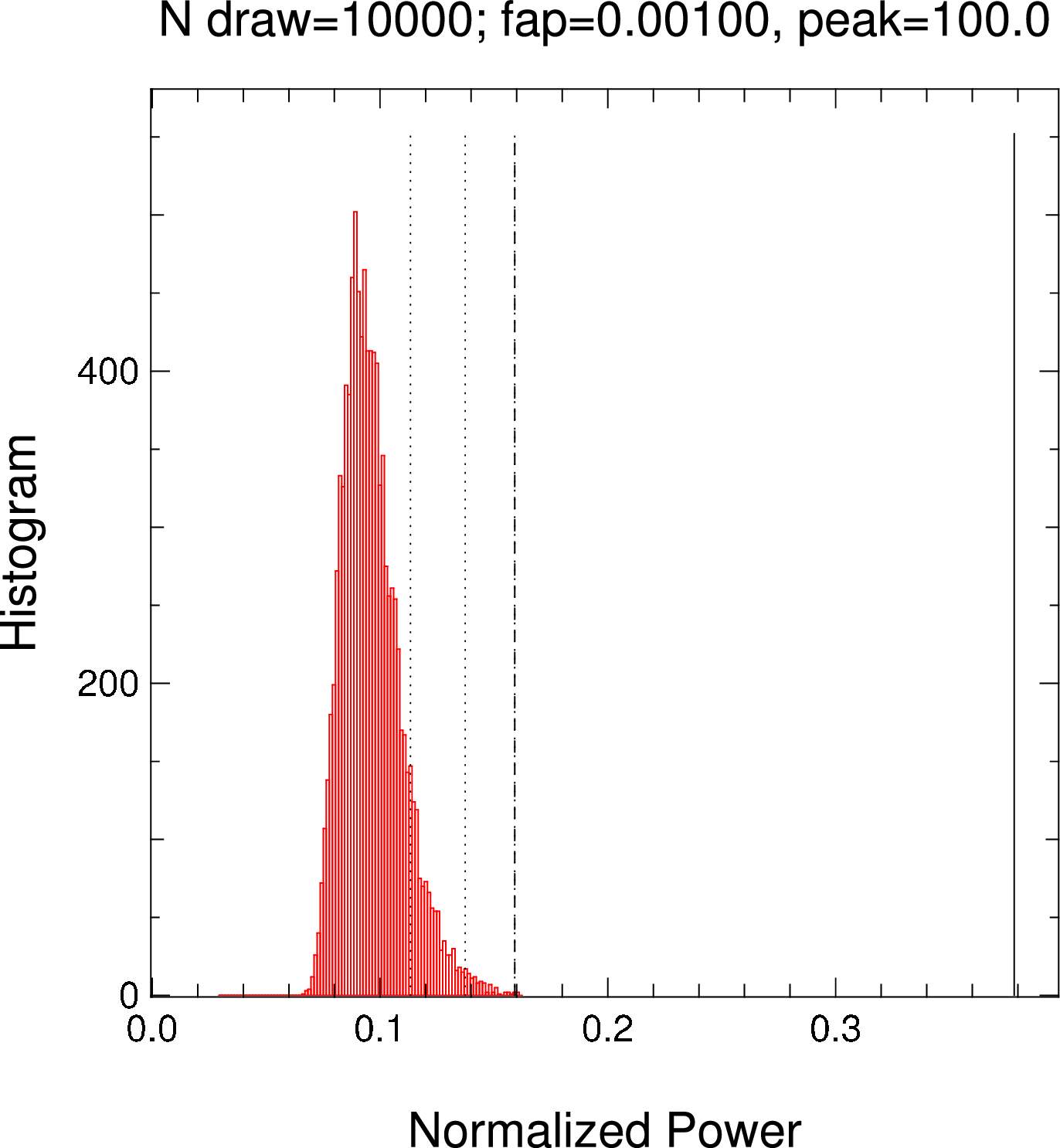}
\includegraphics[bb=30 20 395 400,width=40mm,clip]{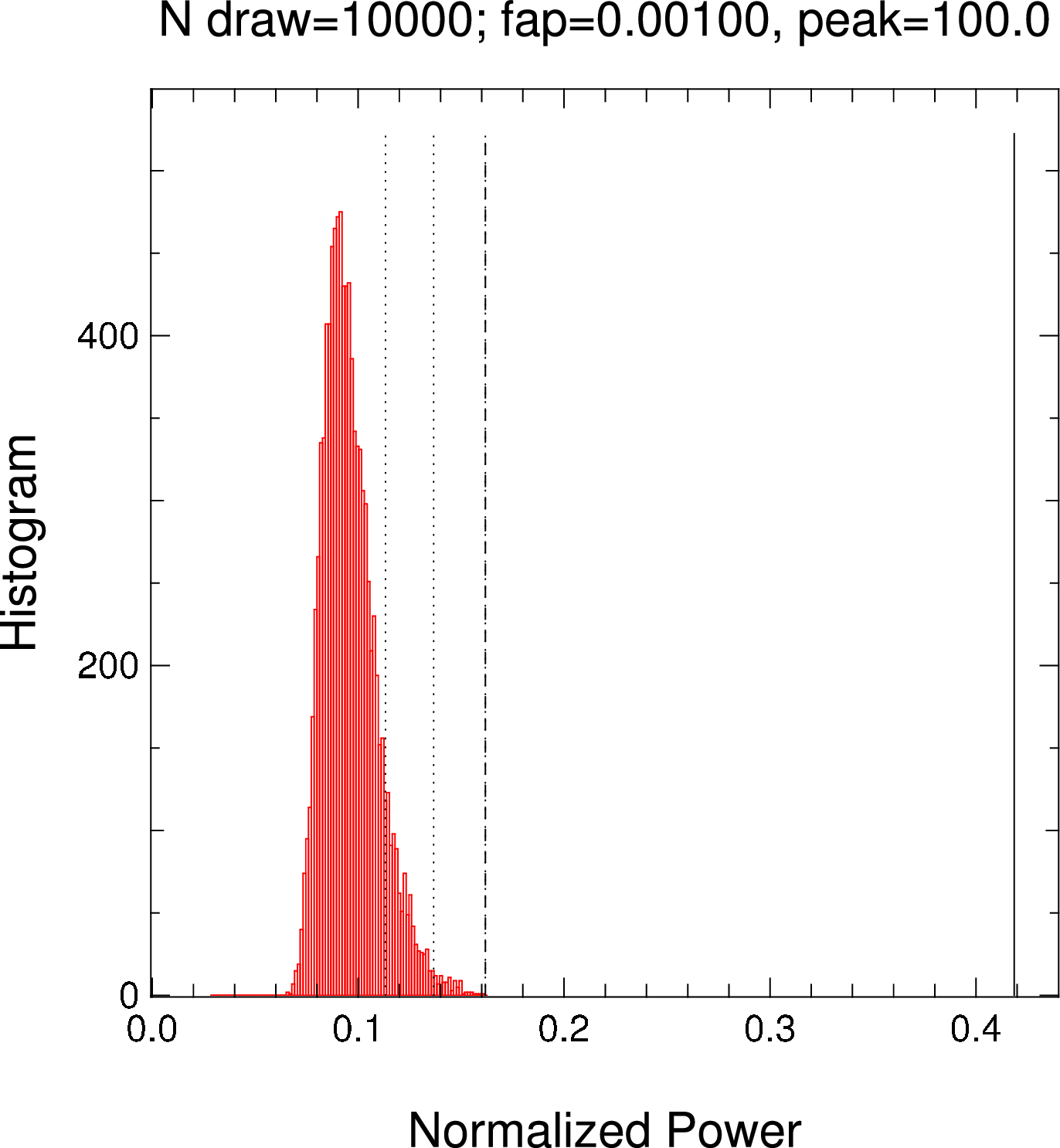}
\includegraphics[bb=30 0 395 400,width=40mm,clip]{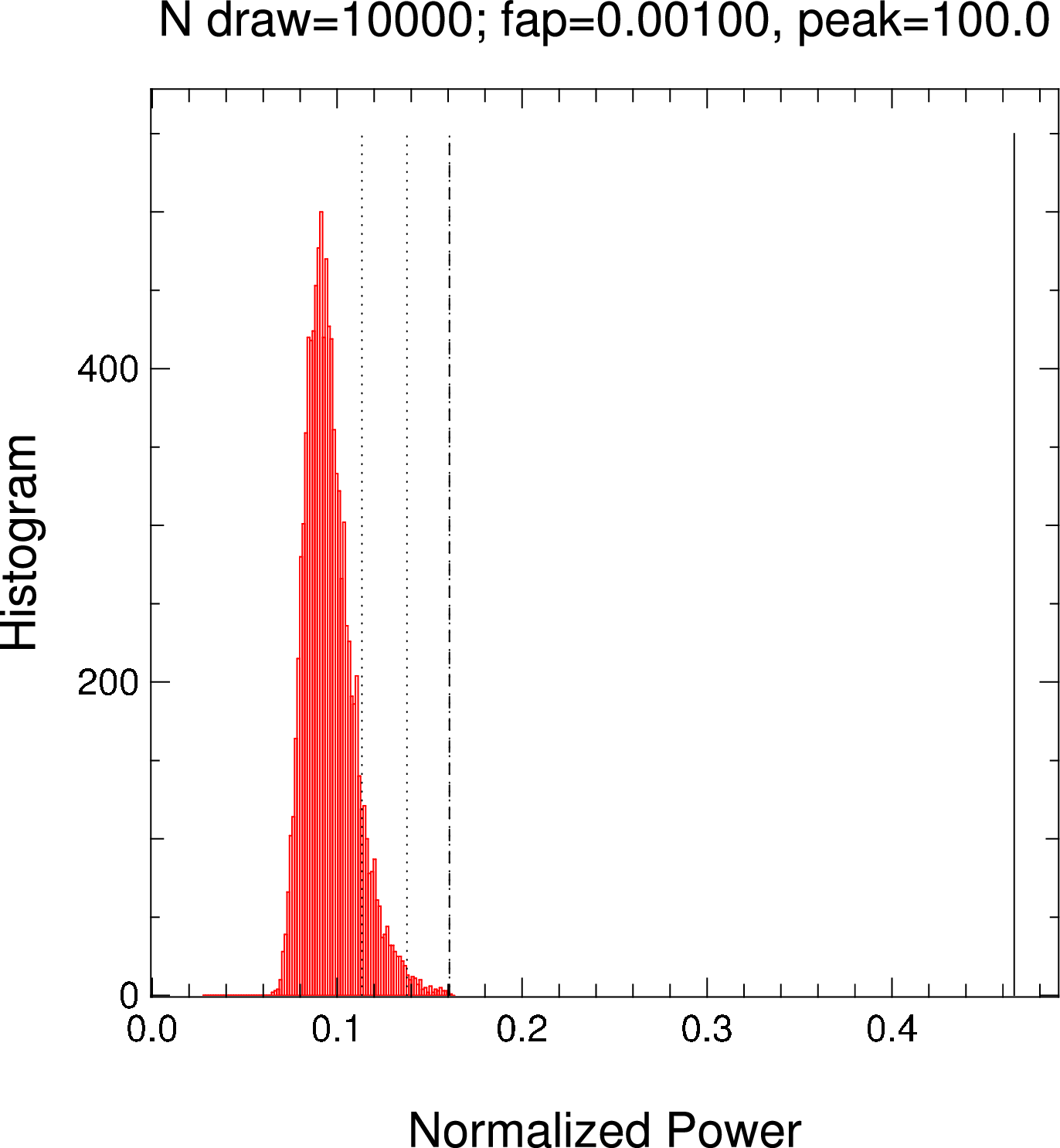}
\includegraphics[bb=30 0 395 400,width=40mm,clip]{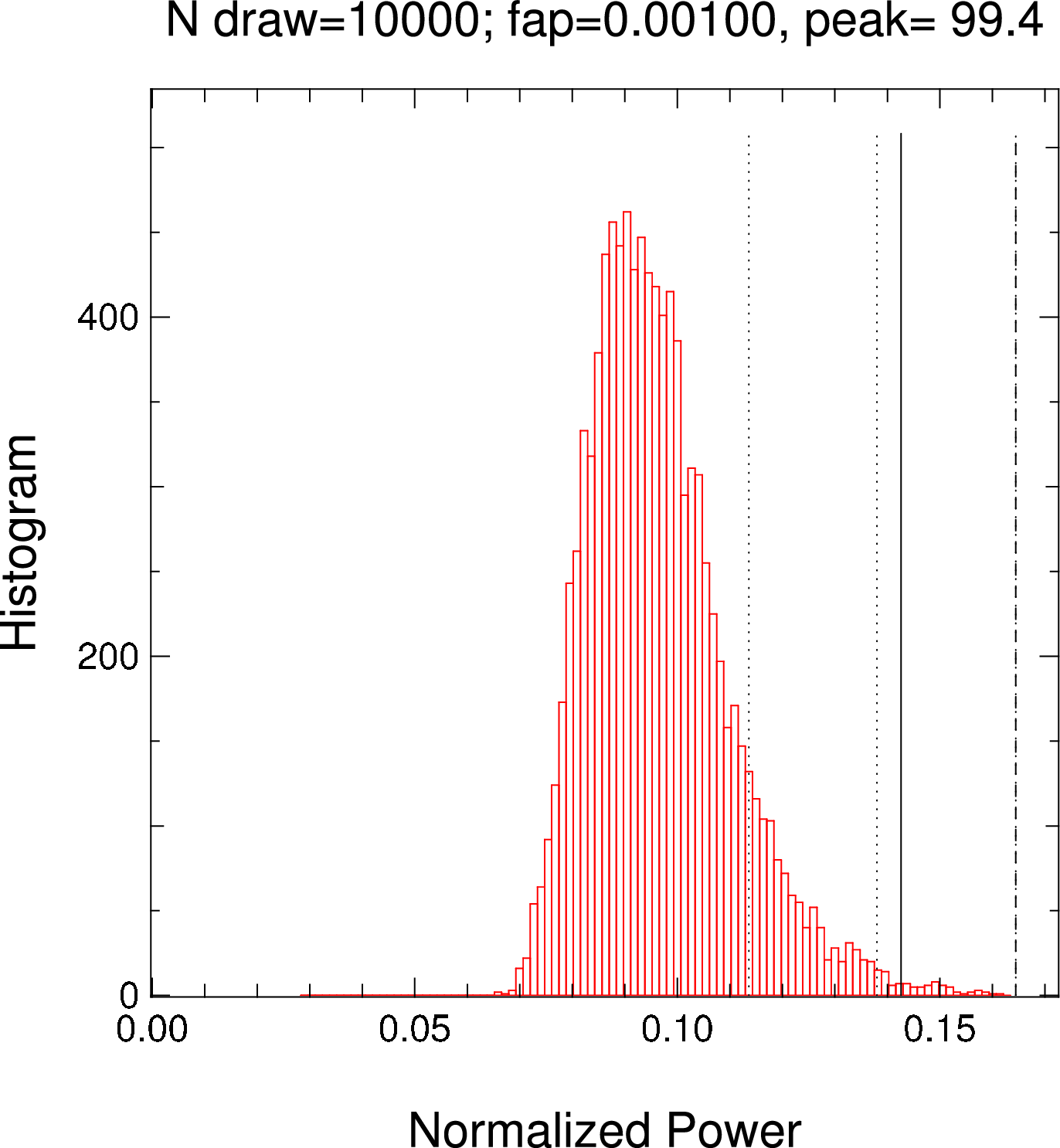}
\includegraphics[bb=30 0 395 400,width=40mm,clip]{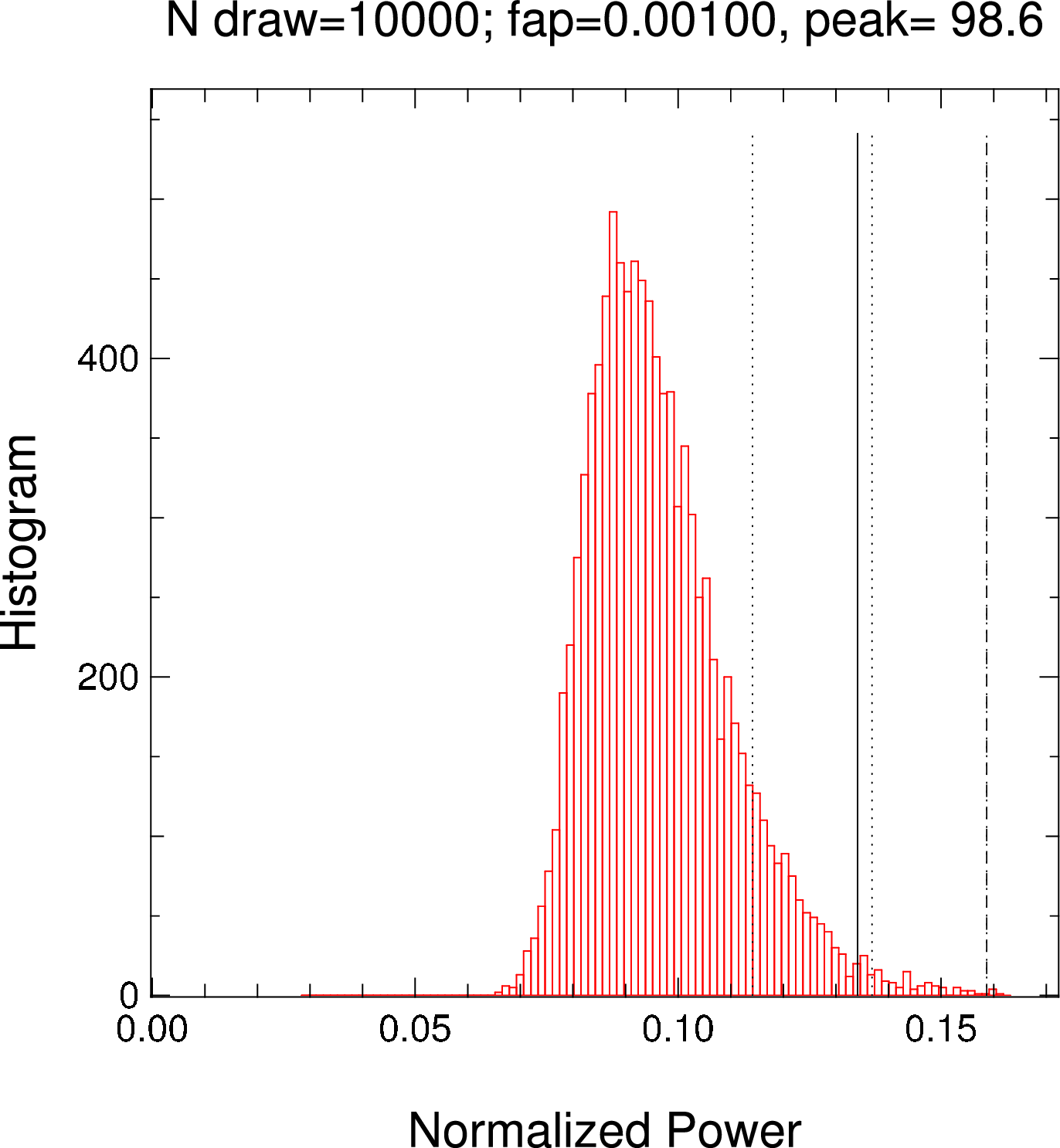}
\includegraphics[bb=30 0 395 400,width=40mm,clip]{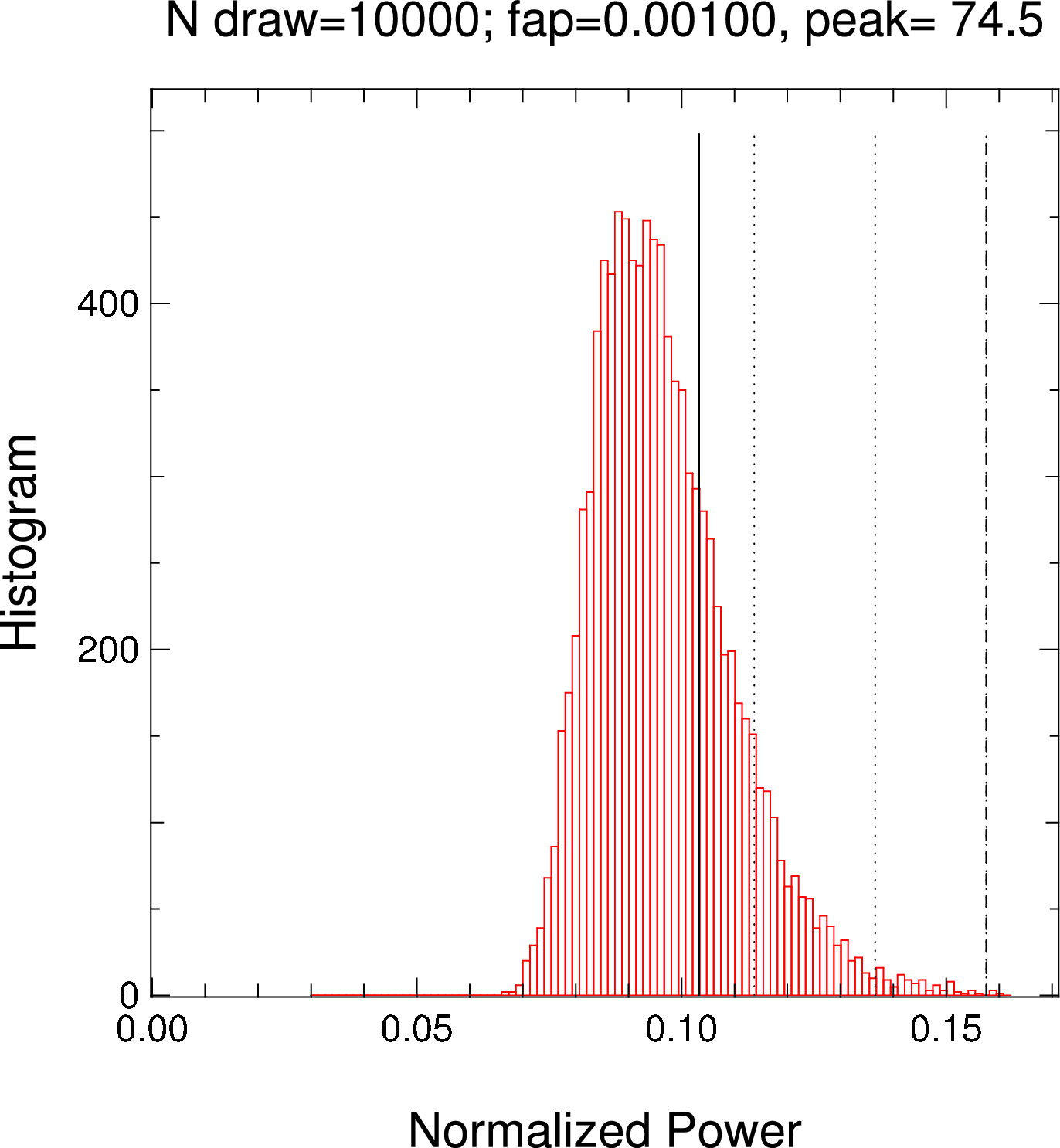}
\caption{Peak power distributions of the random permutations of the residuals to the successive multi-Keplerian models. The peak power of the actual data is shown as a full vertical line, while 10\%, 1\% and 0.1\% thresholds are shown as dashed lines. From left to right and top to bottom, the peak power corresponds to periods of 50 d, 5.8 d, 122 d, 2200 d, 16 d, 600 d and 1.18 d. The first 5 signals have extremely high significance, while the 6th and 7th signals have 0.6\% and 1.4\% FAP, respectively. Nothing significant remains after subtracting the 7-Keplerian model.}
\label{FigDistributions}
\end{figure*}

The raw rms dispersion of the radial velocities is 6.42\,m\,s$^{-1}$, well above instrumental errors and the expected stellar jitter, hinting at the presence of planets orbiting HD~10180. We proceed to an analysis of the data using the {\em Yorbit} software, an analysis package for radial velocity, astrometric and photometric data featuring a genetic algorithm and various tools for exoplanet search (S\'egransan et al. 2011, in prep.). We perform successive multi-Keplerian fits to the RV data, and use Lomb-Scargle periodograms to search for periodic signals of significant amplitude in the residuals to the successive models. We use the generalized Lomb-Scargle (GLS) periodogram as described in \citet{zechmeister09}. False-alarm probabilities (FAPs) are computed by performing random permutations of the data, keeping the observing times fixed, and computing their periodograms. For each trial the peak power is recorded, and then the power of the real signal is compared to the peak power distribution of the permuted datasets. Each time a signal is considered significant, we include it in the multi-Keplerian model and proceed further, assuming the radial velocity signals are due to orbiting planets. To do so, a genetic algorithm is run to efficiently explore the parameter space around suspected orbital periods. Once the population of solutions has converged towards the single best-fitting region of parameter space, a final Levenberg-Marquardt minimization is performed to reach the deepest $\chi^2$ minimum found.

Identification of signals in periodograms is sometimes ambiguous due to the presence of aliases, induced by the non-random time sampling of the observations. This problem has been long known in several fields of astrophysics dealing with ground-based time series analysis, and is also relevant to radial velocity searches for exoplanets \citep[see][for a recent overview]{dawson10}. Aliases occur at frequencies separated from the true peak by a frequency difference at which the window function shows significant power. Fig.~\ref{FigWF} shows 3 close-up views of the window function of the HD~10180 measurements, centered around $f = 0$, $f = 1$ d$^{-1}$ and $f = 2$ d$^{-1}$. These are the regions which contain significant peaks and which could contribute aliases in the frequency range of interest. As expected, the most prominent features are found at $f$ = 0.0027 d$^{-1}$ (1 year), $f$ = 1.0027 d$^{-1}$ (1 sidereal day), $f$ = 1.0000 d$^{-1}$ (1 day), $f$ = 2.0027 d$^{-1}$ and $f$ = 2.0055 d$^{-1}$. In the following we will pay particular attention to possible aliases induced by these peaks.

Fig.~\ref{FigGLS} shows the successive GLS periodograms of the radial velocity data. At each step, the main peak is identified and considered significant if its false-alarm probability (FAP) is smaller than 10$^{-2}$. FAP thresholds of 10$^{-1}$, 10$^{-2}$ and 10$^{-3}$ are indicated by horizontal lines in Fig.~\ref{FigGLS}. The peak power distributions of the shuffled datasets are shown in Fig.~\ref{FigDistributions}. As can be seen from both figures, 5 very significant signals can be successively found in the data, at periods 50 d, 5.8 d, 122 d, 2200 d and 16 d. All the corresponding peaks are already clearly apparent in the periodogram of the raw data. Each of them has a FAP far below 10$^{-3}$, as can be seen from the peak power distributions. We successively subtract these signals by performing multi-Keplerian fits to the data, whose characteristics can be found in Table~\ref{TableModels}. The final, 5-Keplerian fit to the data yields periods of 5.760 d, 16.35 d, 49.74 d, 122.4 d and 2231 d for the 5 signals.

Various aliases of these signals are present in the periodograms. It turns out that the 5 successive highest peaks are always located at periods larger than 2 d, and in each case two other peaks potentially corresponding to their 1-d aliases are clearly seen. Given a highest-peak frequency $f_0$, these are found at $\left|\, f_0 \pm 1.0027 \,\mathrm{d}^{-1} \right|$ and show amplitudes similar to each other but lower than the main peak. Given the properties of the window function, this is the expected pattern if the true signal is the lower-frequency peak at $P$ $>$ 2 d. On the contrary, if the true signal was one of the two peaks near 1 d, the peak at $P$ $>$ 2 d would be a 1-d alias and the other high-frequency peak would be a 0.5-d alias (2.0055 d$^{-1}$). In this case, one would expect a large difference in amplitude between the two peaks near 1 d, and the low-frequency peak should be of intermediate strength. We checked the global pattern of aliases on the periodogram of the raw data for the 5 strong signals, extending the computation to frequencies around 2 d$^{-1}$. For each signal, we verified that the amplitude envelope outlined by the relevant peaks (the main peak plus its supposed 1-d and 0.5-d aliases) corresponds to the amplitude envelope in the window function. In each case we find a symmetric amplitude pattern centered on the lower-frequency peak and thus we conclude that, for the 5 strong signals seen in the raw data, the correct periods are the "long" ones, and the forest of peaks around $P$ = 1 d are aliases.

In summary, we obtain a 5-Keplerian fit with periods 5.760 d, 16.35 d, 49.74 d, 122.4 d and 2231 d. Table~\ref{TableModels} lists the statistical characteristics of the successive multi-Keplerian models.

\begin{table}
\caption{Characteristics of the successive multi-Keplerian models fitted to the data. $P$ is the period of the last signal added to the model.}
\label{TableModels}
\centering
\begin{tabular}{l c c c c c}
\hline\hline
Model & $P$ & $N_{\mathrm{free}}$ & $\chi^2$ & $\chi_r^2$ & $\sigma$(O-C) \\
 & [days] & & & & [m\,s$^{-1}$] \\
\hline
constant & - & 1 & 5741.6 & 30.38 & 6.42 \\
k1 & 50 & 6 & 4065.9 & 22.10 & 5.45 \\
k2 & 5.8 & 11 & 2568.5 & 14.35 & 4.29 \\
k3 & 122 & 16 & 1540.1 & 8.85 & 3.27 \\
k4 & 2200 & 21 & 824.0 & 4.88 & 2.36 \\
k5 & 16 & 26 & 356.4 & 2.17 & 1.57 \\
k6 ($e_5$=0) & 600 & 29 & 276.0 & 1.71 & 1.39 \\
k7 ($e_1$=$e_6$=0) & 1.18 & 32 & 237.1 & 1.50 & 1.27 \\
\hline
\end{tabular}
\end{table}

\subsection{The 600-d signal}

After subtraction of these 5 signals, the periodogram of the residuals still contains appreciable power, with peaks at $P$ = 600 d, 227 d and around 1 d. One immediately sees that the first two peaks are aliases of each other with the 1-year frequency ($1/600 + 1/365 \simeq 1/227$). The FAP of the highest peak ($P$ = 600 d) is 0.6\%, and we thus consider it as significant. Given that the spectral window has relatively high sidelobes at the 1-year frequency, it is not surprising that a fraction of the power is leaking into the 1-year aliases. Here we assume that the correct period is not one of the peaks around 1 d because it is highly unlikely to find a planet with an orbital period so close to 1 d (closer than 0.005 d). To determine which period (600 d or 227 d) is the most likely one, we proceed in the following way. We perform simulations in which we take the residuals from the full 6-Keplerian model (with either $P$ = 600 d or $P$ = 227 d as the 6th signal), shuffle them using permutations, inject an artificial signal at either $P$ = 600 d or $P$ = 227 d, and compare the overall pattern of peaks in the resulting periodograms to the observed one. The periods, amplitudes and phases of the injected signals are drawn from Gaussian distributions centered on the fitted values in the 6-Keplerian model. As a quantitative measure of the similarity of periodograms, we check whether the 3 highest peaks are the same as in the actual data. As actual data to compare to, we take the residuals to the 6-Keplerian model with the fitted 6th signal added (either at $P$ = 600 d or $P$ = 227 d). In this way we compare the simulated data to the 5-Keplerian residuals that are closest to reality under the two assumptions (the 600-d or the 227-d peak is the correct one). We take all these precautions because we are dealing with data on which 5 signals have already been subtracted, and the exact choice of parameters for these 5 signals has a significant impact on the alias pattern of the 6th signal.

The results of the simulations are as follows: when injecting a 600-d signal, the 3 highest peaks seen in the actual data (600 d, 0.9956 d and 0.9989 d) are correctly reproduced in 53\% of the simulated periodograms. When injecting a 227-d signal, the 3 highest peaks (227 d, 1.0017 d and 0.9956 d) are recovered in only 1.3\% of the cases. We thus conclude that a 600-d signal is much more likely to correctly reproduce the data than a 227-d signal.

We proceed to fit a 6-Keplerian model to the data, which yields $\chi_r^2$ = 1.71 and $\sigma$(O-C) = 1.39\,m\,s$^{-1}$, compared to $\chi_r^2$ = 2.17 and $\sigma$(O-C) = 1.57\,m\,s$^{-1}$ for the 5-Keplerian model. We check whether the eccentricity for the 6th, lowest-amplitude signal is constrained by the data by fixing it to zero, refitting and looking at the reduced $\chi^2$. With $\chi_r^2$ = 1.71, as before, the addition of more free parameters in the model is not justified and we adopt the zero-eccentricity solution. Finally, it is also worth mentioning that the 6-Keplerian solution with $P$ = 600 d has a slightly better $\chi_r^2$ (1.71) than the corresponding one with $P$ = 227 d (1.82), reinforcing the case for the longer period.

\subsection{A potential 7th signal}

Continuing the process one step further, we notice two more fairly strong peaks in the periodogram of the residuals to the 6-Keplerian model (see Fig.~\ref{FigGLS}). These are located at periods $P$ = 6.51 d and $P$ = 1.178 d. Again, one of them is clearly the alias of the other one, this time with the 1 sidereal day period ($\left| 1/6.51 - 1.0027 \right| \simeq 1/1.178$). After 10,000 random permutations of the residuals, we obtain a FAP of 1.4\% for the higher peak ($P$ = 6.51 d). This is slightly too high to confidently claim a detection, but it is nevertheless intriguing. We proceed to fit this possible 7th signal, but first the correct period has to be determined. As before, we perform simulations by injecting artificial signals into the 7-Keplerian residuals at both periods. When signals are added at $P$ = 1.178 d, the 3 highest peaks found in the actual data (6.51 d, 1.178 d, 1.182 d) are reproduced in only 0.2\% of the cases. With signals injected at $P$ = 6.51 d, the 3 highest peaks (6.51 d, 1.178 d and 1.182 d, as before) are recovered in just 0.7\% of the simulated periodograms. This slightly favors the 6.51-d period, but only marginally so. Above all, the very low "success" rate of the simulations seems to indicate that this method cannot reliably decide which peak is the correct one. We also note that as much as 11\% of the simulated periodograms with an injected signal at $P$ = 1.178 d yield a highest peak at $P$ = 6.51 d. It would therefore be very speculative to draw any conclusion from the fact that the observed highest peak is at $P$ = 6.51 d. In summary, it is likely that the present data are simply not sufficient to distinguish between aliases. The strong 1-d aliases in the window function (86\% of the main peak) are the main obstacles to overcome, which means that several data points spread within nights will be required in the future to resolve this issue.

In the meantime, we use another kind of argument: if this 7th signal is caused by an orbiting planet, then it is very unlikely that the system would be dynamically stable with two objects at $P$ = 5.76 d and $P$ = 6.51 d, especially considering the 13-$M_\oplus$ minimum mass of the former planet (see Sect.~\ref{SectPlanets} for further discussion on this point). We thus adopt $P$ = 1.178 d as the only viable period from a dynamical point of view. The 7-Keplerian model, with eccentricities of the two lowest-amplitudes signals fixed to zero, has $\chi_r^2$ = 1.50 and $\sigma$(O-C) = 1.27\,m\,s$^{-1}$.

As a last step, an inspection of the periodogram of the residuals to the 7-Keplerian model does not reveal any peak with a FAP below 10\%, thus ending the search for signals in the RV data.

In summary, we firmly detect 6 signals with periods 5.760 d, 16.36 d, 49.75 d, 122.7 d, 602 d and 2248 d. A 7th signal with $P$ = 1.178 d has FAP 1.4\% and requires confirmation. The 6-Keplerian model has 29 free parameters for 190 data points, and $\chi_r^2$ = 1.71. This value goes down to 1.50 for the 7-Keplerian model. The rms dispersion of the residuals is 1.39\,m\,s$^{-1}$, down to 1.27\,m\,s$^{-1}$ in the 7-Keplerian model. We thus have a good fit to the data, confirming that the adopted stellar jitter value (1.0\,m\,s$^{-1}$) is reasonable.

\section{Origin of the radial velocity signals}

\begin{figure*}
\centering
\includegraphics[bb=0 70 595 375,width=75mm,clip]{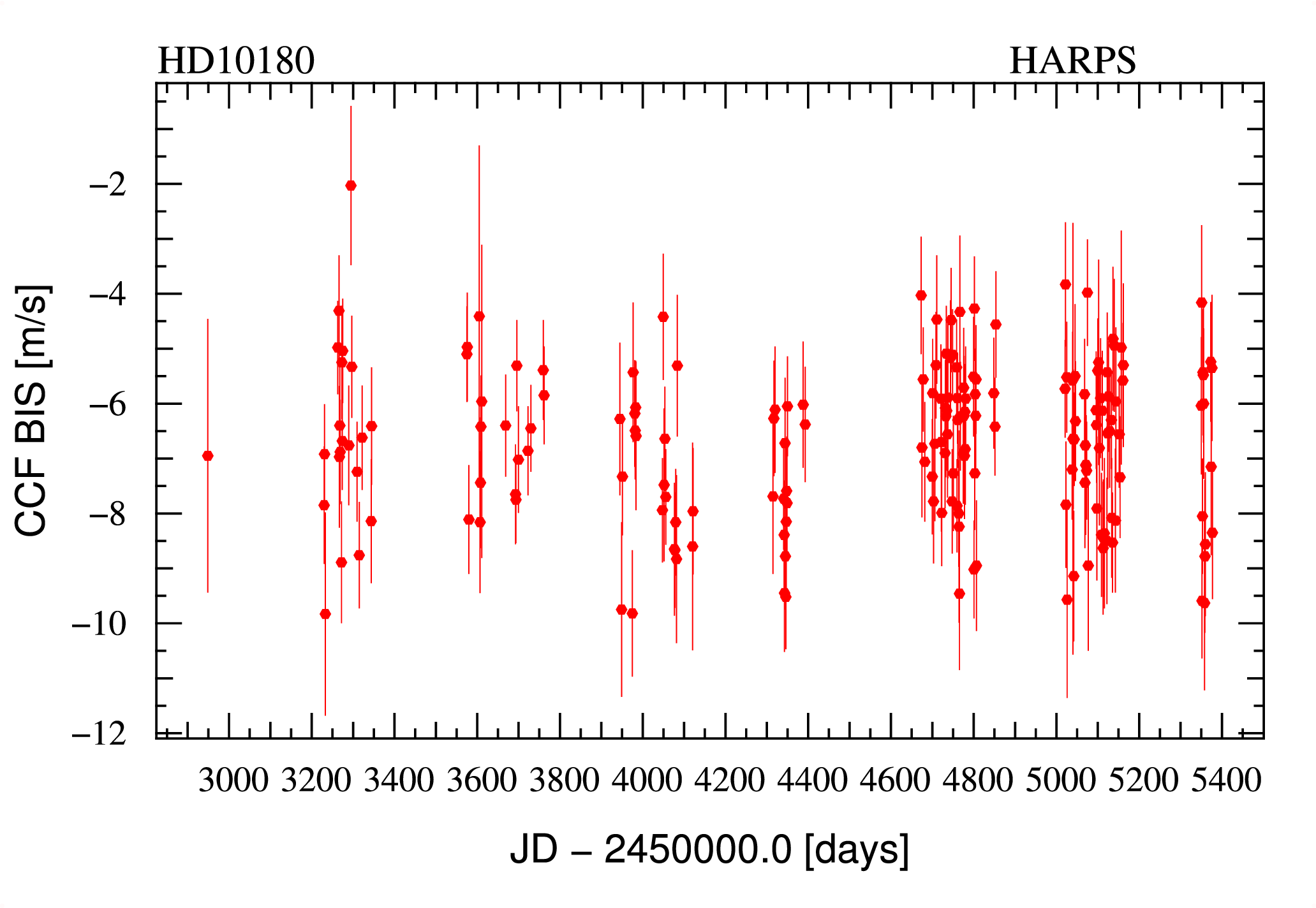}
\includegraphics[bb=0 55 595 257,width=95mm,clip]{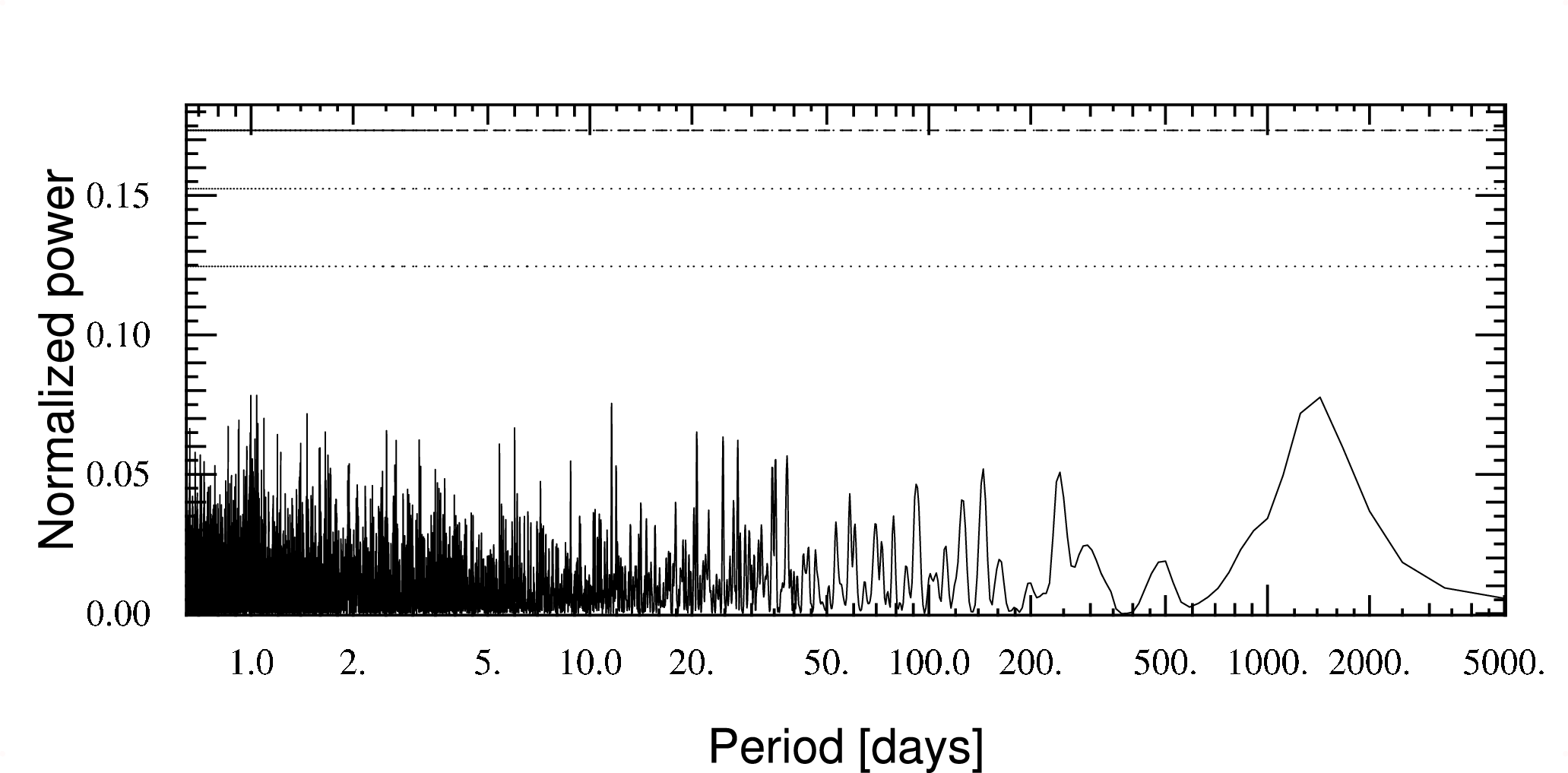}
\includegraphics[bb=0 70 595 375,width=75mm,clip]{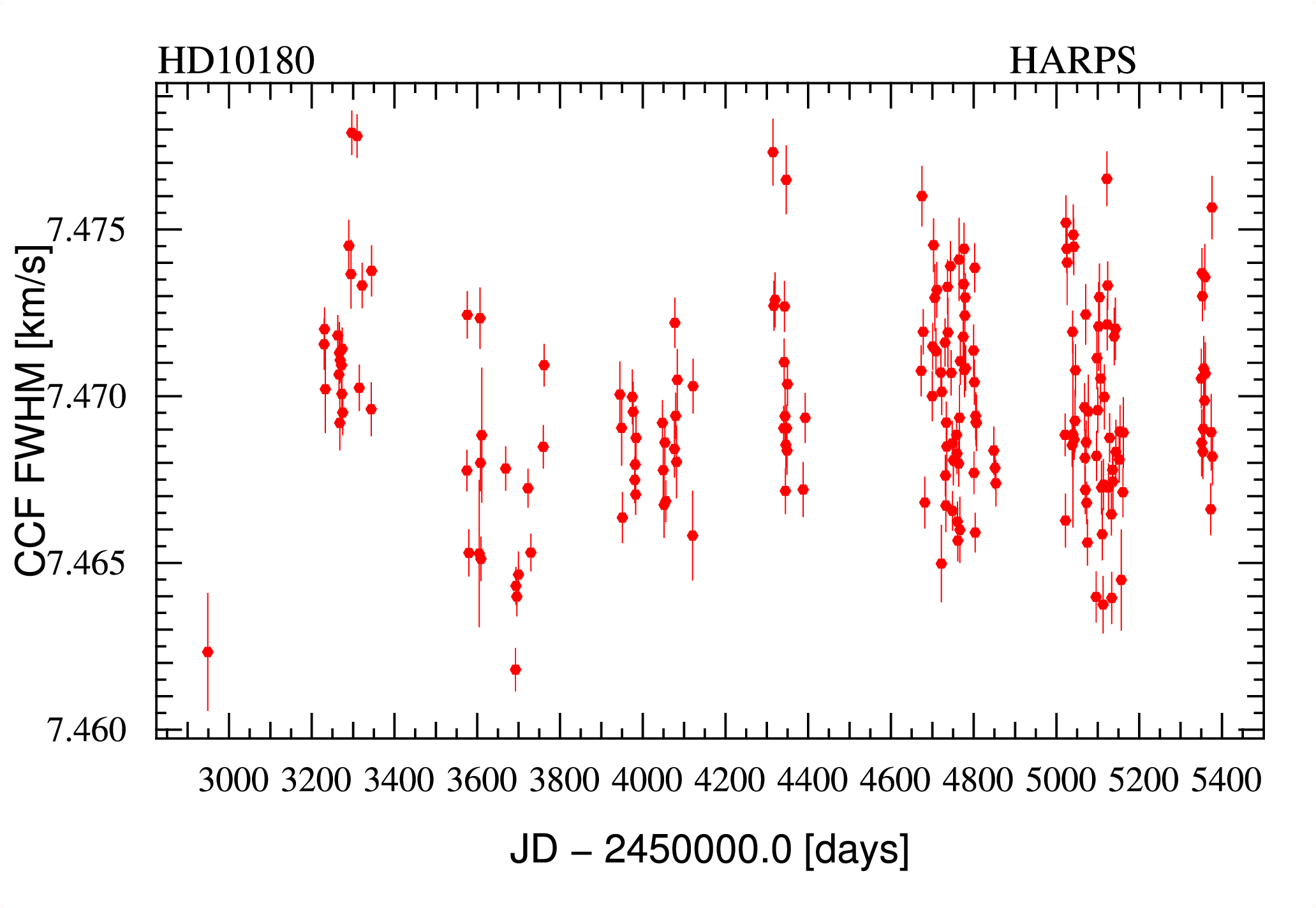}
\includegraphics[bb=0 55 595 257,width=95mm,clip]{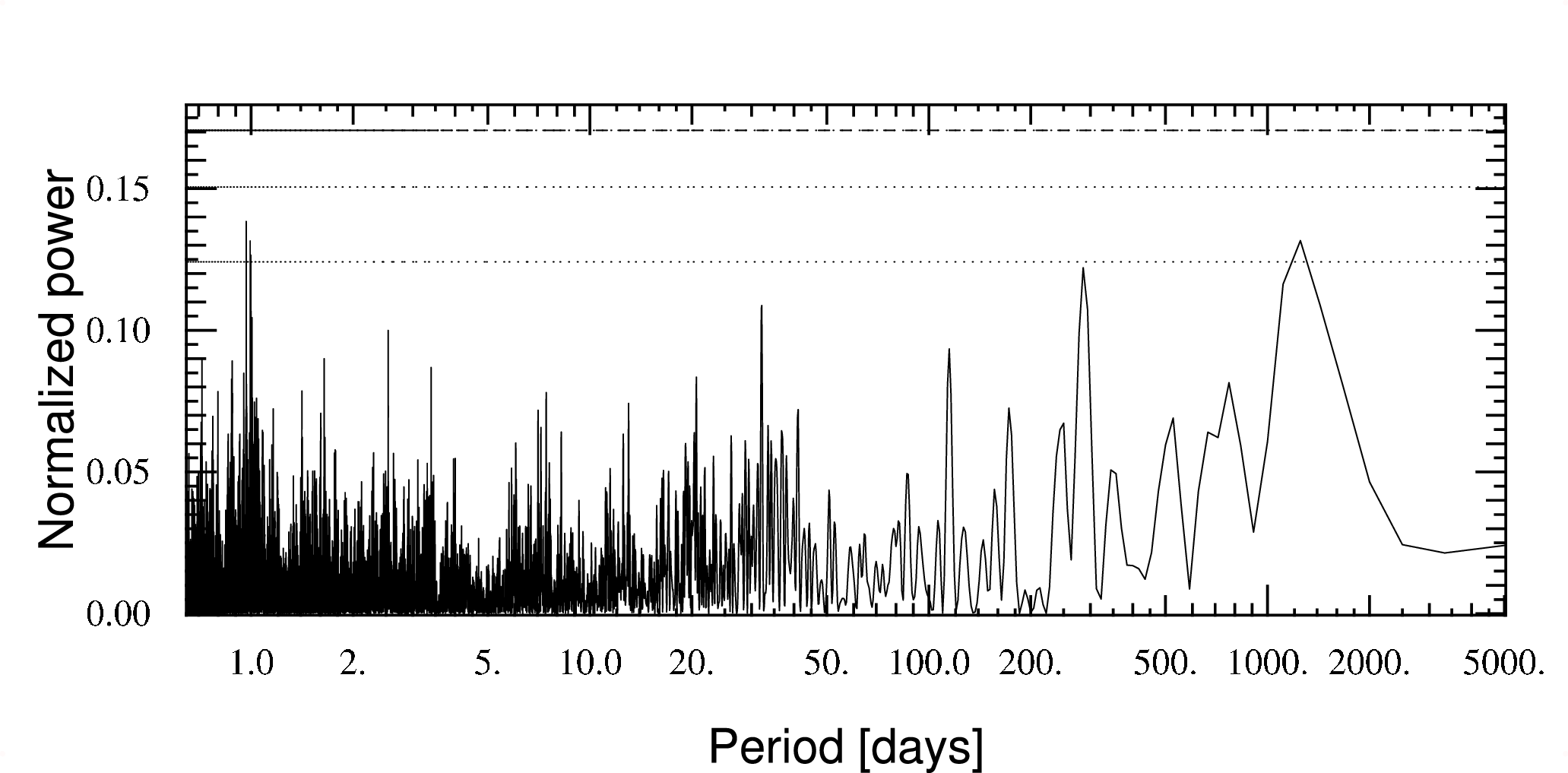}
\includegraphics[bb=0 0 595 375,width=75mm,clip]{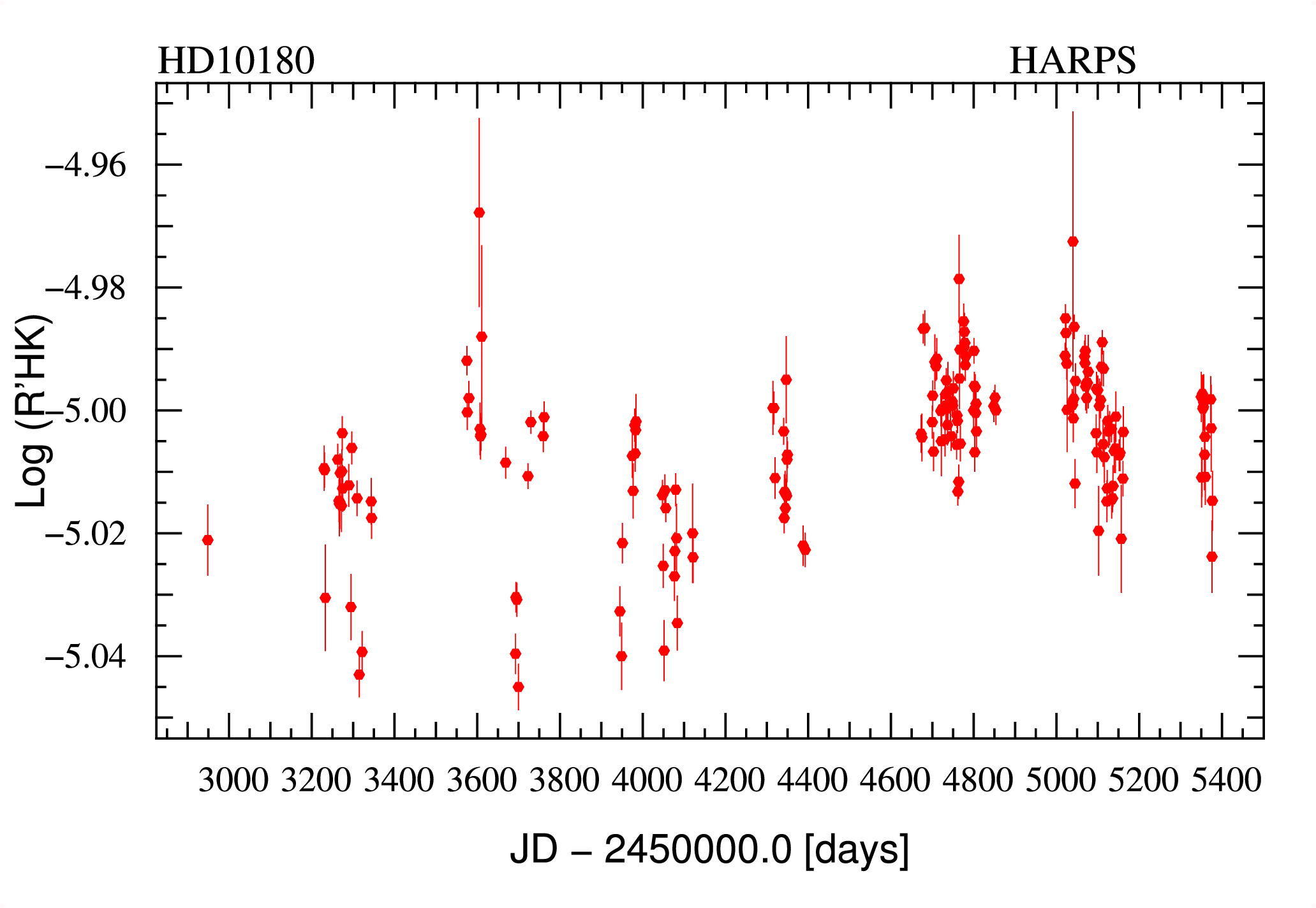}
\includegraphics[bb=0 0 595 257,width=95mm,clip]{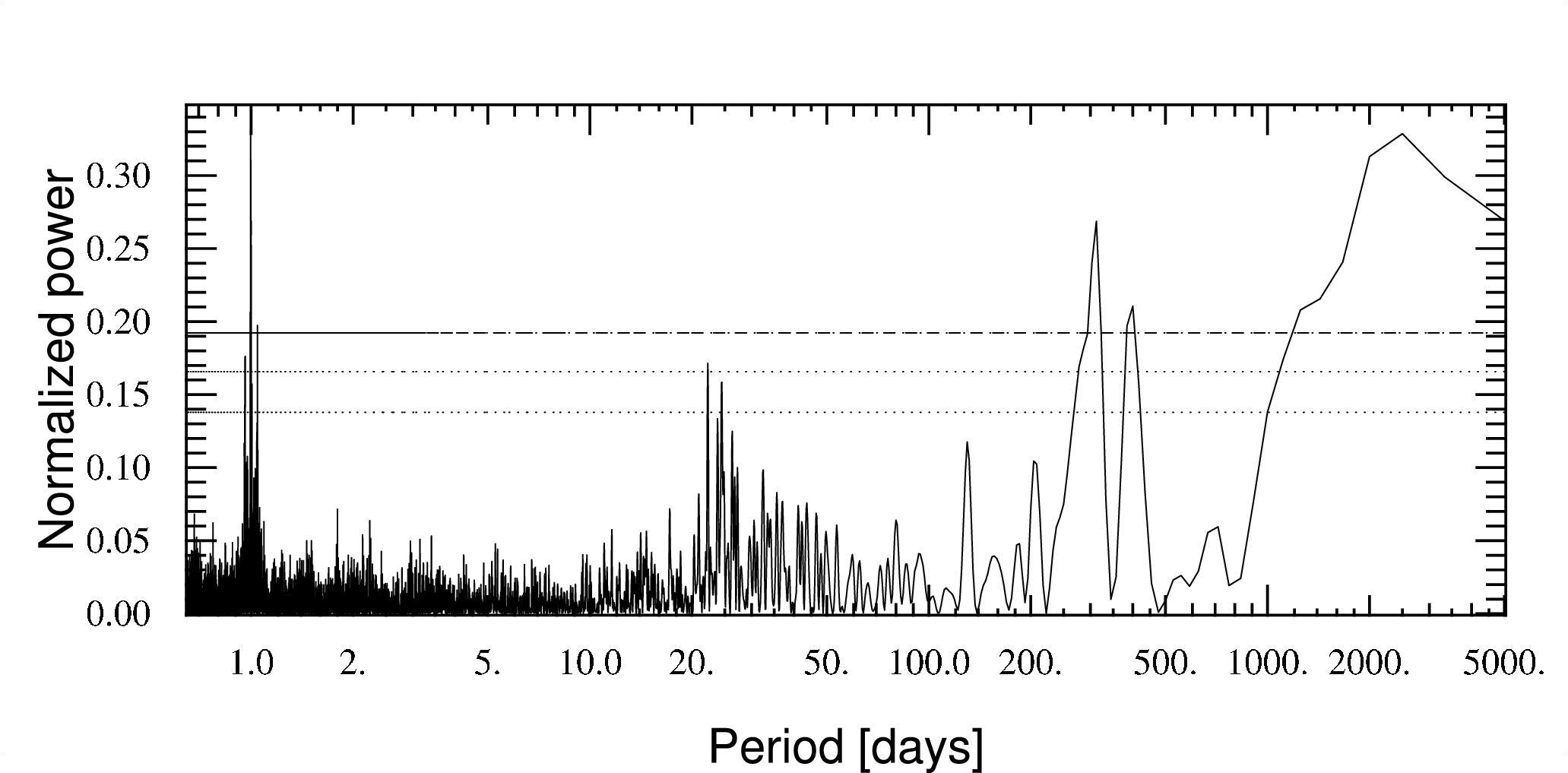}
\caption{Time series (left) and GLS periodogram (right) for the bisector velocity span, CCF FWHM and activity index $\log{R'_{\mathrm{HK}}}$ (from top to bottom).}
\label{FigIndicators}
\end{figure*}

So far we have assumed that all RV signals in the data are caused by orbiting planets. Obviously, with the small semi-amplitudes in play (between 0.81 and 4.5\,m\,s$^{-1}$), this assumption should be further verified. For this purpose we study the behaviour of several spectroscopic indicators: bisector velocity span, FWHM of the cross-correlation function and CaII activity index $\log{R'_{\mathrm{HK}}}$. Time series of these indicators and the corresponding GLS periodograms are shown in Fig.~\ref{FigIndicators}.

\begin{figure}
\centering
\includegraphics[bb=30 100 500 496,width=85mm,clip]{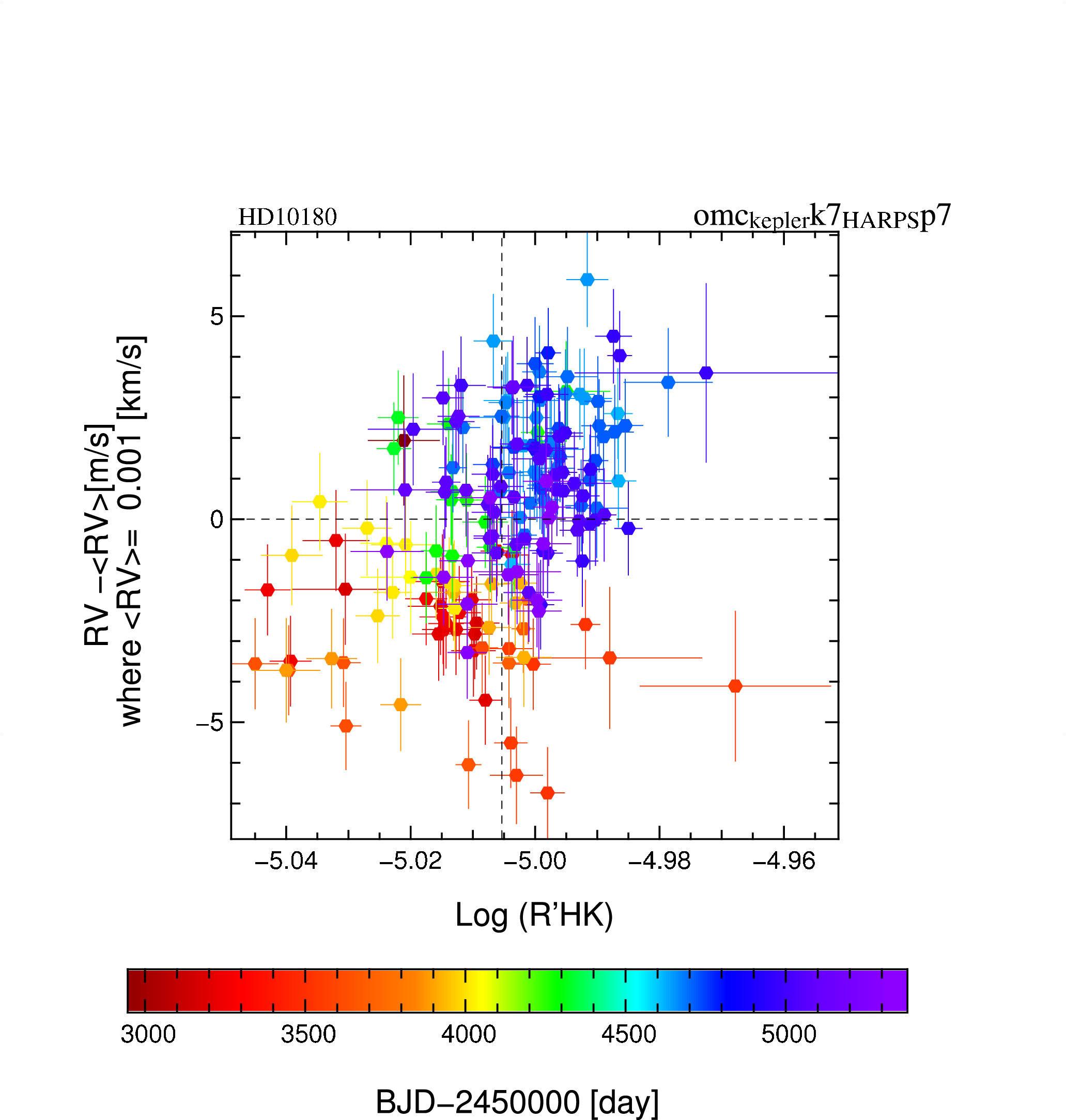}
\caption{Long-period radial velocity signal ($P$ = 2248 d) as obtained from the 7-Keplerian model, as a function of $\log{R'_{\mathrm{HK}}}$. A weak correlation seems to be present between the two quantities ($R$ = 0.44).}
\label{FigCorrelation}
\end{figure}

\subsection{Bisector velocity span and CCF FWHM}

The global rms dispersion of the bisector span is only 1.33\,m\,s$^{-1}$, a remarkable stability over more than 6 years. An analysis of the bisector periodogram reveals no significant power at any frequency, indicating a very quiet star. In particular, no power is seen around the estimated rotation period ($\sim$24 d) or at any of the 7 frequencies seen in the radial velocities. The same is true for the CCF FWHM, although some more structure seems to be present in the data. The FWHM periodogram shows no peak with a FAP below 3\%.

\subsection{CaII activity index}

We now turn to the activity index $\log{R'_{\mathrm{HK}}}$. We see that a long-period modulation is present in these data, although of very low amplitude (a few 0.01 dex) and at the same level as the short-term scatter. This behaviour is not typical of Sun-like magnetic cycles, which have long-term amplitudes an order of magnitude larger. The peak-to-peak variations here are only 0.06 dex, compared to 0.2-0.3 dex for typical activity cycles in solar-type stars. Also, the data seem to show a modulation only in the second half of the observing period. Thus, HD~10180 does not presently show an activity cycle like the Sun and, with an average $\log{R'_{\mathrm{HK}}}$ of -5.00 over 6 years, seems to be in a very quiet state. Its fundamental parameters do not indicate a subgiant status, which could have explained the low and stable activity level. More likely, HD~10180 either has a magnetic cycle with a period much longer than 10 years, or it is in a relatively quiet phase of its main-sequence lifetime, with sporadic, weak variations in activity level. If the latter explanation is true, we might be witnessing an activity state similar to the Maunder minimum of the Sun in the XVIIth century.

As shown by Lovis et al. (2011, in prep.), magnetic cycles do induce RV variations in solar-type stars, at a level that depends on spectral type. Studying a large sample of stars observed with HARPS, they were able to measure the degree of correlation between activity level and radial velocities. It turns out that the sensitivity of RVs to activity variations increases with increasing temperature, early-G dwarfs being much more affected than K dwarfs, which are almost immune to this phenomenon \citep{santos10}. In the case of HD~10180, the $\log{R'_{\mathrm{HK}}}$ periodogram exhibits a very significant peak around 2500 d, as can already be guessed by eye from the time series (see Fig.~\ref{FigIndicators}). A comparison with the raw radial velocity curve shows that the long-period RV signal at $P$ $\sim$ 2200 d might be correlated to the $\log{R'_{\mathrm{HK}}}$ signal. Fig.~\ref{FigCorrelation} shows this long-period RV signal (plus residuals), as obtained from the 7-Keplerian model, as a function of $\log{R'_{\mathrm{HK}}}$. The Pearson's weighted correlation coefficient $R$ is 0.44, indicating a weak positive correlation between the two quantities. The measured slope of the RV-$\log{R'_{\mathrm{HK}}}$ relation is 0.92 $\pm$ 0.13\,m\,s$^{-1}$ per 0.01 dex. This number, although not well constrained, is compatible with the predicted sensitivity to activity of HD~10180. Indeed, for $T_{\mathrm{eff}}$ = 5911 K, the activity-RV relation of Lovis et al. (2011, in prep.) gives 0.99\,m\,s$^{-1}$ per 0.01 dex. It is thus possible that the long-period RV signal is not of planetary origin, but the result of the varying fraction of the stellar surface covered by magnetic regions over time. However, HD~10180 does not show a typical solar-type magnetic cycle, and the quality of the correlation is quite poor. For example, the activity data around JD=53700 exhibit a large scatter instead of a uniformly low value, as would be expected from the RV values in the case of an activity-induced signal. Moreover, the CCF FWHM does not show a clear correlation with $\log{R'_{\mathrm{HK}}}$, in contrast to the stars studied by Lovis et al. (2011, in prep.). Also, the fitted semi-amplitude of the RV signal is 3.11\,m\,s$^{-1}$ while the fitted semi-amplitude on the $\log{R'_{\mathrm{HK}}}$ data is 0.011 dex, which appears to be a factor $\sim$3 too low to account for the RV signal considering a sensitivity of $\sim$1\,m\,s$^{-1}$ per 0.01 dex. Such a factor is large compared to the scatter in the Lovis et al. (2011, in prep.) relation and it thus seems unlikely that HD~10180 could be so sensitive to activity.

In conclusion, we favor the planetary interpretation for this long-period RV signal, although some doubt remains on its origin. Future observations will likely solve this issue. In particular, it will be interesting to see whether radial velocities follow a downward trend in the near future, as would be expected in the planetary case, or whether they follow the more chaotic variations of the activity index.

There is no indication that the 6 other RV signals might be due to stellar activity. None of them are related to the expected stellar rotation period (24 $\pm$ 3 d) or to its harmonics. We tried to detect the rotation period in the $\log{R'_{\mathrm{HK}}}$ data, but found no convincing signal. This confirms that HD~10180 is a very quiet star. Another strong argument in favor of the planetary interpretation is the very high significance of the signals in the periodograms. It is a clear indication that the signals are coherent in time, as expected from the clock-like signature of an orbiting planet. Activity-related phenomena like spots and plages have short coherence times in inactive stars (of the order of the rotation period), and are not able to produce such well-defined, high-significance peaks in a dataset spanning more than 6 years. We thus conclude that a planetary origin is the only viable interpretation for the 6 RV signals between 1 d and 600 d.

\section{The HD 10180 planetary system}
\label{SectPlanets}

\begin{table*}
\caption{Orbital and physical parameters of the planets orbiting HD 10180, as obtained from a 7-Keplerian fit to the data. Error bars are derived using Monte Carlo simulations. $\lambda$ is the mean longitude ($\lambda$ = $M$ + $\omega$) at the given epoch.}
\label{TablePlanetsHD10180}
\begin{center}
\begin{tabular}{l l c c c c c c c}
\hline \hline
\noalign{\smallskip}
{\bf Parameter}	& {\bf [unit]}		& {\bf HD 10180 b}	& {\bf HD 10180 c}	& {\bf HD 10180 d}	& {\bf HD 10180 e}	& {\bf HD 10180 f}	& {\bf HD 10180 g}	& {\bf HD 10180 h} \\
\hline 
\noalign{\smallskip}
Epoch		& [BJD]			& \multicolumn{7}{c}{2,454,000.0 (fixed)}    \\ 
$i$			& [deg]			& \multicolumn{7}{c}{$ 90 $ (fixed) }  \\  
$V$			& [km\,s$^{-1}$]		& \multicolumn{7}{c}{$ 35.53015\,(\pm 0.00045) $}  \\
\hline 
\noalign{\smallskip}
$P$			& [days]			& $1.17765$		& $5.75962$		& $16.3567$		& $49.747 $		& $122.72 $		& $602 $			& $2248 $ \\ 
			&				& $(\pm 0.00018)$	& $(\pm 0.00028)$	& $(\pm 0.0043)$	& $(\pm 0.024)$	& $(\pm 0.20) $		& $(\pm 11) $		& $(_{-106}^{+102}) $ \\ 
$\lambda$	& [deg]			& $ 186 $			& $ 238.1 $		& $ 197.9 $		& $ 102.3 $		& $ 251.4 $		& $ 321 $			& $ 238.8 $ \\ 
			&				& $(\pm 11) $		& $ (\pm 1.9)  $		& $(\pm 3.3)  $		& $(\pm 2.2)  $		& $(\pm 3.2)  $		& $(\pm 11) $		& $(\pm 4.1)$ \\ 
$e$			&				& $ 0.0 $			& $ 0.077 $		& $ 0.143 $		& $ 0.065 $		& $ 0.133 $		& $ 0.0 $			& $ 0.151 $ \\ 
			&				& (fixed)			& $(\pm 0.033)$	& $(\pm 0.058)$	& $(\pm 0.035)$	& $(\pm 0.066)$	& (fixed)			& $(\pm 0.072)$ \\ 
$\omega$		& [deg]			& $  0.0 $			& $ 279 $			& $ 292 $			& $ 174 $			& $ 265 $			& $ 0.0 $			& $ 194 $ \\ 
			&				& (fixed)			& $(_{-25}^{+63}) $	& $(_{-10}^{+42}) $	& $(_{-59}^{+63}) $	& $(_{-209}^{+78}) $	& (fixed)			& $(_{-55}^{+59}) $ \\ 
$K$			& [m\,s$^{-1}$]		& $   0.81 $		& $   4.54 $		& $   2.93 $		& $   4.25 $		& $   2.95 $		& $   1.56 $		& $   3.11 $ \\  
			&				& $(\pm 0.15)   $	& $(\pm 0.15)  $	& $(\pm 0.16)  $	& $(\pm 0.18)  $	& $(\pm 0.18)  $	& $(\pm 0.21)  $	& $(\pm 0.21)  $ \\
\hline
\noalign{\smallskip}
$m \sin i$		& [$M_\oplus$]		& $ 1.39 $			& $ 13.17 $		& $ 11.94 $		& $ 25.4 $			& $ 23.6 $			& $ 21.4 $			& $ 65.3 $ \\
			&				& $(\pm 0.26)   $	& $(\pm 0.63)  $	& $(\pm 0.75)  $	& $(\pm 1.4)  $		& $(\pm 1.7)  $		& $(\pm 3.0)  $		& $(\pm 4.6)  $ \\
$a$			& [AU]			& $ 0.02225 $		& $ 0.0641 $		& $ 0.1285 $		& $ 0.2699 $		& $ 0.4928 $		& $ 1.423 $		& $ 3.42 $ \\
			&				& $(\pm 0.00038)$	& $(\pm 0.0011)$	& $(\pm 0.0022)$	& $(\pm 0.0045)$	& $(\pm 0.0083) $	& $(\pm 0.030) $	& $(_{-0.13}^{+0.12}) $ \\
\hline
\noalign{\smallskip}
$N_\mathrm{meas}$ &			& \multicolumn{7}{c}{190}  \\
Span		& [days]			& \multicolumn{7}{c}{2428} \\
rms			& [m\,s$^{-1}$]		& \multicolumn{7}{c}{1.27} \\
$\chi_r^2$	&				& \multicolumn{7}{c}{1.50} \\
\hline
\end{tabular}
\end{center}
\end{table*}

\begin{figure}
\centering
\includegraphics[bb=0 0 595 490,width=85mm,clip]{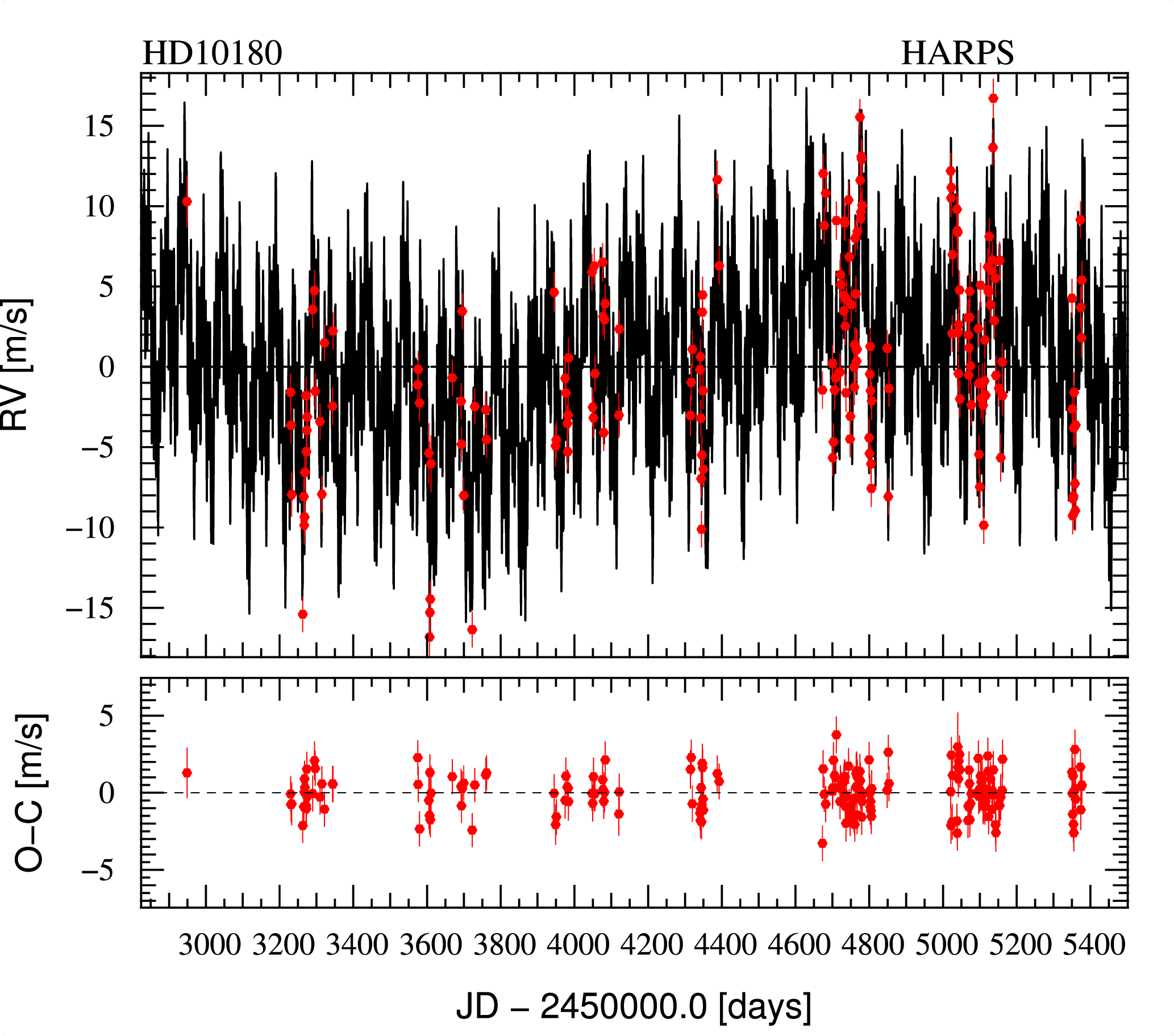}
\caption{Radial velocity time series with the 7-Keplerian model overplotted. The lower panel shows the residuals to the model.}
\label{FigRV}
\end{figure}

\begin{figure}
\centering
\includegraphics[bb=0 75 595 375,width=70mm,clip]{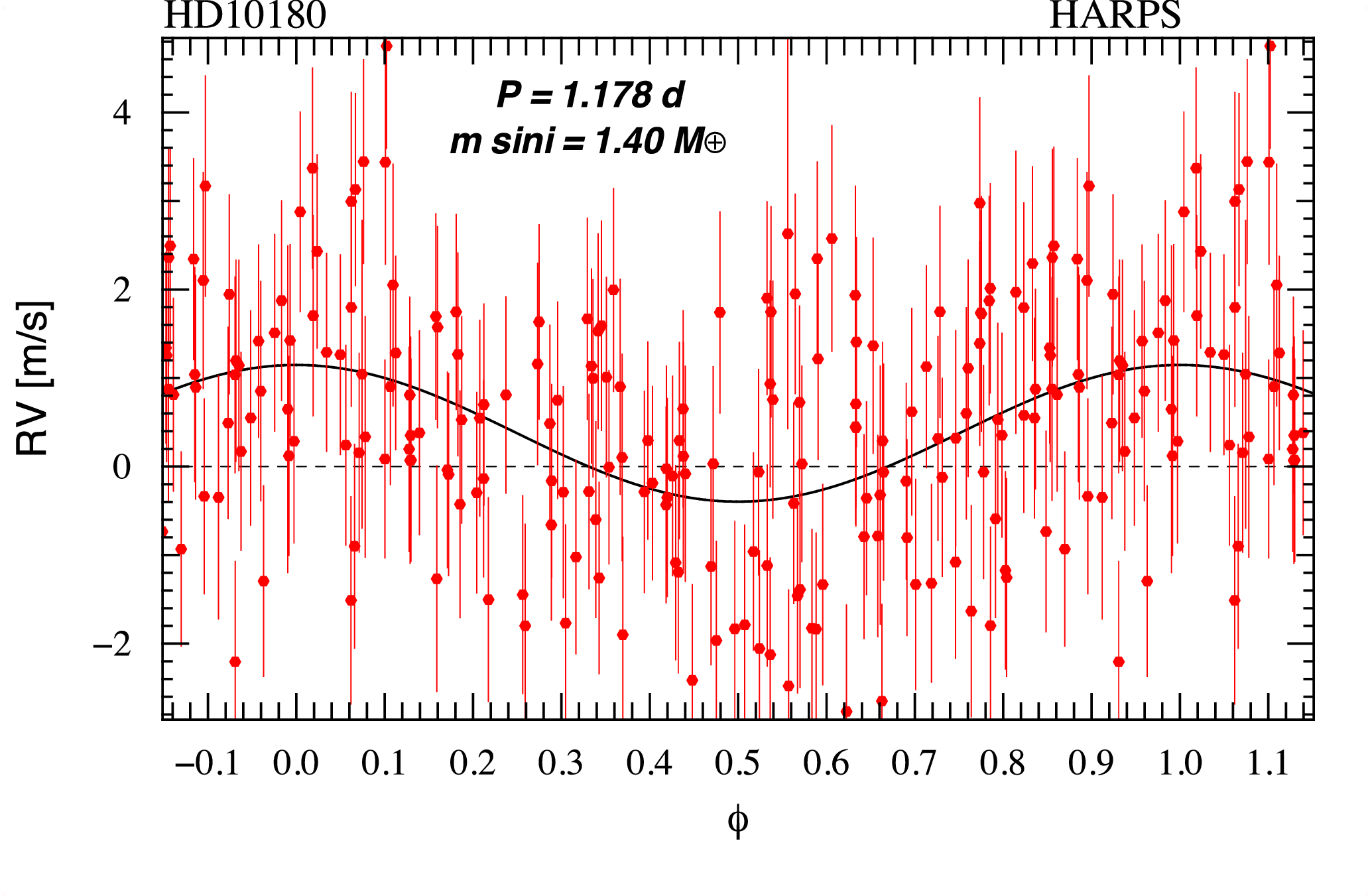}
\includegraphics[bb=0 75 595 375,width=70mm,clip]{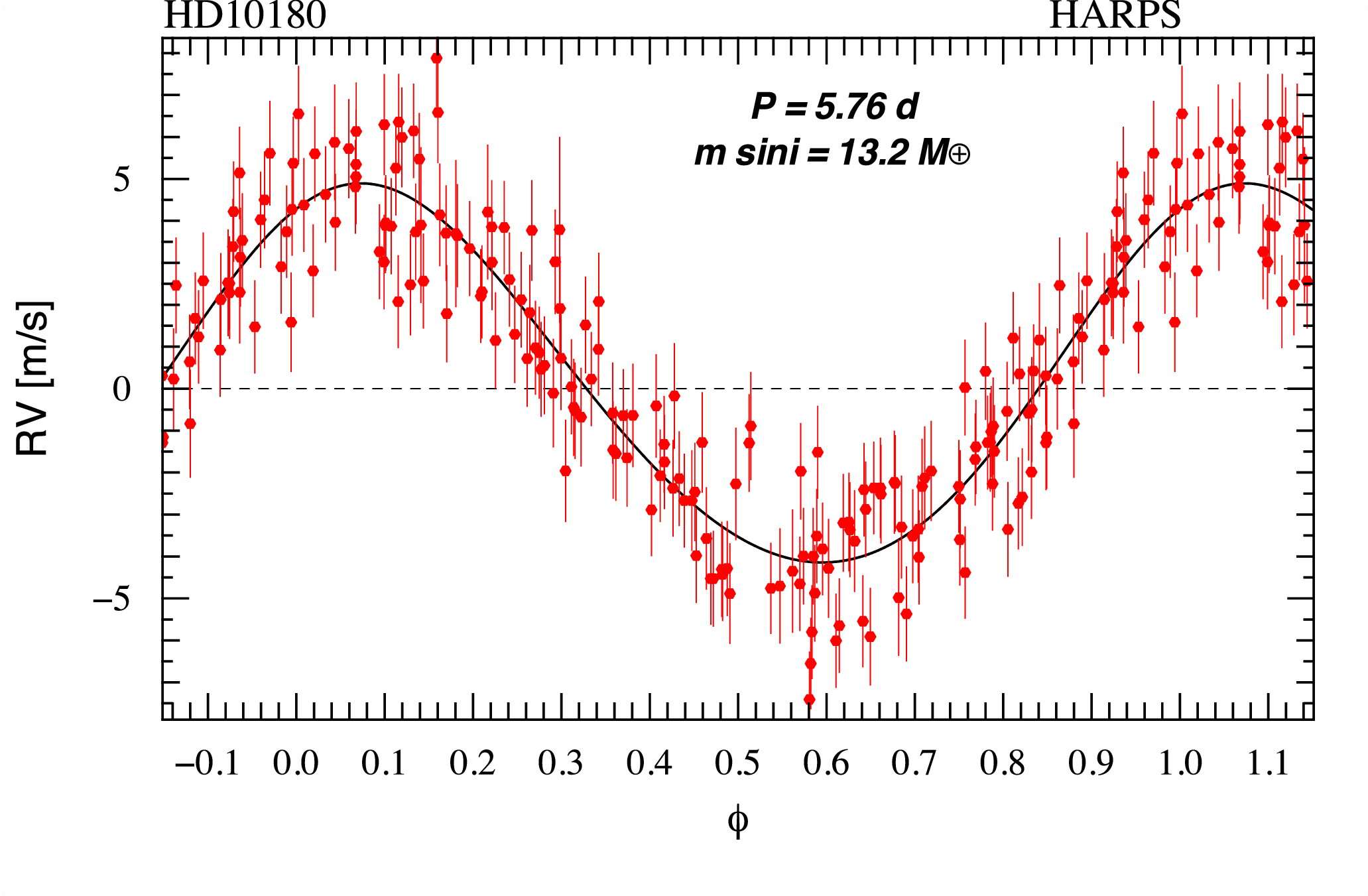}
\includegraphics[bb=0 75 595 375,width=70mm,clip]{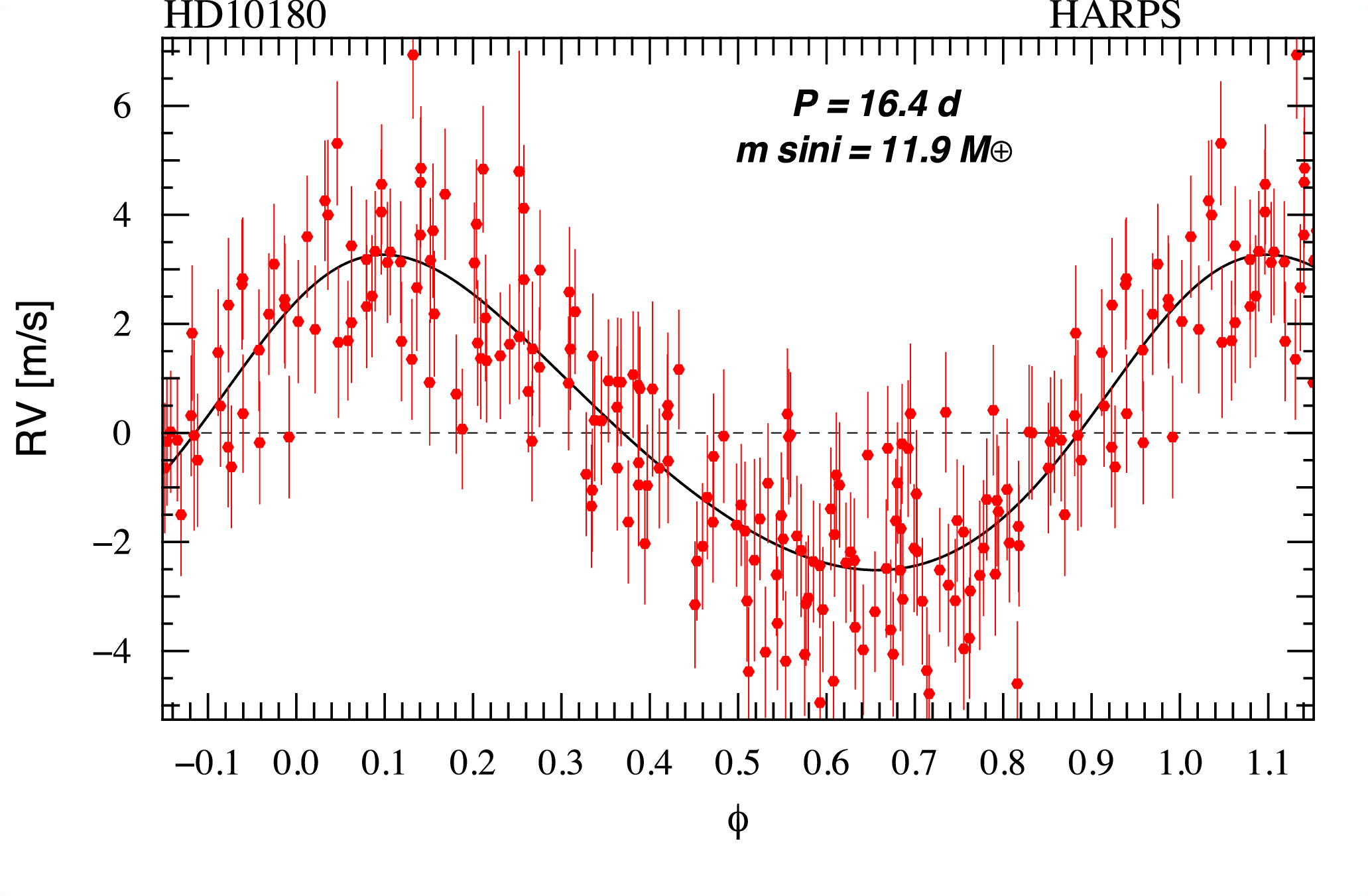}
\includegraphics[bb=0 75 595 375,width=70mm,clip]{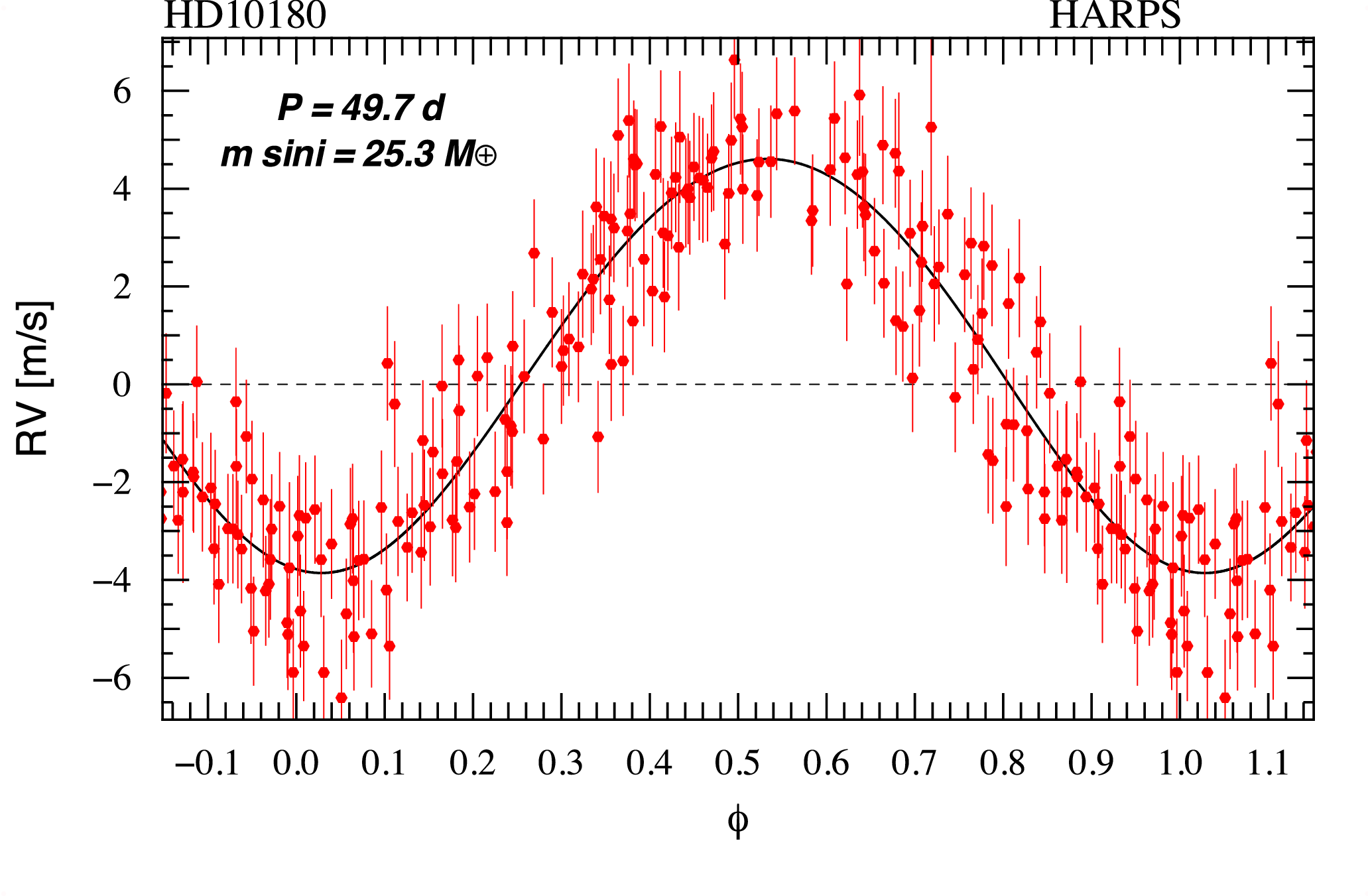}
\includegraphics[bb=0 75 595 375,width=70mm,clip]{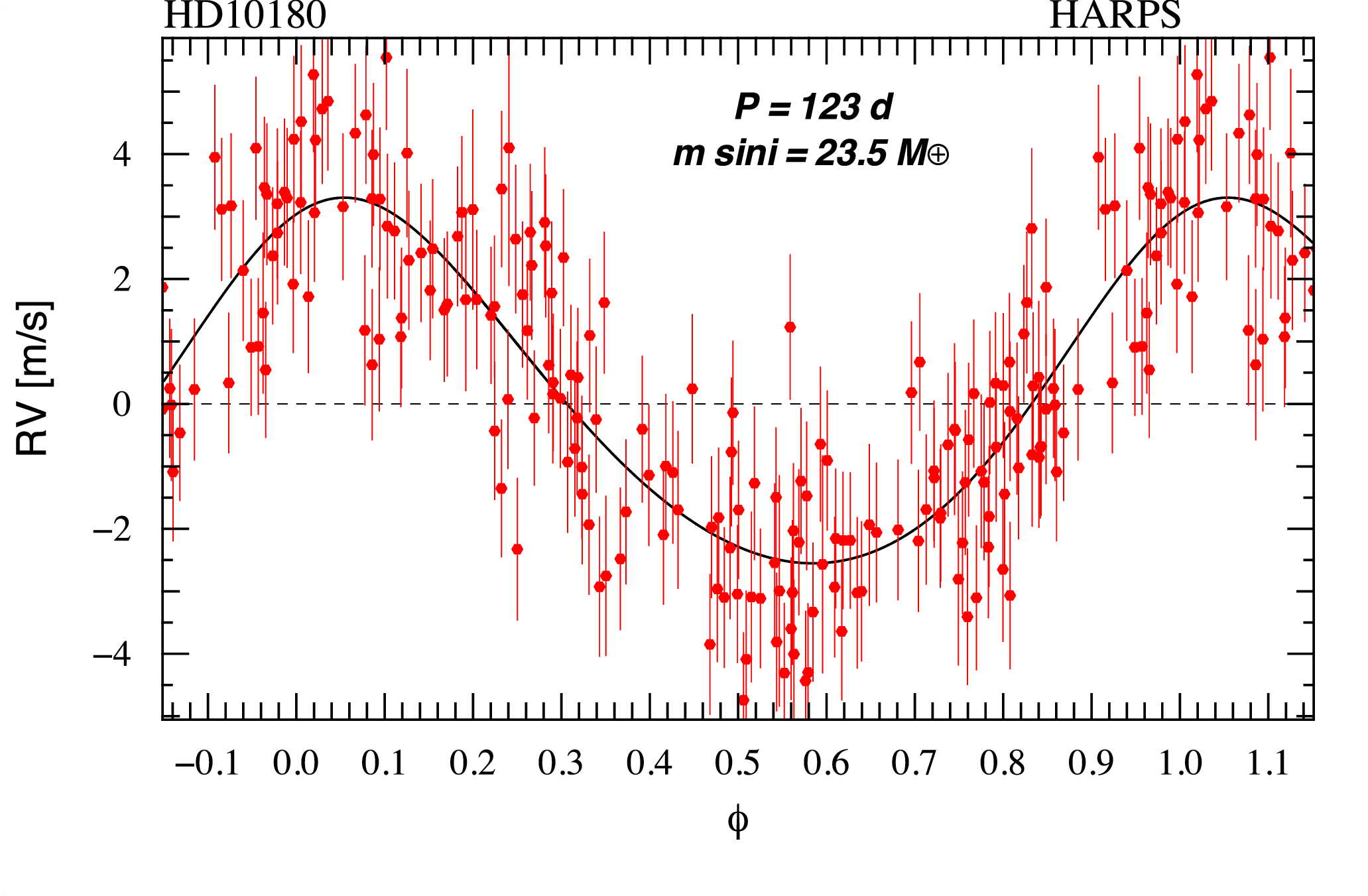}
\includegraphics[bb=0 75 595 375,width=70mm,clip]{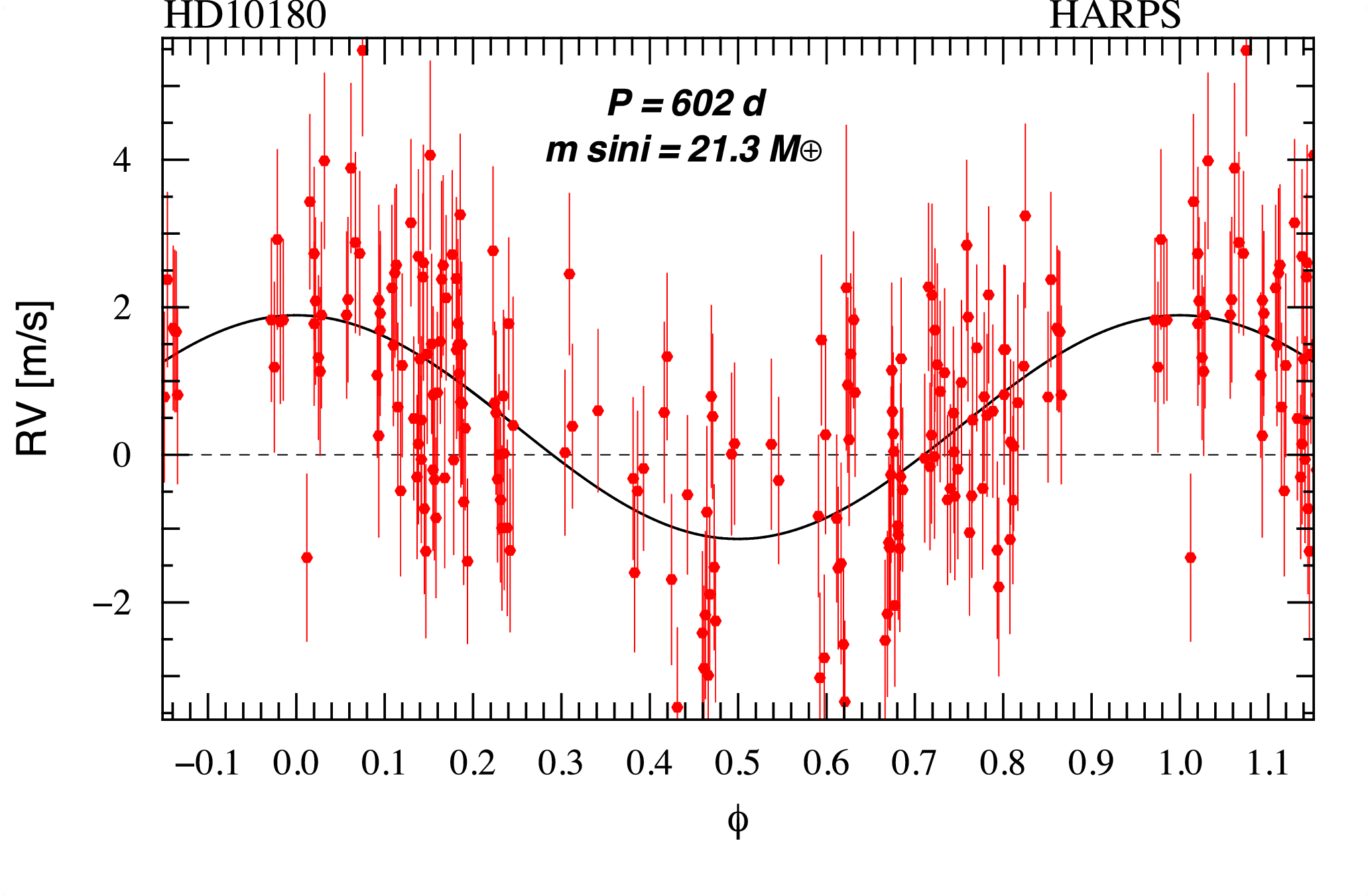}
\includegraphics[bb=0 20 595 375,width=70mm,clip]{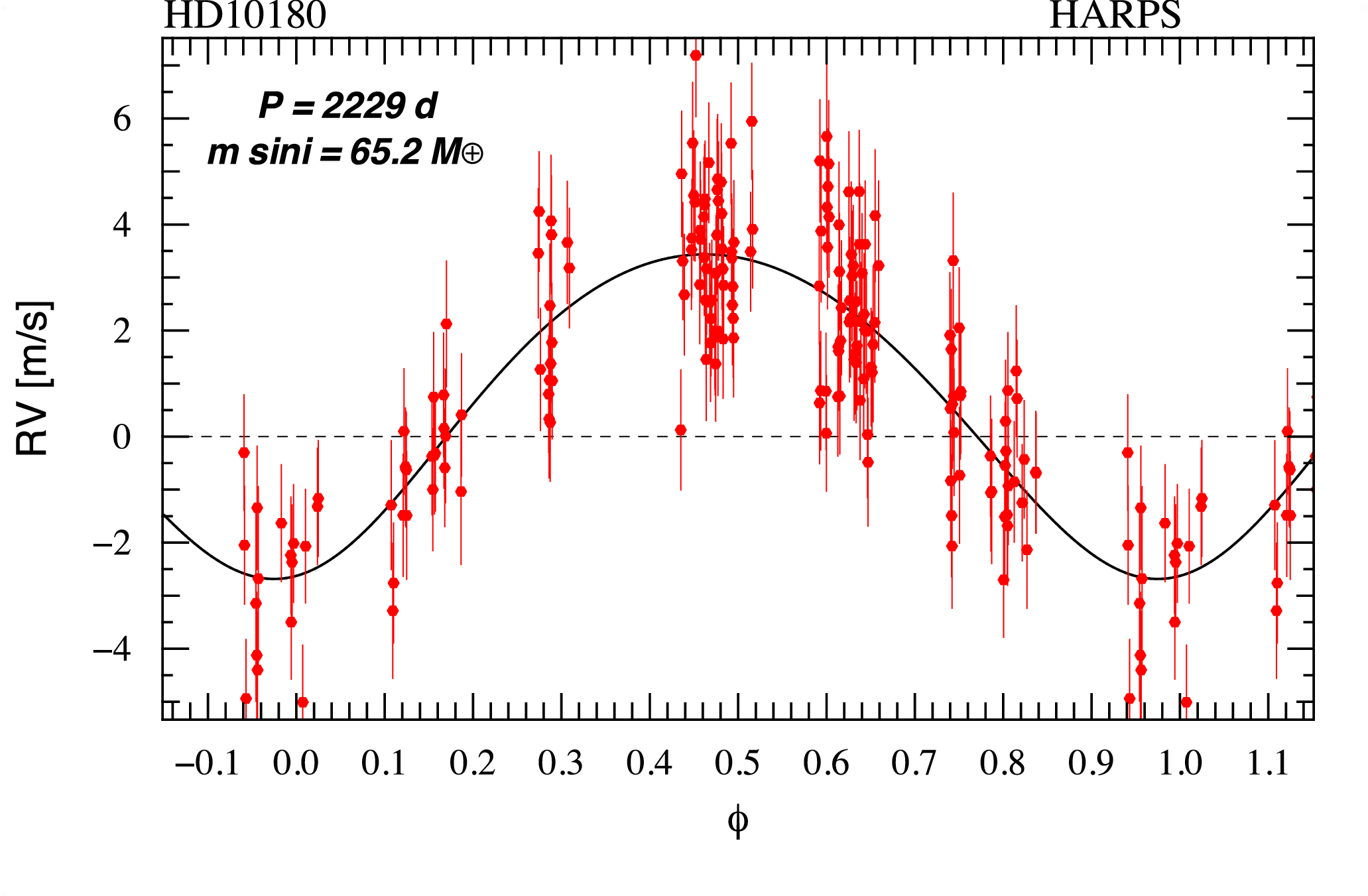}
\caption{Phased RV curves for all signals in the 7-Keplerian model. In each case, the contribution of the other 6 signals has been subtracted.}
\label{FigPhasedCurves}
\end{figure}

From now on we will assume that all 7 RV signals are real and of planetary origin. We thus keep the 7-Keplerian model - and the fit - as they were obtained in Sect.~\ref{SectSignals}. Table~\ref{TablePlanetsHD10180} gives the orbital parameters of all planets in the system. Fig.~\ref{FigRV} shows the full radial velocity curve as a function of time, while the phased RV curves for all signals are shown in Fig.~\ref{FigPhasedCurves}. Uncertainties on the fitted parameters have been obtained with Monte Carlo simulations, where actual data points are modified by drawing from a Gaussian distribution with a standard deviation equal to the error bar on each point. The modified datasets are then re-fitted, initializing the non-linear minimization with the nominal solution, and confidence intervals are derived from the obtained distributions of orbital elements.

The orbital solution given in Table~\ref{TablePlanetsHD10180} can be further improved by using N-body integrations and adding constraints on the orbital elements obtained from dynamical considerations (see Sect.~\ref{SectDynamics}). To decouple things, we first show here the solution obtained with a multi-Keplerian model and based on the RV data alone. Its main advantage is its much higher computational simplicity, which allows us to derive reliable error bars through Monte Carlo simulations.

The HD 10180 planetary system is unique in several respects. First of all, it presents five Neptune-like planets, orbiting between 0.06 and 1.4 AU from the central star. With minimum masses between 12 and 25 $M_{\oplus}$, this represents a relatively large total planetary mass within the inner region of the system, and multi-body migration processes are likely needed to explain this. Besides these Neptune-mass objects, the system has no massive gas giant. At most, it contains a small Saturn ($m \sin{i}$ = 65 $M_{\oplus}$) at 3.4 AU. In fact, the present data show no detectable long-term drift and allow us to exclude any Jupiter-mass planet within $\sim$10 AU for an edge-on system. At the other extreme of the mass and semi-major axis scales, the system also probably contains an Earth-mass object ($m \sin{i}$ = 1.4 $M_{\oplus}$) orbiting at only 0.022 AU. This is the planet with the lowest minimum mass found to date, and may represent another member of a hot rocky planet population that is starting to emerge (e.g. CoRoT-7b, GJ~581~e, HD~40307~b). The discovery of the HD~10180 planets highlights once again how diverse the outcome of planet formation can be.

\section{Dynamical analysis}
\label{SectDynamics}

With such a complex system at hand, various dynamical studies are in order. For the first time, an extrasolar planetary system comes close to the Solar System as far as the number of bodies involved is concerned. An obvious aspect to be checked is whether the fitted orbital solution is dynamically stable over Gyr timescales (the age of the star). The stability of such a system is not straightforward, in particular taking into account that the minimum masses of the planets are of the same order as Neptune's mass and the fitted eccentricities are relatively high when compared with the eccentricities of the planets in the Solar System. As a consequence, mutual gravitational interactions between planets in the {HD}\,10180 system cannot be neglected and may give rise to instability. This said, it must be noted that the fitted eccentricities of all planets are different from zero by less than 2.5\,$\sigma$ (their probability distributions are close to Gaussians), thus making it difficult to discuss this issue based on the RV data only. Considering the low radial velocity amplitudes induced by these objects, it will be challenging to better constrain these eccentricities in the future with the RV method.

\subsection{The secular planetary equations}

Over long times, and in absence of strong mean motion resonances, the variations of the planetary elliptical elements are well described by the secular equations, that is the equations obtained after averaging over the longitudinal motion of the planets \citep[see][]{Laskar_1990}. The secular system can even  be limited to its  first order and linear part, which is usually called the Laplace-Lagrange system \citep[see][]{Laskar_1990} which can be written using the classical complex notation $ z_k = e_k \mathrm{e}^{\mathrm{i} \omega_k} $ for $ k = b,\, c, \, ..., h $

\begin{equation}
{d\over dt}\left(\begin{array}{c} z_b\\ \vdots \\z_h \end{array}\right)= 
\mathrm{i} \LL
\left(\begin{array}{c} z_b\\ \vdots \\z_h \end{array}\right) \ .
\label{eq.laplag}
\end{equation}

This linear equation is classically solved by diagonalizing the real matrix $\LL$ through the linear transformation to the proper modes 

\begin{equation}
\left(\begin{array}{c} z_b\\ \vdots \\z_h \end{array}\right)= 
\SS
\left(\begin{array}{c} u_1\\ \vdots \\u_7 \end{array}\right) \ .
\label{eq.lape}
\end{equation}

\begin{table*}
\caption{The orthogonal matrix $\SS$ of transformation to the proper modes (Eq. \ref{eq.lape}).} 
\label{tab:S}
\begin{center}
\begin{tabular}{rrrrrrr} 
\hline
   0.993728744 &  -0.095494900 &   0.054659690 &  -0.019554658 &   0.003718962 &  -0.000102684 &   0.000014775 \\ 
  -0.105551000 &  -0.630354828 &   0.710320992 &  -0.288029266 &   0.063179205 &  -0.001823646 &   0.000265194 \\ 
   0.036827945 &   0.763187708 &   0.531792964 &  -0.347823478 &   0.111312700 &  -0.003535982 &   0.000525407 \\ 
  -0.002421676 &  -0.104900945 &  -0.438298758 &  -0.741307569 &   0.497011613 &  -0.017867723 &   0.002723007 \\ 
   0.000090033 &   0.008592412 &   0.132463357 &   0.496102923 &   0.857137172 &  -0.039224786 &   0.006228615 \\ 
  -0.000000038 &  -0.000001984 &  -0.000555340 &  -0.004533518 &  -0.043581594 &  -0.986552482 &   0.157461056 \\ 
  -0.000000002 &   0.000000033 &  -0.000012877 &  -0.000099417 &   0.000096168 &   0.157608762 &   0.987501625 \\ 
\hline
\end{tabular}
\end{center}
\end{table*}

Using the initial conditions of the fit in Table\,\ref{ta:tides}, we have  computed analytically the (real) Laplace-Lagrange matrix $\LL$, and derived the (real) matrix $\SS$ of its eigenvectors which gives the relation from the original eccentricity variables $z_k$ to the proper modes $u_k$. The matrix $\SS$ is given in Table~\ref{tab:S}. After the transformation (\ref{eq.lape}), the linear system (\ref{eq.laplag}) becomes diagonal 
\begin{equation}
{d\over dt}(u_k) = Diag(g_1, \dots g_7) (u_k)
\label{eq.diag}
\end{equation}
where $g_1,\dots,g_7$ are the eigenvalues of the Laplace-Lagrange matrix $\LL$. The solution is then trivially given for all $k=1,\dots,7$ by 
\begin{equation}
u_k = u_k(0) \mathrm{e}^{\mathrm{i} \,g_k\, t} \ ,
\end{equation}
while the secular solution is obtained through (\ref{eq.lape}). It can be noted that as the matrix $\LL$, and thus $\SS$,  only depend on the masses and semi-major axis of the planets, they do not change much for different fits to the data because  the periods and masses are well constrained (for a given inclination of the system).

The secular frequencies $g_k$ that are responsible for the precession of the orbits and for most of the eccentricity variations are given in Table\,\ref{tab.freq}.

\begin{table}
 \caption{Fundamental secular period and frequencies. 
 \label{tab.freq}} 
 \begin{center}
 \begin{tabular}{ccccc}
 \hline\hline
     $k$& Period     & $g_k$  (num)&  $g_k$  (ana)&  $\tilde g_k$ \\
            & yr     & deg yr$^{-1}$      &  deg yr$^{-1}$      & yr$^{-1}$ \\
 \hline
     1 &    1029.34 &   0.349739  &     0.358991 & -3.68 $\times 10^{-07}$  \\
     2 &    1453.39 &   0.247696  &     0.245229 & -3.40 $\times 10^{-09}$  \\
     3 &    3020.08 &   0.119202  &     0.118471 & -1.11 $\times 10^{-09}$  \\
     4 &    4339.70 &   0.082955  &     0.079644 & -1.43 $\times 10^{-10}$  \\
     5 &   13509.96 &   0.026647  &     0.025290 & -5.16 $\times 10^{-12}$  \\
     6 &   61517.43 &   0.005852  &     0.005581 & -3.93 $\times 10^{-15}$  \\
     7 &  473061.76 &   0.000761  &     0.000663 & -8.14 $\times 10^{-17}$  \\
 \hline
\end{tabular}
\end{center}
{\bf Note}. The period and $g_k$  (num) are computed numerically from the integrated solution of Table\,\ref{ta:tides} through frequency analysis. $g_k$  (ana) are the corresponding frequencies computed  from the Laplace-Lagrange linear approximation (Eq.\ref{eq.lape}), and $\tilde g_k$ is the value of the damping term in the proper mode amplitudes resulting from the tidal dissipation of planet $b$ ($u_k=u_k(0) e^{-\tilde g_k t} e^{i g_k t}$).
\end{table}

\subsection{Stability of the short-period planet $b$}

\begin{figure}
    \includegraphics*[bb=40 120 554 770,width=8.5cm,clip]{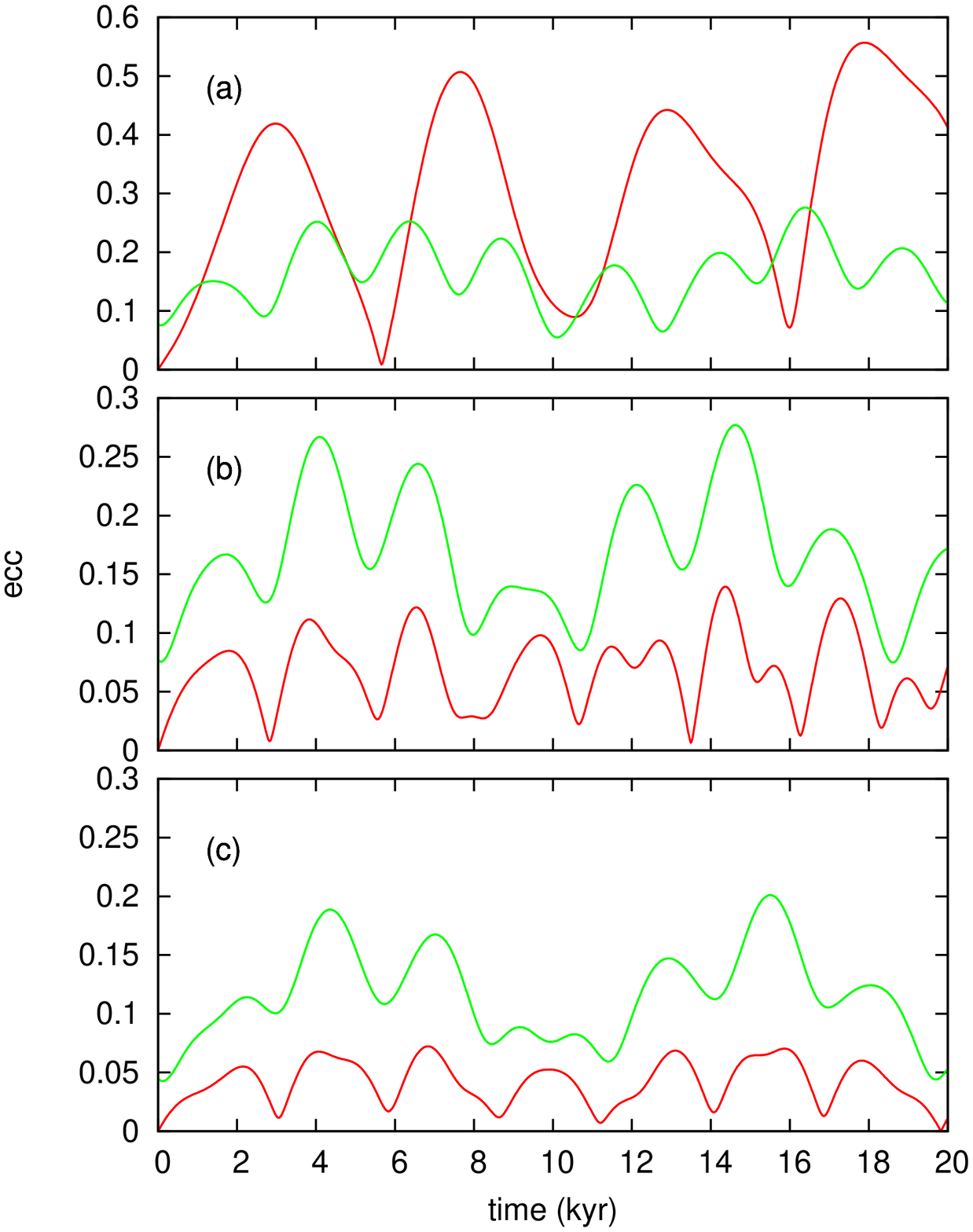}   
\caption{Evolution of the eccentricities of planets $b$ (red) and $c$ (green) during 20\,kyr for three different models. In the top picture the initial eccentricity of planet $b$ is set at zero, but mutual gravitational perturbations increase its value to 0.4 in less than 2~kyr (Table\,\ref{TablePlanetsHD10180}). In the middle figure we included general relativity, which calms down the eccentricity variations of the innermost planet, but still did not prevent the eccentricity of planet $c$ to reach high values. In the bottom figure we use a model where the eccentricities of both planets were previously damped by tidal dissipation (Table\,\ref{ta:tides}). This last solution is stable at least for 1\,Myr. 
\label{eccsfig}}
\end{figure}

Despite the proximity of planet $b$ to the star, it undergoes strong gravitational perturbations from the remaining bodies, due to secular interactions. Even for a model where the initial eccentricity of the planet $b$ is initially set at zero (Table\,\ref{TablePlanetsHD10180}), its eccentricity shows a rapid increase, which can reach up to 0.8 in only 1\,Myr (Fig.\,\ref{eccsfig}a). 
The inclusion of general relativity in the model calms down the eccentricity evolution, but still did not prevent the eccentricity of planet $b$ to attain values above 0.4, and planet $c$ to reach values around 0.3, which then destabilizes the whole system (Fig.\,\ref{eccsfig}b).

Most of the variations observed in Fig.\,\ref{eccsfig} are well described by  the secular system (\ref{eq.laplag},\ref{eq.lape},\ref{eq.diag}). In particular, the effect of general relativity will be to largely increase the first diagonal terms in the Laplace-Lagrange matrix $\LL$, which will increase in $z_1$ the contribution of the proper mode $u_1$ and thus decrease 
the long time oscillations due to the contribution of the modes.

At this stage, one may question the existence of the innermost short-period planet. However, since the mass of this planet should be in the Earth-mass regime, we may assume that it is mainly a rocky planet. As a consequence, due to the proximity of the star, this planet will undergo intense tidal dissipation, that may continuously damp its orbital eccentricity.

\subsection{Tidal contributions}

Using a simplified model \citep{Correia_2009} the tidal variation of the eccentricity is 
\begin{equation}
\dot e = - K n f_6(e) (1-e^2) e 
\ , \label{090522b}
\end{equation}
where 
$ f_6 (e) = (1 + 45e^2/14 + 8e^4 + 685e^6/224 + 255e^8/448 + 25e^{10}/1792) (1-e^2)^{-15/2} / (1 + 3e^2 + 3e^4/8) $, and
\begin{equation}
K = \frac{21}{2} \left( \frac{M_\star}{m}\right)
\left(\frac{R}{a}\right)^5 \frac{k_2}{Q} \ . \label{090514m}
\end{equation}
$M_\star$ is the mass of the star, $m$ and $R$ are the mass and the radius of the planet respectively, $k_2$ is the second Love number, and $Q$ the dissipation factor.

As for the Laplace-Lagrange linear system, we can linearize the tidal contribution from expression (\ref{090522b}), and we obtain for each planet $k$ a contribution  

\begin{equation}
\dot e_k = - \gamma_k e_k  \quad \hbox{with} \quad \gamma_k = K_k n_k \ ,
\label{eq.mar1}
\end{equation}
that is, an additional real contribution on each diagonal term of the Laplace-Lagrange matrix $\mathrm{i}\LL$ 
\begin{equation}
\dot z_k = - \gamma_k z_k \ ,
\label{eq.mar2}
\end{equation}
which thus adds an imaginary part $\mathrm{i}\gamma_k$ to the diagonal terms of $\LL$. In fact, since apart from planet $b$, all planetary masses are relatively large (Table \ref{TablePlanetsHD10180}), the dissipation in these planets may be small, and we will uniquely consider the tidal dissipation in the innermost planet $b$. It should nevertheless be stressed that all the eigenvalues of the  matrix $\LL$ will be modified, and will present an imaginary part that will induce an exponential term in the amplitude of the proper modes (Table \ref{tab.freq}). 

\begin{figure}
    \includegraphics*[width=8.8cm]{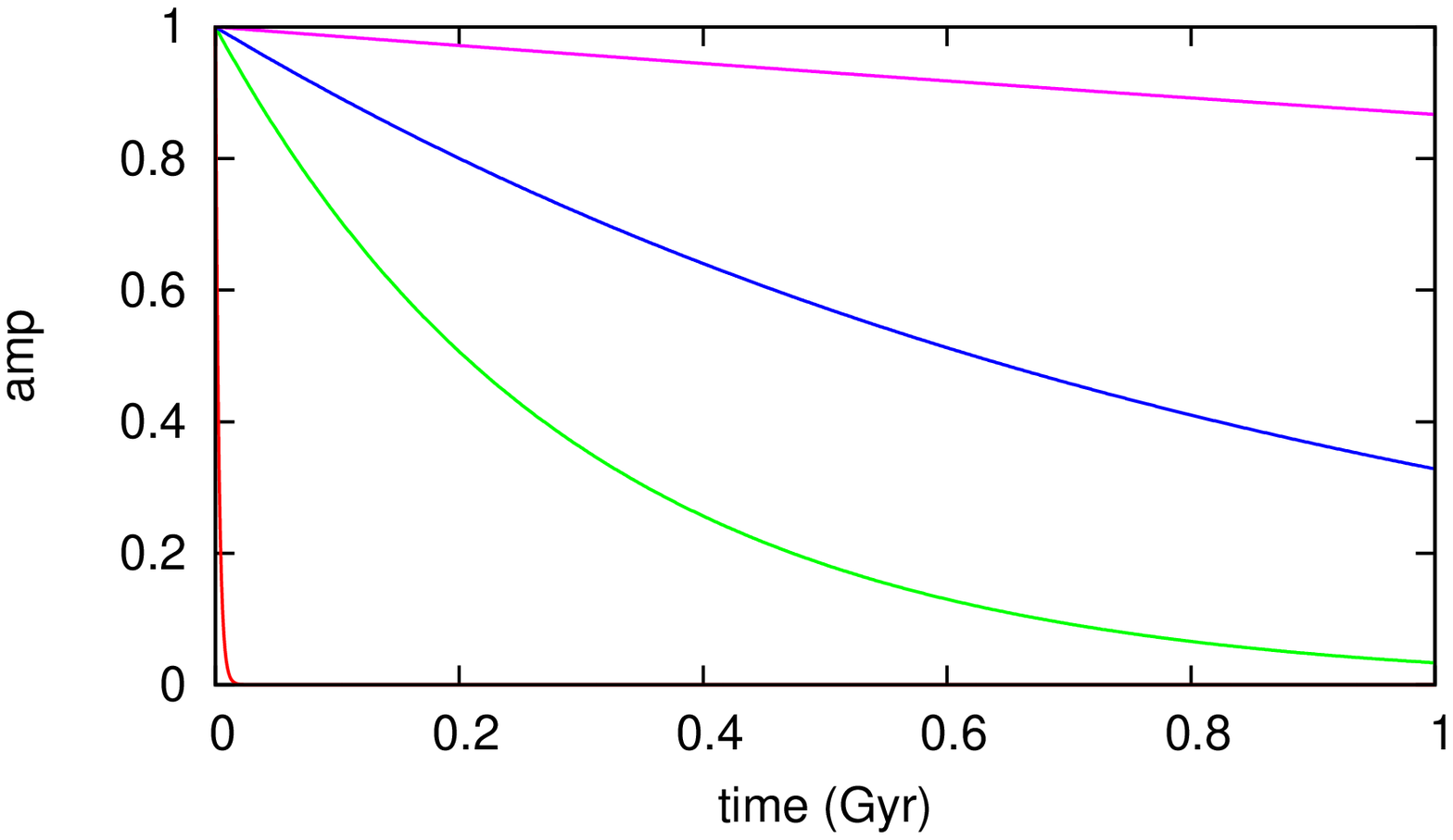}   
\caption{Tidal evolution of the amplitude of the proper modes  $u_1$ (red),  $u_2$ (green), $u_3$ (blue), and $u_4$ (pink)  resulting from the tidal dissipation on planet $b$
with  $ k_2 / Q = 0.0015$. 
\label{Tides1}}
\end{figure}

Adopting  a value similar to Mars $ k_2 / Q = 0.0015 $ (which is a minimal estimation for the dissipation), we  can see in Fig.~\ref{Tides1} that the amplitude of the proper mode $u_1$ will be rapidly damped in a few tens of Myr, whatever its original value. Over more than 1 Gyr,  the amplitude of the second proper mode $u_2$ is also most probably damped to a small value.

In order to obtain a realistic solution that is the result of the tidal evolution of the system, it is thus not sufficient to impose that the innermost planets have small eccentricity, as this may only be realized at the origin of time (Fig.~\ref{eccsfig}). It is also necessary that the amplitude of the first proper  modes, and  particularly $u_1$ are set to small values. This will be the way to obtain a solution with moderate variations of the eccentricities on the innermost planets, which will then increase its stability. 

\begin{table*}
\caption{Orbital parameters for the seven bodies orbiting {HD}\,10180, obtained with a 8-body Newtonian fit to observational data, including the effect of tidal dissipation. Uncertainties are estimated from the covariance matrix, and $ \lambda $ is the mean longitude at the given epoch ($\lambda$ = $M$ + $\omega$). The orbits are assumed co-planar.}
\label{ta:tides}
\begin{center}
\begin{tabular}{l l c c c c c c c}
\hline \hline
\noalign{\smallskip}
{\bf Parameter}	& {\bf [unit]}		& {\bf HD 10180 b}	& {\bf HD 10180 c}	& {\bf HD 10180 d}	& {\bf HD 10180 e}	& {\bf HD 10180 f}	& {\bf HD 10180 g}	& {\bf HD 10180 h} \\
\hline 
\noalign{\smallskip}
Epoch		& [BJD]			& \multicolumn{7}{c}{2,454,000.0 (fixed)}    \\ 
$i$			& [deg]			& \multicolumn{7}{c}{$ 90 $ (fixed) }  \\  
$V$			& [km\,s$^{-1}$]		& \multicolumn{7}{c}{$ 35.52981\,(\pm 0.00012) $}  \\
\hline 
\noalign{\smallskip}
$P$			& [days]			& $1.17768$		& $5.75979$		& $16.3579$		& $49.745 $		& $122.76 $		& $601.2 $		& $2222 $ \\ 
			&				& $(\pm 0.00010)$	& $(\pm 0.00062)$	& $(\pm 0.0038)$	& $(\pm 0.022)$	& $(\pm 0.17) $		& $(\pm 8.1) $		& $(\pm 91) $ \\ 
$\lambda$	& [deg]			& $ 188 $			& $ 238.5 $		& $ 196.6 $		& $ 102.4 $		& $ 251.2 $		& $ 321.5 $		& $ 235.7 $ \\ 
			&				& $(\pm 13) $		& $ (\pm 2.3)  $		& $(\pm 3.8)  $		& $(\pm 2.4)  $		& $(\pm 3.6)  $		& $(\pm 9.9) $		& $(\pm 6.0)$ \\ 
$e$			&				& $ 0.0000 $		& $ 0.045 $		& $ 0.088 $		& $ 0.026 $		& $ 0.135 $		& $ 0.19 $			& $ 0.080 $ \\ 
			&				& $(\pm 0.0025) $	& $(\pm 0.026)$	& $(\pm 0.041)$	& $(\pm 0.036)$	& $(\pm 0.046)$	& $(\pm 0.14)$		& $(\pm 0.070)$ \\ 
$\omega$		& [deg]			& $  39 $			& $ 332 $			& $ 315 $			& $ 166 $			& $ 332 $			& $ 347 $			& $ 174 $ \\ 
			&				& $(\pm 78)  $		& $(\pm 43) $		& $(\pm 33) $		& $(\pm 110)$		& $(\pm 20) $		& $(\pm 49) $		& $(\pm 74) $ \\ 
$K$			& [m\,s$^{-1}$]		& $   0.78 $		& $   4.50 $		& $   2.86 $		& $   4.19 $		& $   2.98 $		& $   1.59 $		& $   3.04 $ \\  
			&				& $(\pm 0.13)   $	& $(\pm 0.12)  $	& $(\pm 0.13)  $	& $(\pm 0.14)  $	& $(\pm 0.15)  $	& $(\pm 0.25)  $	& $(\pm 0.19)  $ \\
\hline
\noalign{\smallskip}
$m \sin i$		& [$M_\oplus$]		& $ 1.35 $			& $ 13.10 $		& $ 11.75 $		& $ 25.1 $			& $ 23.9 $			& $ 21.4 $			& $ 64.4 $ \\
			&				& $(\pm 0.23)   $	& $(\pm 0.54)  $	& $(\pm 0.65)  $	& $(\pm 1.2)  $		& $(\pm 1.4)  $		& $(\pm 3.4)  $		& $(\pm 4.6)  $ \\
$a$			& [AU]			& $ 0.02225 $		& $ 0.0641 $		& $ 0.1286 $		& $ 0.2699 $		& $ 0.4929 $		& $ 1.422 $		& $ 3.40 $ \\
			&				& $(\pm 0.00035)$	& $(\pm 0.0010)$	& $(\pm 0.0020)$	& $(\pm 0.0042)$	& $(\pm 0.0078) $	& $(\pm 0.026) $	& $(\pm 0.11) $ \\
\hline
\noalign{\smallskip}
$N_\mathrm{meas}$ &			& \multicolumn{7}{c}{190}  \\
Span		& [days]			& \multicolumn{7}{c}{2428} \\
rms			& [m\,s$^{-1}$]		& \multicolumn{7}{c}{1.28} \\
$\chi_r^2$	&				& \multicolumn{7}{c}{1.54} \\
\hline
\end{tabular}
\end{center}
\end{table*}

\subsection{Orbital solution with dissipative constraint}

As the proper modes of the two innermost planets, $u_1$ and $u_2$ are damped after about 1\,Gyr, we expect them to be small for the present observations (the age of the star is estimated to be 4\,Gyr, Table\,\ref{TableHD10180}). The initial conditions for the {HD}\,10180 planetary system should then take into account the tidal damping. We have thus chosen to modify our fitting procedure in order to include a constraint for the tidal damping of the proper modes $u_1$ and $u_2$. For that purpose, we added to the $ \chi^2 $ minimization an additional term corresponding to these proper modes:
\begin{equation}
\chi^2_R = R \left( u_1^2 + u_2^2 \right) \ ,
\label{chiprop}
\end{equation}
where $R$ is a positive constant, that is chosen arbitrarily in order to obtain a small value for $ u_1 $ and $ u_2 $ simultaneously, without any significant decrease of the quality of the fit.

Using $ R = 350 $ we get $ u_1 \sim 0.0017 $ and $ u_2 \sim 0.044 $ and obtain a final $\chi_r^2 = 1.54$, which is nearly identical to the results obtained without this additional constraint ($R = 0$, $\chi_r^2 = 1.50$). The best-fit solution obtained by this method is listed in Table\,\ref{ta:tides}. We believe that this solution is a more realistic representation of the true system than the one listed in Table\,\ref{TablePlanetsHD10180}, and we will adopt it henceforward for the remaining of the paper. Actually, with this constraint, the variation of the eccentricity of the two innermost planets are now smaller (Fig.\ref{eccsfig}c).

\subsection{Stability of the system}

\begin{figure*}
\begin{center}$
\begin{array}{ccc}
\includegraphics[width=\figw]{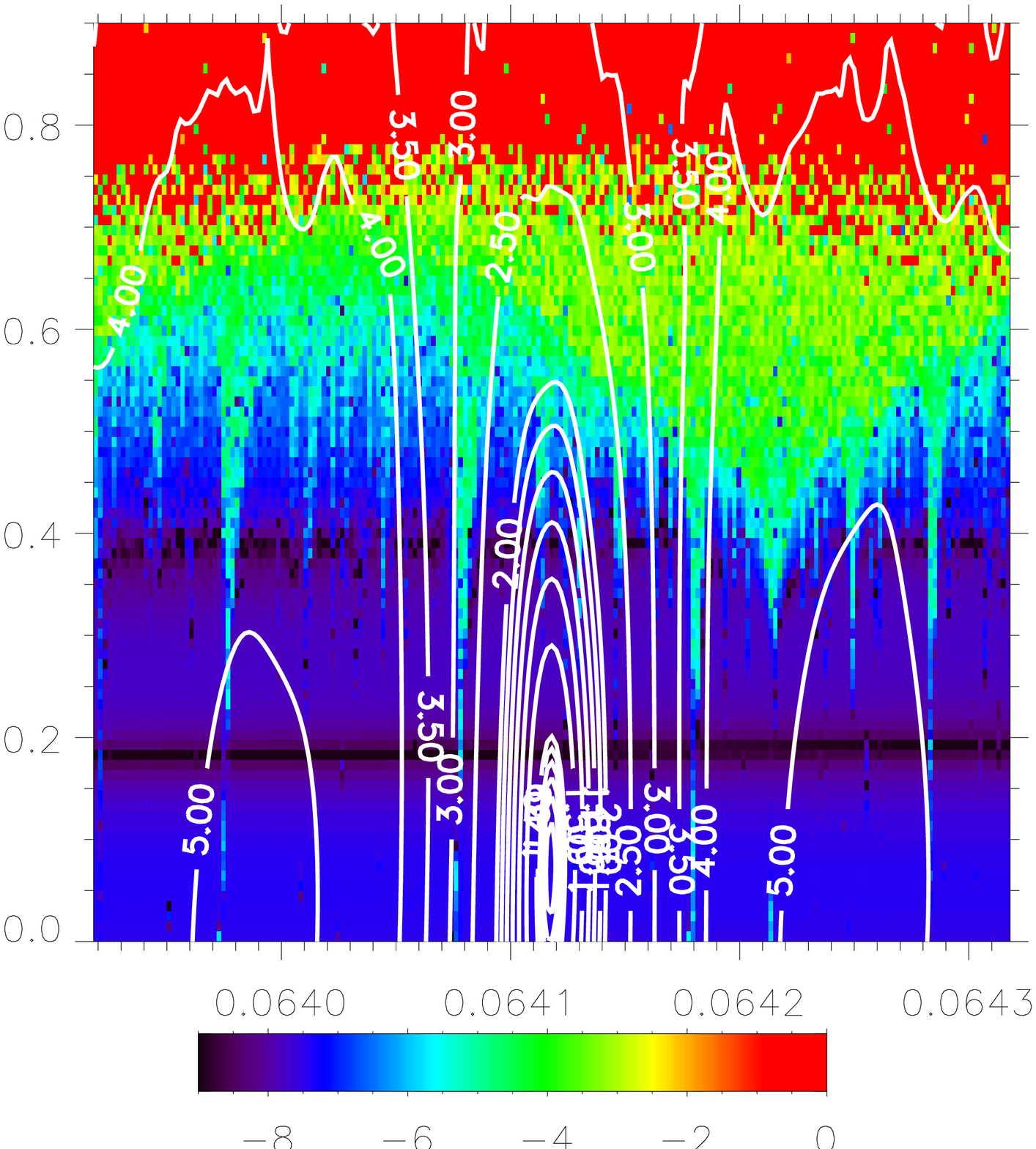}&
\includegraphics[width=\figw]{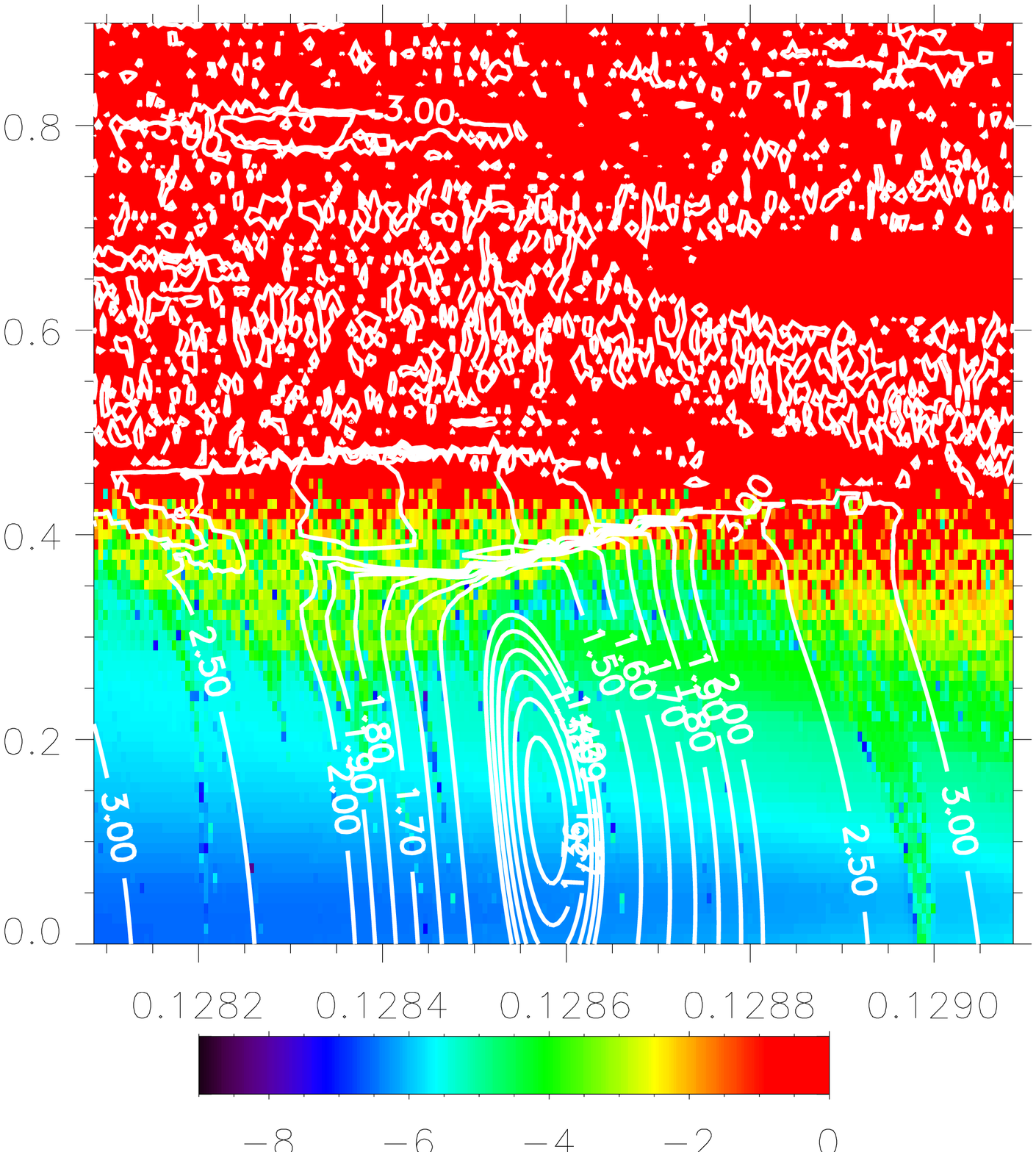}&
\includegraphics[width=\figw]{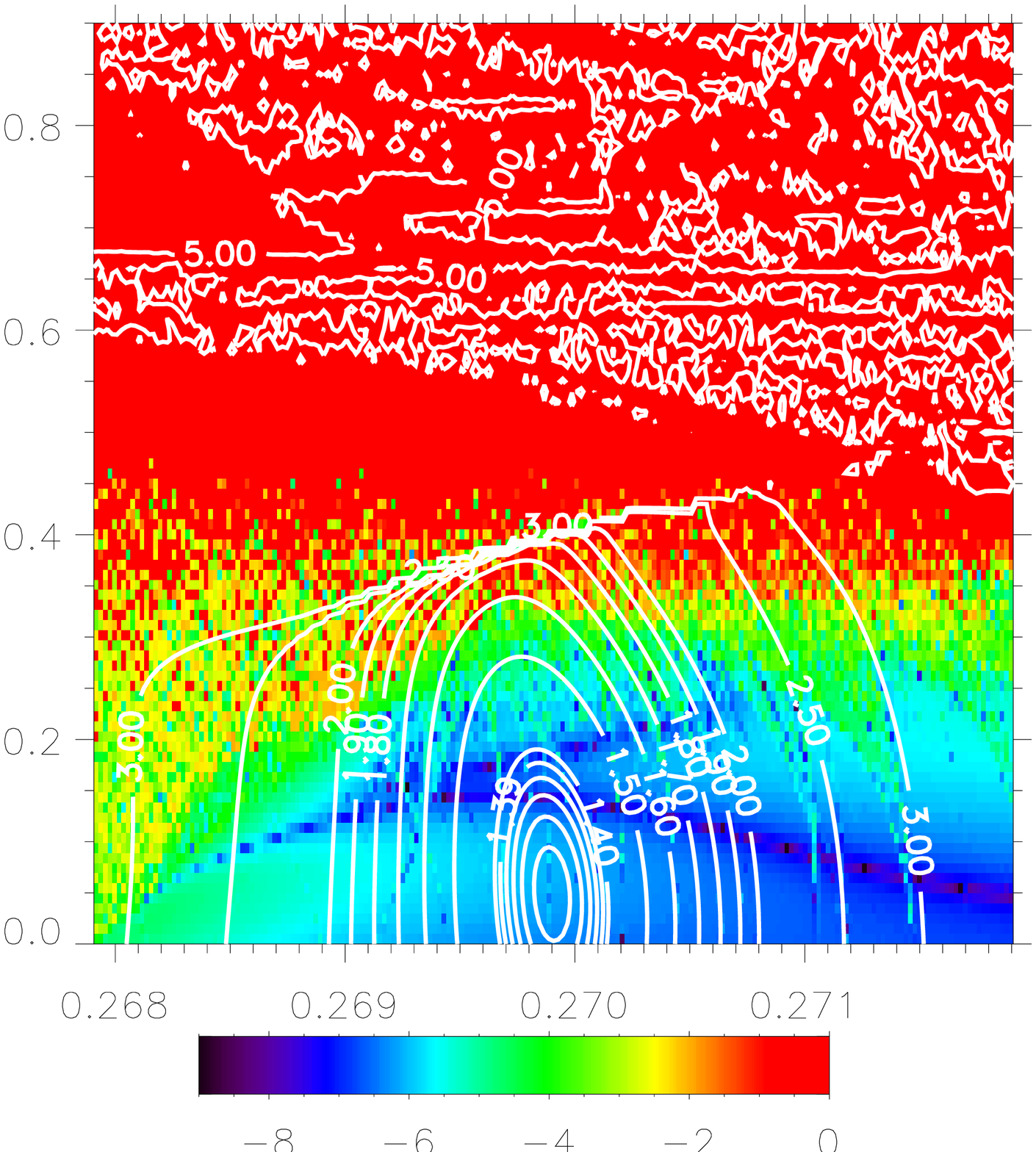}\\
\includegraphics[width=\figw]{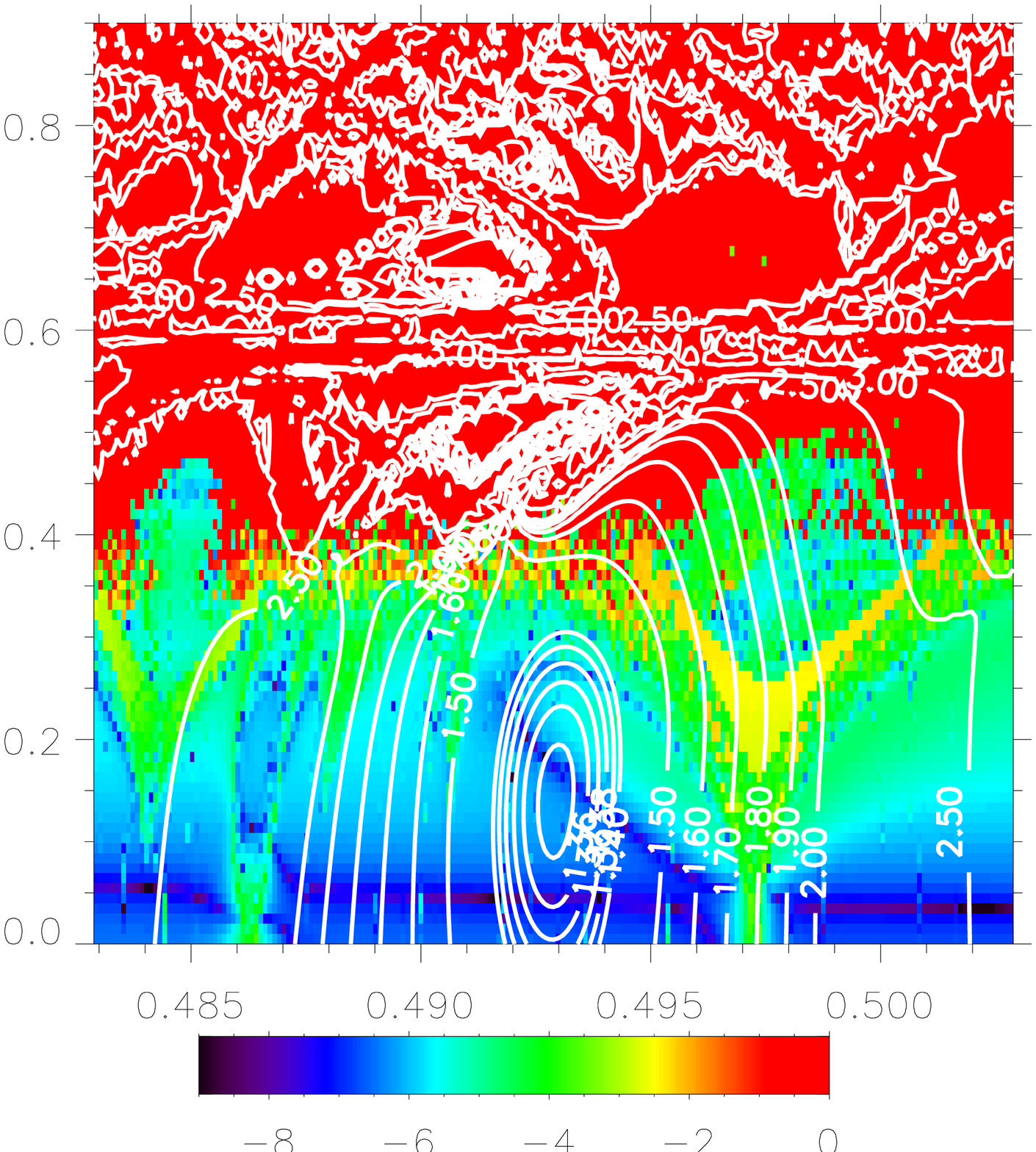}&
\includegraphics[width=\figw]{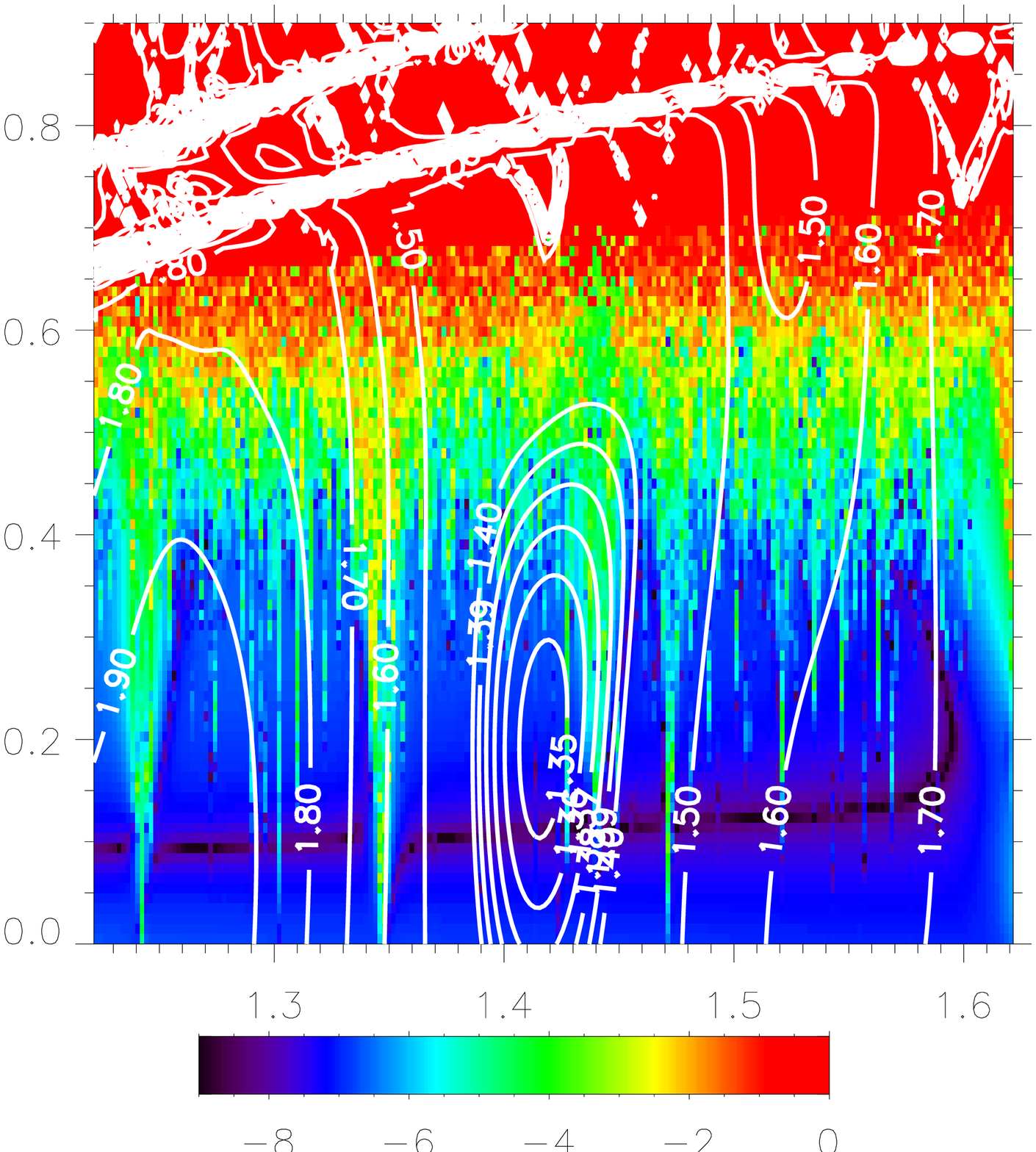}&
\includegraphics[width=\figw]{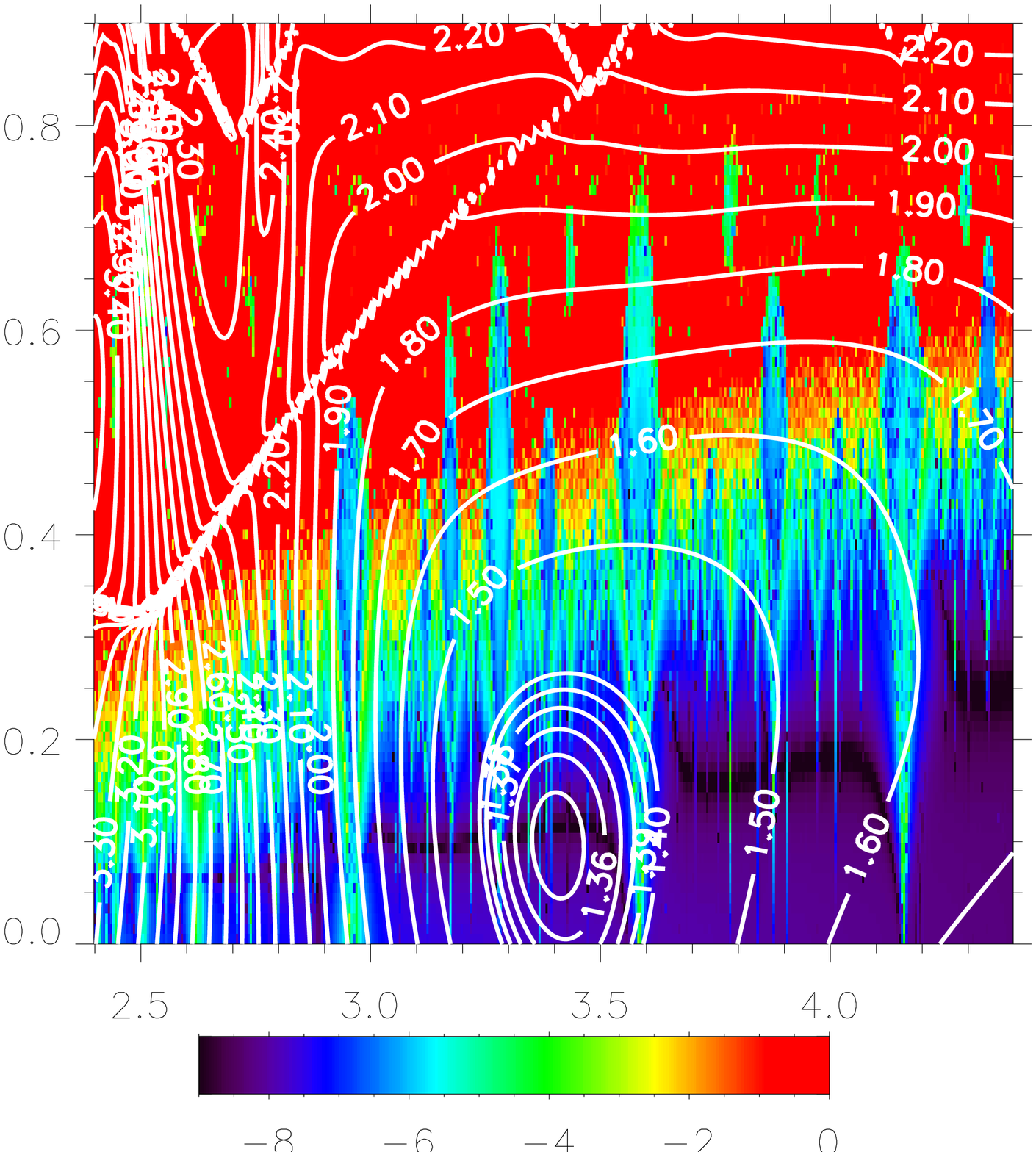}\\
\end{array}$
\end{center}
\caption{Global view of the dynamics of the {HD}\,10180 system for variations of the eccentricity and the semi-major axis of the six outermost planets. The color scale is the stability index obtained through a frequency analysis of the longitude of the planets over two consecutive time intervals. Red areas correspond to high orbital diffusion (instability) and the blue ones to low diffusion (stable orbits). Labeled lines give the value of $ \chi^2 $ obtained for each choice of parameters.}
\label{Dyn1}
\end{figure*}

To analyze the stability of the nominal solution in Table\,\ref{ta:tides} we performed a global frequency analysis \citep{Laskar_1993PD} in the vicinity of the six outermost planets (Fig.\,\ref{Dyn1}), in the same way as achieved for the {HD}\,202206 or the {HD}\,45364 planetary systems \citep{Correia_etal_2005,Correia_etal_2009}. For each planet, the system is integrated on a regular 2D mesh of initial conditions, with varying semi-major axis and eccentricity, while the other parameters are retained at their nominal values. The solution is integrated over 10~kyr for each initial condition and a stability indicator is computed to be the variation in the measured mean motion over the two consecutive 5~kyr intervals of time. For regular motion, there is no significant variation in the mean motion along the trajectory, while it can vary significantly for chaotic trajectories \citep[for more details see][]{Couetdic_etal_2010}. The result is reported in color in Fig.\,\ref{Dyn1}, where ``red'' represents the strongly chaotic trajectories, and ``dark blue'' the extremely stable ones. 

In all plots of Fig.\ref{Dyn1}, the  zones  of minimal $\chi^2$ obtained in comparing with the observations appear to belong to stable ``dark blue'' areas. This picture thus presents a coherent view of dynamical analysis and radial velocity measurements, and reinforces the confidence that the sub-system formed by the six outermost planets given in Table\,\ref{ta:tides} is stable for long timescales.

Nevertheless, due to the large number of planets in the system, many mean motion resonances can be observed, several of them being unstable. None of the planets determined by the solution in Table\,\ref{ta:tides} are in resonance, but some of them lie in between. In particular, the pair $d$ and $e$ is close to a 3:1 mean motion resonance, and the pair $e$ and $f$ is close to a 5:2 mean motion resonance (similar to Jupiter and Saturn).

\subsection{Long-term orbital evolution}

The estimated age of the {HD}\,10180 system is about 4.3\,Gyr (Table\,\ref{TableHD10180}), indicating that the present planetary system had to survive during such a long timescale. We tested the stability of the system by performing a numerical integration of the orbits of the 6 outermost planets in Table\,\ref{ta:tides} over 150~Myr using the symplectic integrator SABAC4 of \citet{Laskar_Robutel_2001} with a step size of $10^{-3}$~years, including general relativity.

\begin{figure}
    \includegraphics*[width=8.5cm]{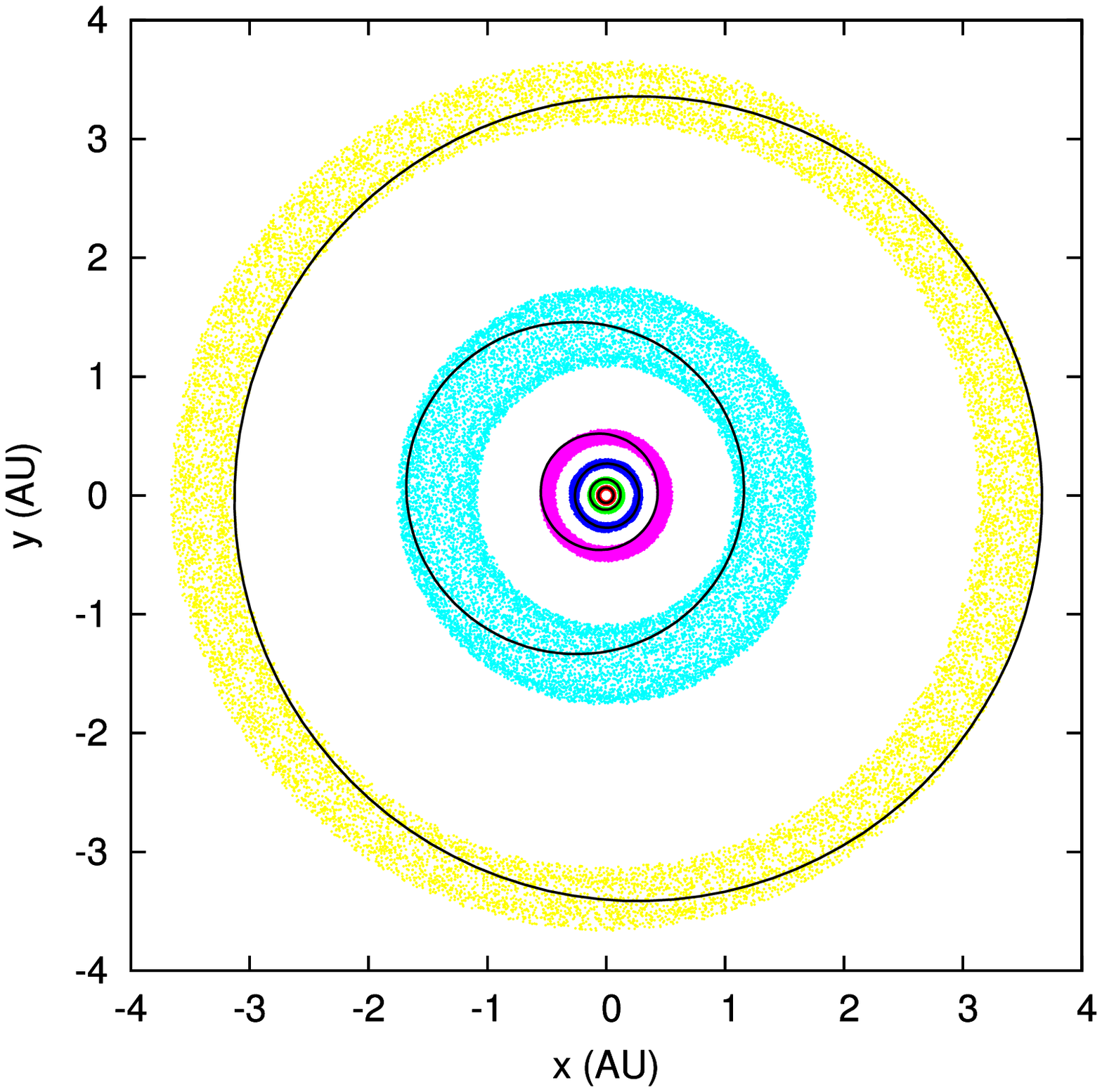} 
  \caption{Long-term evolution of the {HD}\,10180 planetary system over {\bf 150\,Myr} starting with the orbital solution from Table\,\ref{ta:tides}. The panel shows a face-on view of the system. $x$ and $y$ are spatial coordinates in a frame centered on the star. Present orbital solutions are traced with solid lines and each dot corresponds to the position of the planet every 10~kyr. The semi-major axes (in AU) are almost constant, but the eccentricities undergo significant variations (Table \ref{ta:meca}). 
\label{Dyn3}}   
\end{figure}

The orbits, displayed in Fig.\,\ref{Dyn3}, remain very stable throughout the simulation. We have verified by frequency analysis that these orbits evolve in a very regular way. We have also integrated the full 7-planet system over 1 Myr with a step size of $10^{-4}$~years, without any sign of strong instability, although the frequency analysis of the solutions with 7 planets shows that these solutions are not as well approximated by quasiperiodic series as the 6-planet solutions. This will have to be analyzed further as, like in our solar system, the presence of this innermost planet seems to be critical for the long-term stability of the system \citep{Laskar_Gastineau_2009}.

The fact that we are able to find a stable solution compatible with the observational data can still be considered as a good indicator of the reliability of the determination of the {HD}\,10180 planetary system. 

\begin{table}
 \caption{Semi-major axes and eccentricity minima and maxima observed over 1 Myr in the 7-planet solution of Table\,\ref{ta:tides}.}
 \begin{center}
 \begin{tabular}{crr r c}
 \hline\hline
     $k$ &  $a_{min}$ &  $a_{max}$& $e_{min}$     & $e_{max}$ \\
 \hline
     1 &    0.022253 &   0.022253 &        0.000 &      0.082 \\
     2 &    0.064114 &   0.064122 &        0.010 &      0.203 \\
     3 &    0.128536 &   0.128626 &        0.000 &      0.179 \\
     4 &    0.269814 &   0.270092 &        0.000 &      0.156 \\
     5 &    0.492348 &   0.493184 &        0.023 &      0.137 \\
     6 &    1.419645 &   1.424347 &        0.188 &      0.242 \\
     7 &    3.387207 &   3.402716 &        0.044 &      0.081 \\
 \hline
\end{tabular}
\end{center}
\label{ta:meca}
\end{table}

Because of the strong gravitational secular interactions between the planets, their orbital eccentricities present significant variations, while their semi-major axes are almost constant (Table~\ref{ta:meca}), which is also the signature that the system is far from strong resonances. As the secular frequencies $g_k$ (Table\,\ref{tab.freq}) are relatively large, the secular variations of the orbital parameters are more rapid than in our Solar System, which may allow us to detect them directly from observations, and hence access the true masses and mutual inclinations of the planets as it was done for the {GJ}\,876 system \citep{Correia_etal_2010}.

\subsection{Additional constraints}

The stability analysis summarized in Fig.\,\ref{Dyn1} shows a good agreement between the ``dark blue'' stable areas and the $\chi^2$ contour curves. We can thus assume that the dynamics of the seven planets is not disturbed much by the presence of an additional body close-by.

\begin{figure}
  \begin{center}
    \begin{tabular}{c}
    \includegraphics*[width=8.5cm]{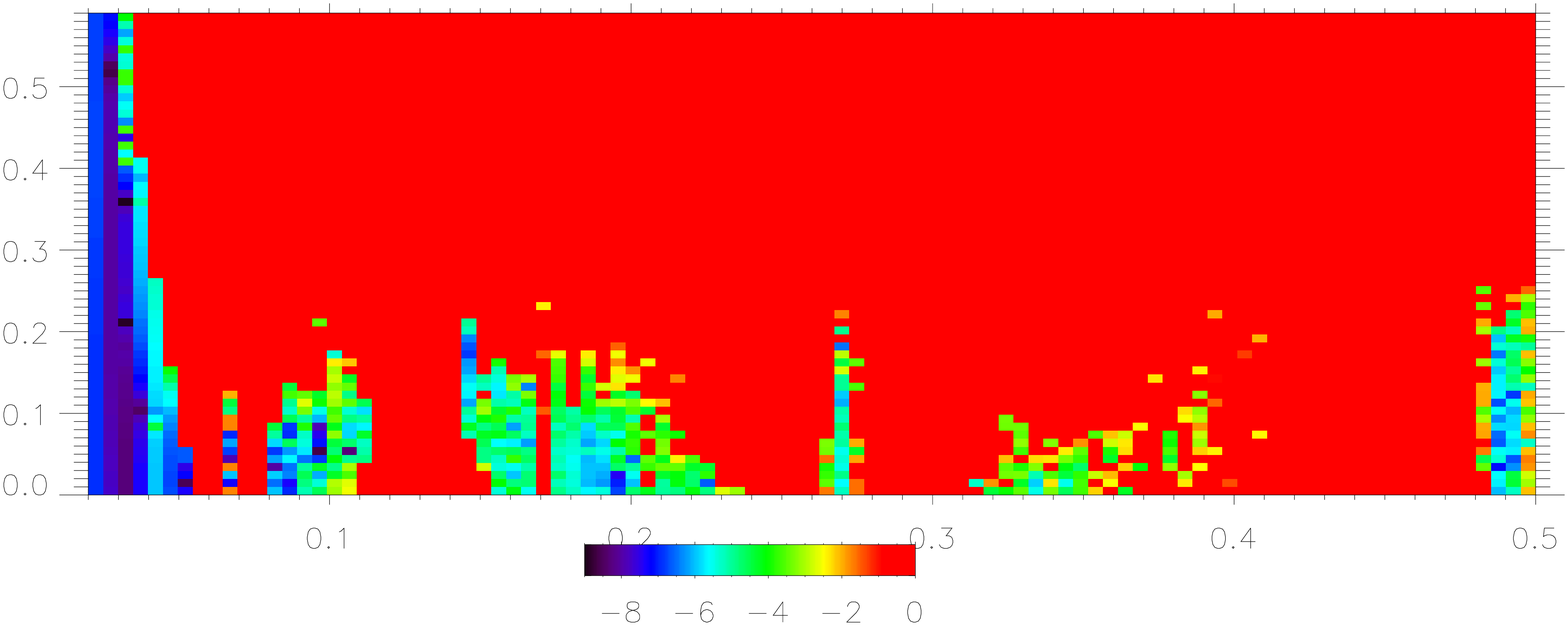} \\
    \includegraphics*[width=8.5cm]{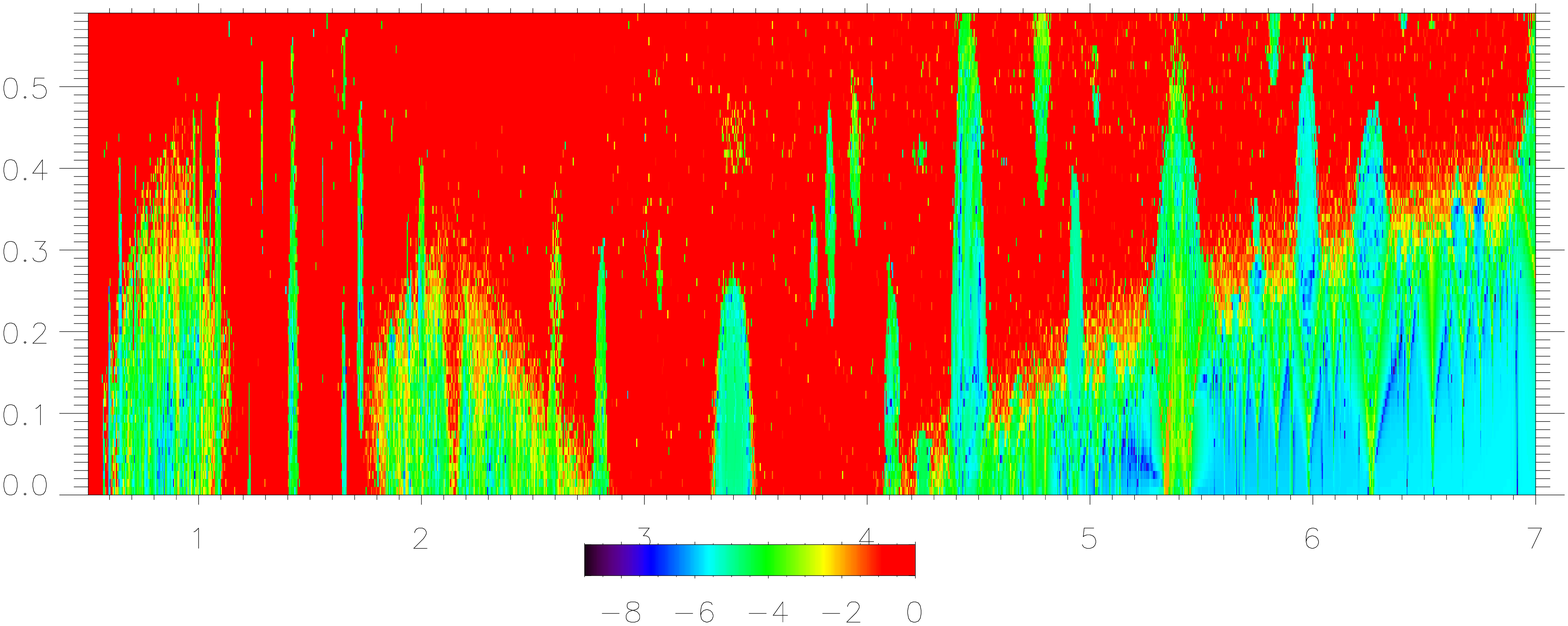} \\
    \end{tabular}
  \caption{Possible location of an additional eighth planet in the {HD}\,10180 system. The stability of a small-mass particle in the system is analyzed, for various semi-major axes and eccentricities, and for $K$ = 0.78 m\,s$^{-1}$. The stable zones where additional planets could be found are the ``dark blue'' regions.
    \label{Dyn2}}   
  \end{center}
  \end{figure}

We then tested the possibility of an additional eighth planet in the system by varying the semi-major axis and the eccentricity over a wide range, and performing a stability analysis (Fig.\,\ref{Dyn2}). The test was done  for a fixed $K$ value ($K$ = 0.78 m\,s$^{-1}$), similar to planet $b$.

From this analysis (Fig.\,\ref{Dyn2}), one can see that stable orbits are possible beyond 6~AU (outside the outermost planet's orbit). More interestingly, stability appears to be also possible around 1~AU, which corresponds to orbital periods within $300-350$~days, between the orbits of planets $f$ and $g$, exactly at the habitable zone of {HD}\,10180. Among the already known planets, this is the only zone where additional planetary mass companions can survive. With the current HARPS precision of $\sim$1~m\,s$^{-1}$, we estimate that any object with a minimum mass $M >$ 10 $M_\oplus$ would already be visible in the data. Since this does not seem to be the case, if we assume that a planet exists in this stable zone, it should be at most an Earth-sized object.

We can also try to find constraints on the maximum masses of the current seven-planet system if we assume co-planarity of the orbits. Indeed, up to now we have been assuming that the inclination of the system to the line-of-sight is $90^\circ$, which gives minimum values for the planetary masses (Table\,\ref{ta:tides}).

By decreasing the inclination of the orbital plane of the system, we increase the mass values of all planets and repeat a stability analysis of the orbits, as in Figure\,\ref{Dyn1}. As we decrease the inclination, the stable ``dark-blue'' areas become narrower, to a point that the minimum $\chi^2$ of the best fit solution lies outside the stable zones. At that point, we conclude that the system cannot be stable anymore. It is not straightforward to find a transition inclination between the two regimes, but we can infer from our plots that stability of the whole system is still possible for an inclination of $30^\circ$, but becomes impossible for an inclination of $10^\circ$. Therefore, we conclude that the maximum masses of the planets are most probably obtained for an inclination around $20^\circ$, corresponding to a scaling factor of about 3 with respect to minimum masses.

\section{On the properties of low-mass planetary systems}

\subsection{Dynamical architecture}

The increasing number of multi-planet systems containing at least three known planets greatly extends the possibilities to study the orbital architectures of extrasolar planetary systems and compare them to our Solar System. Although there are already 15 systems with at least three planets as of May 2010, one should recognize that our knowledge of many of them is still highly incomplete due to observational biases. The RV technique finds the most massive, close-in planets first in each system, and the secure detection of multiple planets requires a large number of observations, roughly proportional to the number of planets for RV signals well above the noise floor. Lower-amplitude signals like the ones induced by ice giants and super-Earths require a noise floor at or below $\sim$1\,m\,s$^{-1}$ to keep the number of observations within a reasonable range. Moreover, the phase of each signal must be sufficiently well covered, which requires a large enough time span and appropriate sampling. As a consequence, the planet detection limits in many of the 15 systems mentioned above probably do not reach down to the Neptune-mass range yet, preventing us from having a sufficiently complete picture of them. Nevertheless, considering the well-observed cases followed at the highest precision, the RV technique shows here its ability to study the structure of planetary systems, from gas giants to telluric planets.

The dynamical architecture of planetary systems is likely to convey extremely useful information on their origins. The dominant planet formation scenario presently includes several physical processes that occur on similar timescales in protoplanetary disks: formation of cores, preferentially beyond the ice line, through accretion of rocky and icy material; runaway gas accretion on cores having reached a critical mass, rapidly forming giant planets; inward migration of cores (type I) and giant planets (type II) through angular momentum exchange with the gaseous disk; disk evolution and dissipation within a few Myr; dynamical interactions between protoplanets leading to eccentricity pumping, collisions or ejections from the system. It is extremely challenging to build models that include all these effects in a consistent manner, given the complicated physics involved and the scarce observational constraints available on the early stages of planet formation. In particular, attempts to simultaneously track the formation, migration and mutual interactions of {\it several} protoplanets are still in their infancy. Observational results on the global architecture of planetary systems may therefore provide important clues to determine the relative impact of each process.

\begin{figure}
\centering
\includegraphics[width=90mm]{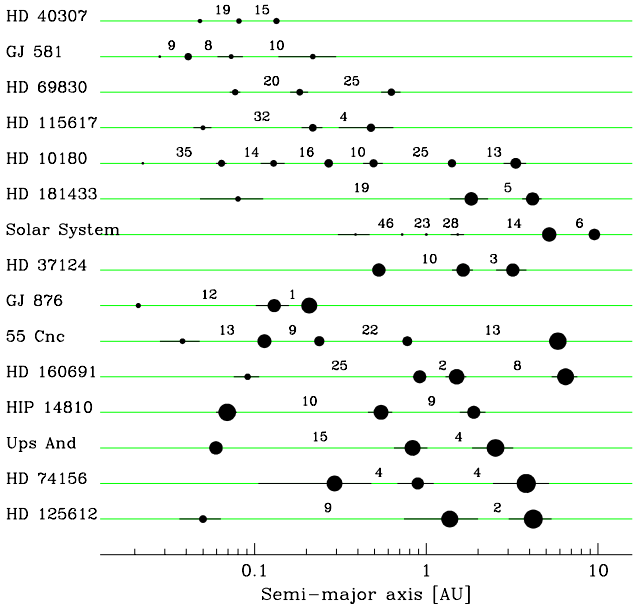}
\caption{The 15 planetary systems with at least three known planets as of May 2010. The numbers give the minimal distance between adjacent planets expressed in mutual Hill radii. Planet sizes are proportional to $\log{(m\sin{i})}$.}
\label{FigMultiPlanetSystems}
\end{figure}

From the observational point of view, the least that can be said is that planetary systems display a huge diversity in their properties, which after all is not surprising given their complex formation processes. Fig.~\ref{FigMultiPlanetSystems} shows planet semi-major axes on a logarithmic scale for the 15 systems known to harbour at least three planets. Systems are shown in increasing order of their mean planetary mass, while individual planet masses are illustrated by varying dot size. The full range of distances covered by each planet between periastron and apastron is denoted by a black line. The minimal distance between each neighbouring pair of planets is also given, expressed in units of the mutual Hill radius which is defined as:

\begin{equation}
\label{EquHill}
R_{\mathrm{H,M}} = \frac{a_1 + a_2}{2} \left(\frac{m_1 + m_2}{3\,M_*}\right)^{1/3} .
\end{equation}

As shown by \citet{chambers96}, the instability timescale of a coplanar, low-eccentricity multi-planet system is related to the distance between planets, expressed in mutual Hill radii, in a relatively simple manner, and approximate 'life expectancies' of planetary systems can be estimated based on these separations, the number of planets, and their masses. The Chambers et al. simulations do not extend to masses above $\sim$3\,$M_{\oplus}$ or to timescales above 10$^8$ years, but a moderate extrapolation of their results shows that for systems with 3--5 planets and masses between a few $M_{\oplus}$ and a few $M_{\mathrm{J}}$, separations between adjacent planets should be of at least 7--9 mutual Hill radii to ensure stability on a 10-Gyr timescale. These numbers should not be taken too exactly since they were obtained assuming regularly-spaced, equal-mass bodies. They are also not applicable to eccentric orbits and dynamical configurations such as mean-motion resonances, where stability 'islands' do exist at reduced spacings (e.g. GJ~876). However, the global picture emerging from Fig.~\ref{FigMultiPlanetSystems} shows that many known planetary systems are dynamically 'packed', with little or no space left for additional planets \citep[e.g.][]{barnes04a,barnes04b}. This result was already noted by several authors, giving rise for example to the 'packed planetary system' hypothesis \citep{barnes04b}. Here we show that this seems to be true also for several low-mass systems, i.e. those which do not contain gas giants (or only distant ones), as illustrated by HD~40307, GJ~581 and HD~10180. Indeed, several planets in these systems are separated from each other by typically less than 15 mutual Hill radii. There are still a few 'empty' places, however, and further observations will tell if smaller planets are hiding between the known ones.

\subsection{Extrasolar Titius-Bode-like laws?}

It is intriguing that many gas giant {\em and} low-mass systems seem to share the property of being dynamically packed. An attractive explanation would be that at each moment of their history, many planetary systems are 'saturated' with planets and exhibit dynamical configurations whose lifetime is of the same order of magnitude as the age of the system. This would point to a major role for dynamical interactions in the shaping of planetary systems, at least since the dissipation of the gaseous disk. The observed packing may support the view that close-in low-mass systems could be primarily the result of strong interactions (collisions and ejections) between several large protoplanets {\em after} these were brought to the inner regions of the disk through type I migration, i.e. after disk dissipation. These systems would then naturally evolve towards planets separated from each other by a roughly constant number of mutual Hill radii \citep[e.g.][]{laskar00,raymond09}. Since Hill radii are proportional to the semi-major axes, the orbital distances of successive planets with similar masses will tend to obey an approximate exponential law, much like the century-long debated and polemical Titius-Bode law in the Solar System. Indeed, \citet{hayes98} have shown that any planetary system subject to some 'radius-exclusion' law such as the Hill criterion is likely to have its planets distributed according to a geometric progression. \citet{laskar00} presents a simplified model of planetary accretion focusing on the evolution of the angular momentum deficit (AMD) of the averaged system. Starting from a given density of planetesimals $\rho(a)$, the final state of the system, defined by the end of the collision phase, can be derived analytically and the spacing between adjacent planets can be predicted for different functional forms of $\rho(a)$. Interestingly, an exponential law $\log{a_n} = c_1 + c_2 n$ is obtained when the initial density $\rho(a)$ goes as $a^{-3/2}$, while a constant density $\rho(a)$ yields a semi-major axis relation of the form $\sqrt{a_n} = c_1 + c_2 n$. Studies of the orbital arrangements in a large number of systems may thus eventually allow us to constrain the original density of planetesimals.

\begin{figure}
\centering
\includegraphics[width=85mm]{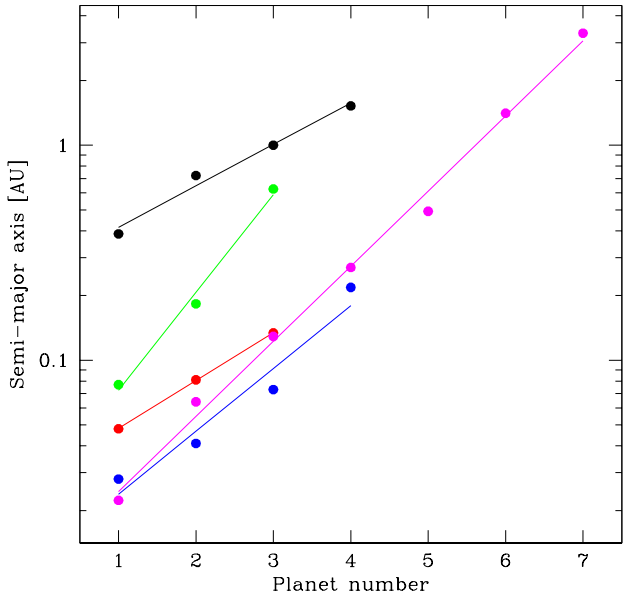}
\caption{Fit of exponential laws to semi-major axes as a function of planet number for the inner Solar System (black), HD 40307 (red), GJ 581 (blue), HD 69830 (green) and HD 10180 (magenta).}
\label{FigTB}
\end{figure}

\begin{table}
\caption{Exponential fits to semi-major axis distributions.}
\label{TableTB}
\centering
\begin{tabular}{l c c c c c}
\hline\hline
System & $N_{\mathrm{pl}}$ & Average mass & $c_1$ & $c_2$ & rms \\
 & & ($M_{\oplus}$) & & & (\%) \\
\hline
Inner Solar System & 4 & 0.49 & 0.267 & 1.56 & 8.03 \\
HD 40307 & 3 & 6.76 & 0.029 & 1.67 & 0.57 \\
GJ 581 & 4 & 7.50 & 0.012 & 1.96 & 21.0 \\
HD 69830 & 3 & 12.5 & 0.025 & 2.85 & 10.2 \\
HD 10180 & 7 & 23.4 & 0.011 & 2.24 & 12.0 \\
\hline
\end{tabular}
\end{table}

Looking more closely at Fig.~\ref{FigMultiPlanetSystems}, we see that exponential laws may indeed exist in some planetary systems. However, a meaningful test requires that {\em all} successive planets have been discovered, especially low-mass ones. This is far from being guaranteed in the presently-known systems, and we therefore limit ourselves to those observed with the HARPS spectrograph, and to planets within 1 AU, to minimize observational biases. We do not want to speculate on 'missing' planets introducing gaps in the Titius-Bode-like relations, since almost anything can be fitted to the present, limited datasets if more than two free parameters are allowed. As can be seen by eye in Fig.~\ref{FigMultiPlanetSystems}, a somewhat regular spacing between adjacent planets seems to exist in the low-mass systems HD 40307, HD 69830 and HD 10180, but less so in GJ 581. Among massive systems, 55 Cnc also shows a somewhat regular spacing \citep{proveda08}, but the presence of close-in gas giants in this system makes planet-planet interactions much stronger and hints at a different formation history (e.g. type II instead of type I migration) compared to low-mass systems.

Focusing on low-mass systems, Fig.~\ref{FigTB} shows an exponential fit $a_n = c_1\,c_2^n$ to the observed semi-major axes as a function of planet number, starting at $n$ = 1. A reasonable fit is obtained for HD~40307, HD~69830 and HD~10180, with a relative standard deviation of the residuals of 0.57\%, 10.2\% and 12.0\%, respectively. The fit to the inner Solar System is also shown, with a relative standard deviation of 8.0\%. The fit to GJ~581 is less convincing, with a dispersion of 21.0\%. It is tempting to conjecture that there exists an additional body between the third and fourth planets in this system, which would make the exponential fit significantly better, and at the same time provide an exciting candidate for a habitable world. Table~\ref{TableTB} gives the values of the best-fit parameters $c_1$ and $c_2$ for each system, together with the average planetary masses. Interestingly, a positive correlation between $c_2$ and mass may be present, possibly illustrating the fact that more massive planetary systems tend to be more widely spaced, as would be expected in the context of Hill stability. The $c_1$ values show how 'special' the inner Solar System is, with the first planet (Mercury) very distant from the central star compared to the other systems.

We emphasize that we do not consider these Titius-Bode-like 'laws' as having any other meaning than a possible signature of formation processes. As such, we would expect them to hold only in certain types of planetary systems, e.g. close-in, low-mass, many-body configurations. The presently-known massive systems, on the other hand, likely experienced a more chaotic history. Moreover, not all low-mass systems satisfy such exponential relations (e.g. GJ 581) and the physics of planet formation is so diverse and complex that we do not expect any universal rule on planet ordering to exist.

\subsection{Formation and evolution}

These emerging patterns, if confirmed by further discoveries of planetary systems, may provide clues on how the observed systems of close-in super-Earths and Neptunes were formed. These systems appear to be quite common, but their formation history remains a puzzle. On the one hand, it seems unlikely that they formed {\em in situ} given the very high inner disk densities that would be required. However, little is known about statistical properties of protoplanetary disks and their density profiles, and this possibility can probably not be completely rejected at this point. On the other hand, such systems may be the result of convergent type I migration of planetary cores formed at or beyond the ice line \citep[e.g.][]{terquem07,kennedy08}. But how can several protoplanets grow to masses in the super-Earth/Neptune range while migrating together during the disk lifetime, and end up in a configuration which is not necessarily close to mean-motion resonances? Near-commensurability of the orbits would be expected according to \citet{terquem07}. Loss of commensurability could occur through orbital decay due to stellar tides, but this is probably efficient only for the planets closest to the star. So this scenario still has difficulties in explaining a system such as HD~10180.

Nevertheless, as a testable prediction of this type I migration scenario, \citet{kennedy08} suggest that the masses of close-in planets will increase as stellar mass increases, and will even reach the gas giant range for stars above $\sim$1\,$M_{\odot}$. This is due the combined effects of increased disk mass, higher isolation mass, a more distant snow line and a mass-dependent migration timescale to the inner regions that favors more massive planets as stellar mass increases. Interestingly, we note that the average masses of close-in planets do increase between HD 40307 ($M$ = 0.77\,$M_{\odot}$, [Fe/H] = -0.31), HD 69830 ($M$ = 0.82\,$M_{\odot}$, [Fe/H] = -0.06) and HD 10180 ($M$ = 1.06\,$M_{\odot}$, [Fe/H] = 0.08), hinting at some correlation between stellar mass, metallicity and masses of close-in planetary systems (see also Sect.~\ref{SectMetal} below). In any case, there is great hope that these systems will allow for a much better characterization of type I migration in the near future.

Other mechanisms have been proposed to produce close-in low-mass planets, many of which involve the influence of (migrating) gas giant(s) further out in the system \citep[e.g.][]{fogg05, zhou05, raymond08}. In this context, it is interesting to note that the present RV data can exclude the presence of Jupiter-mass objects within $\sim$10 AU in the HD~40307, GJ~581, HD~69830 and HD~10180 systems. It is therefore unlikely that gas giants played a major role in the shaping of these systems. Their {\em absence} may actually be the factor that favored the formation and survival of many lower-mass objects.

\subsection{Correlations with stellar mass and metallicity}
\label{SectMetal}

\begin{figure*}
\centering
\includegraphics[bb=18 165 592 380,width=170mm,clip]{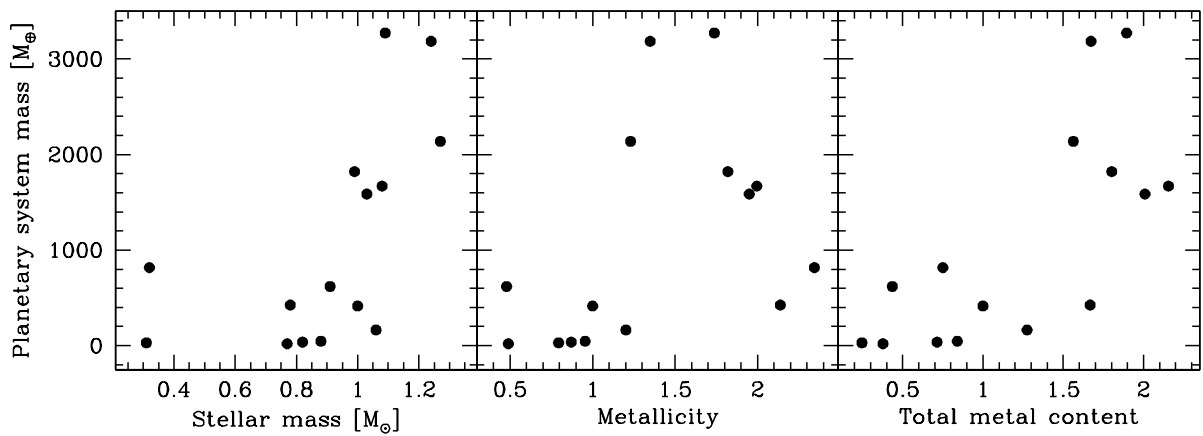}
\caption{Total planetary system mass as a function of stellar mass (left), stellar metallicity (middle) and the overall metal content of the star (right). All 3 quantities are relative to the Sun. The 15 systems with at least 3 known planets are shown.}
\label{FigMM}
\end{figure*}

Finally, we may also investigate the impact of stellar mass and metallicity on planet formation by further considering the 15 systems with at least 3 known planets. Fig.~\ref{FigMM} shows the total planetary mass in these systems as a function of stellar mass alone, stellar metallicity alone, and the total amount of heavy elements in the star given by $M_{star} 10^{\mathrm{[Fe/H]}}$. We note two obvious facts: 1) all very massive systems are found around massive {\em and} metal-rich stars; 2) the 4 lowest-mass systems are found around lower-mass {\em and} metal-poor stars. It thus appears that both quantities independently impact the mass of formed planets. When both effects of stellar mass and metallicity are combined (right panel), we obtain an even stronger correlation between total planetary system mass and total metal content in the star. The latter quantity can be seen as a proxy for the total amount of heavy elements that was present in the protoplanetary disk. These findings confirm previous trends already observed for the whole sample of planet-host stars, and are well explained by formation scenarios based on the core-accretion model.

\section{Conclusion}

In this paper we have presented a new, very rich planetary system with planets ranging from Saturn-mass to Earth-mass, and comprising 5 Neptune-like objects. Long-term radial velocity monitoring at 1\,m\,s$^{-1}$ precision was necessary to detect the low RV amplitudes of these planets. The dynamical architecture of this system reveals a compact configuration, with planets roughly equally spaced on a logarithmic scale, and with significant secular interactions. The presence of an Earth-mass body at 0.022 AU has important implications for the dynamics of the system and highlights the role of tidal dissipation to guarantee stability. Future measurements will allow us to confirm the existence of this planet. The HD~10180 system shows the ability of the RV technique to study complex multi-planet systems around nearby solar-type stars, with detection limits reaching rocky/icy objects within habitable zones. Future instruments like VLT-ESPRESSO will build on the successful HARPS experience and carry out a complete census of these low-mass systems in the solar neighborhood, pushing towards planets of a few Earth masses at 1 AU.

With the advent of the space observatories CoRoT and Kepler, low-mass planets have also become accessible to transit searches. According to early announcements, the Kepler mission will soon confirm what radial velocity surveys are already starting to find: rocky/icy planets are very common in the Universe. The combination of both techniques is likely to bring rapid progress in our understanding of the formation and composition of this population.

The HD~10180 system represents an interesting example of the various outcomes of planet formation. No massive gas giant was formed, but instead a large number of still relatively massive objects survived, and migrated to the inner regions. Building a significant sample of such low-mass systems will show what are the relative influences of the different physical processes at play during planet formation and evolution.

\begin{acknowledgements}
We are grateful to all technical and scientific collaborators of the HARPS Consortium, ESO Headquarters and ESO La Silla who have contributed with their extraordinary passion and valuable work to the success of the HARPS project. We would like to thank the Swiss National Science Foundation for its continuous support. We acknowledge support from French PNP and GENCI-CINES supercomputer facilities. FB wish to thank the French National Research Agency (ANR-08-JCJC-0102-01) for its continuous support to our planet-search programs. NCS would like to thank the support by the European Research Council/European Community under the FP7 through a Starting Grant, as well from Funda\c{c}\~ao para a Ci\^encia e a Tecnologia (FCT), Portugal, through a Ci\^encia\,2007 contract funded by FCT/MCTES (Portugal) and POPH/FSE (EC), and in the form of grants reference PTDC/CTE-AST/098528/2008 and PTDC/CTE-AST/098604/2008.
\end{acknowledgements}

\bibliographystyle{aa}
\bibliography{biblio}

\longtab{1}{
\begin{longtable}{ccccccccc}
\caption{HARPS radial velocities, CCF FWHM, bisector span and $\log{R'_{\mathrm{HK}}}$ measurements for HD~10180.}
\label{TableData} \\
\hline\hline
BJD-2,400,000 & RV & $\sigma_{\mathrm{RV}}$ & FWHM & $\sigma_{\mathrm{FWHM}}$ & BIS & $\sigma_{\mathrm{BIS}}$ & $\log{R'_{\mathrm{HK}}}$ & $\sigma_{\log{R'_{\mathrm{HK}}}}$ \\
 & [km\,s$^{-1}$] & [km\,s$^{-1}$] & [km\,s$^{-1}$] & [km\,s$^{-1}$] & [km\,s$^{-1}$] & [km\,s$^{-1}$] & [dex] & [dex] \\
\hline
\endfirsthead
\caption{Continued.} \\
\hline\hline
BJD-2,400,000 & RV & $\sigma_{\mathrm{RV}}$ & FWHM & $\sigma_{\mathrm{FWHM}}$ & BIS & $\sigma_{\mathrm{BIS}}$ & $\log{R'_{\mathrm{HK}}}$ & $\sigma_{\log{R'_{\mathrm{HK}}}}$ \\
 & [km\,s$^{-1}$] & [km\,s$^{-1}$] & [km\,s$^{-1}$] & [km\,s$^{-1}$] & [km\,s$^{-1}$] & [km\,s$^{-1}$] & [dex] & [dex] \\
\hline
\endhead
\hline
\endfoot
52948.700543 & 35.54006 & 0.00124 & 7.46233 & 0.00146 & -0.00695 & 0.00124 & -5.0211 & 0.0057 \\ 
53229.865153 & 35.52818 & 0.00053 & 7.47156 & 0.00115 & -0.00785 & 0.00098 & -5.0094 & 0.0036 \\ 
53230.885388 & 35.52613 & 0.00045 & 7.47201 & 0.00094 & -0.00692 & 0.00080 & -5.0097 & 0.0028 \\ 
53232.827037 & 35.52182 & 0.00092 & 7.47021 & 0.00212 & -0.00983 & 0.00180 & -5.0305 & 0.0086 \\ 
53262.832674 & 35.51437 & 0.00042 & 7.47182 & 0.00087 & -0.00498 & 0.00074 & -5.0080 & 0.0025 \\ 
53265.827803 & 35.52169 & 0.00050 & 7.47065 & 0.00108 & -0.00431 & 0.00092 & -5.0147 & 0.0034 \\ 
53266.810718 & 35.51991 & 0.00064 & 7.47130 & 0.00144 & -0.00697 & 0.00122 & -5.0152 & 0.0052 \\ 
53267.798078 & 35.52042 & 0.00057 & 7.46920 & 0.00127 & -0.00640 & 0.00108 & -5.0149 & 0.0043 \\ 
53268.738793 & 35.52320 & 0.00047 & 7.47108 & 0.00101 & -0.00688 & 0.00086 & -5.0102 & 0.0031 \\ 
53271.800840 & 35.52798 & 0.00055 & 7.47093 & 0.00122 & -0.00889 & 0.00104 & -5.0155 & 0.0042 \\ 
53272.765661 & 35.52449 & 0.00051 & 7.47007 & 0.00111 & -0.00525 & 0.00094 & -5.0099 & 0.0035 \\ 
53273.756743 & 35.52582 & 0.00044 & 7.47142 & 0.00094 & -0.00668 & 0.00080 & -5.0037 & 0.0027 \\ 
53274.766046 & 35.52666 & 0.00047 & 7.46951 & 0.00104 & -0.00504 & 0.00088 & -5.0127 & 0.0037 \\ 
53289.726717 & 35.53334 & 0.00054 & 7.47451 & 0.00118 & -0.00676 & 0.00100 & -5.0122 & 0.0034 \\ 
53294.718402 & 35.53452 & 0.00072 & 7.47366 & 0.00162 & -0.00203 & 0.00138 & -5.0320 & 0.0053 \\ 
53296.731043 & 35.52825 & 0.00046 & 7.47790 & 0.00099 & -0.00533 & 0.00084 & -5.0061 & 0.0026 \\ 
53309.672955 & 35.52636 & 0.00045 & 7.47780 & 0.00097 & -0.00724 & 0.00082 & -5.0143 & 0.0028 \\ 
53314.739453 & 35.52184 & 0.00048 & 7.47025 & 0.00104 & -0.00876 & 0.00088 & -5.0430 & 0.0036 \\ 
53321.730886 & 35.53126 & 0.00047 & 7.47332 & 0.00101 & -0.00662 & 0.00086 & -5.0393 & 0.0033 \\ 
53343.650435 & 35.52732 & 0.00056 & 7.46961 & 0.00122 & -0.00814 & 0.00104 & -5.0148 & 0.0037 \\ 
53344.635203 & 35.53200 & 0.00053 & 7.47376 & 0.00115 & -0.00641 & 0.00098 & -5.0175 & 0.0033 \\ 
53574.883736 & 35.52865 & 0.00043 & 7.46777 & 0.00089 & -0.00510 & 0.00076 & -4.9919 & 0.0023 \\ 
53575.902574 & 35.52962 & 0.00049 & 7.47244 & 0.00104 & -0.00497 & 0.00088 & -5.0003 & 0.0028 \\ 
53579.892152 & 35.52752 & 0.00049 & 7.46530 & 0.00104 & -0.00811 & 0.00088 & -4.9980 & 0.0027 \\ 
53604.864112 & 35.52439 & 0.00155 & 7.46528 & 0.00363 & -0.00441 & 0.00308 & -4.9678 & 0.0153 \\ 
53606.878560 & 35.51295 & 0.00064 & 7.47234 & 0.00141 & -0.00816 & 0.00120 & -5.0030 & 0.0042 \\ 
53607.862386 & 35.51448 & 0.00059 & 7.46800 & 0.00130 & -0.00744 & 0.00110 & -5.0042 & 0.0037 \\ 
53608.843526 & 35.51531 & 0.00046 & 7.46512 & 0.00097 & -0.00642 & 0.00082 & -5.0039 & 0.0026 \\ 
53610.851301 & 35.52374 & 0.00142 & 7.46883 & 0.00332 & -0.00596 & 0.00282 & -4.9880 & 0.0148 \\ 
53668.682097 & 35.52909 & 0.00046 & 7.46783 & 0.00097 & -0.00640 & 0.00082 & -5.0085 & 0.0025 \\ 
53692.577848 & 35.52763 & 0.00045 & 7.46180 & 0.00097 & -0.00765 & 0.00082 & -5.0396 & 0.0032 \\ 
53693.539827 & 35.52495 & 0.00039 & 7.46431 & 0.00080 & -0.00775 & 0.00068 & -5.0304 & 0.0024 \\ 
53695.546933 & 35.53323 & 0.00041 & 7.46399 & 0.00085 & -0.00531 & 0.00072 & -5.0308 & 0.0027 \\ 
53699.598690 & 35.52176 & 0.00048 & 7.46465 & 0.00104 & -0.00702 & 0.00088 & -5.0450 & 0.0037 \\ 
53722.597783 & 35.51341 & 0.00040 & 7.46724 & 0.00080 & -0.00686 & 0.00068 & -5.0107 & 0.0020 \\ 
53729.533673 & 35.52730 & 0.00039 & 7.46531 & 0.00075 & -0.00645 & 0.00064 & -5.0019 & 0.0018 \\ 
53759.542249 & 35.52710 & 0.00045 & 7.46848 & 0.00092 & -0.00539 & 0.00078 & -5.0042 & 0.0025 \\ 
53761.531857 & 35.52524 & 0.00044 & 7.47093 & 0.00092 & -0.00585 & 0.00078 & -5.0011 & 0.0025 \\ 
53944.920185 & 35.53441 & 0.00069 & 7.47005 & 0.00134 & -0.00628 & 0.00114 & -5.0327 & 0.0040 \\ 
53948.939084 & 35.52485 & 0.00079 & 7.46905 & 0.00162 & -0.00975 & 0.00138 & -5.0400 & 0.0054 \\ 
53950.928829 & 35.52522 & 0.00053 & 7.46636 & 0.00113 & -0.00733 & 0.00096 & -5.0216 & 0.0032 \\ 
53974.824611 & 35.52905 & 0.00057 & 7.46998 & 0.00113 & -0.00982 & 0.00096 & -5.0074 & 0.0035 \\ 
53976.807559 & 35.52814 & 0.00063 & 7.46953 & 0.00132 & -0.00543 & 0.00112 & -5.0131 & 0.0044 \\ 
53980.828169 & 35.52625 & 0.00048 & 7.46749 & 0.00101 & -0.00618 & 0.00086 & -5.0024 & 0.0026 \\ 
53981.858869 & 35.52449 & 0.00044 & 7.46795 & 0.00092 & -0.00649 & 0.00078 & -5.0070 & 0.0029 \\ 
53982.853704 & 35.52676 & 0.00042 & 7.46705 & 0.00085 & -0.00607 & 0.00072 & -5.0032 & 0.0024 \\ 
53983.819994 & 35.53033 & 0.00067 & 7.46875 & 0.00148 & -0.00659 & 0.00126 & -5.0018 & 0.0044 \\ 
54047.593747 & 35.53566 & 0.00047 & 7.46920 & 0.00097 & -0.00794 & 0.00082 & -5.0138 & 0.0024 \\ 
54049.609607 & 35.52725 & 0.00057 & 7.46778 & 0.00122 & -0.00442 & 0.00104 & -5.0253 & 0.0035 \\ 
54051.639934 & 35.52659 & 0.00069 & 7.46674 & 0.00153 & -0.00748 & 0.00130 & -5.0391 & 0.0049 \\ 
54053.688938 & 35.53604 & 0.00047 & 7.46861 & 0.00097 & -0.00664 & 0.00082 & -5.0130 & 0.0025 \\ 
54055.676278 & 35.52935 & 0.00043 & 7.46685 & 0.00087 & -0.00770 & 0.00074 & -5.0159 & 0.0022 \\ 
54076.684878 & 35.53628 & 0.00060 & 7.46842 & 0.00130 & -0.00865 & 0.00110 & -5.0270 & 0.0039 \\ 
54077.618155 & 35.53289 & 0.00052 & 7.47220 & 0.00108 & -0.00867 & 0.00092 & -5.0229 & 0.0030 \\ 
54079.624460 & 35.52567 & 0.00048 & 7.46941 & 0.00099 & -0.00816 & 0.00084 & -5.0129 & 0.0026 \\ 
54081.656910 & 35.53270 & 0.00076 & 7.46803 & 0.00170 & -0.00883 & 0.00144 & -5.0208 & 0.0055 \\ 
54083.637673 & 35.53370 & 0.00064 & 7.47049 & 0.00141 & -0.00531 & 0.00120 & -5.0346 & 0.0044 \\ 
54120.586525 & 35.52676 & 0.00094 & 7.46582 & 0.00214 & -0.00860 & 0.00182 & -5.0200 & 0.0080 \\ 
54121.545387 & 35.53212 & 0.00057 & 7.47030 & 0.00122 & -0.00796 & 0.00104 & -5.0239 & 0.0041 \\ 
54314.833837 & 35.52673 & 0.00070 & 7.47732 & 0.00148 & -0.00769 & 0.00126 & -4.9996 & 0.0043 \\ 
54316.846569 & 35.52880 & 0.00052 & 7.47271 & 0.00101 & -0.00627 & 0.00086 & -4.9996 & 0.0026 \\ 
54319.864920 & 35.53086 & 0.00057 & 7.47289 & 0.00115 & -0.00611 & 0.00098 & -5.0110 & 0.0033 \\ 
54340.789644 & 35.52960 & 0.00046 & 7.46904 & 0.00085 & -0.00773 & 0.00072 & -5.0034 & 0.0021 \\ 
54341.857588 & 35.53040 & 0.00049 & 7.47102 & 0.00092 & -0.00839 & 0.00078 & -5.0175 & 0.0024 \\ 
54342.791757 & 35.52657 & 0.00053 & 7.47269 & 0.00106 & -0.00945 & 0.00090 & -5.0133 & 0.0028 \\ 
54343.884005 & 35.52280 & 0.00059 & 7.46940 & 0.00122 & -0.00672 & 0.00104 & -5.0133 & 0.0034 \\ 
54344.804342 & 35.51966 & 0.00048 & 7.46716 & 0.00092 & -0.00878 & 0.00078 & -5.0159 & 0.0024 \\ 
54345.824377 & 35.52429 & 0.00047 & 7.46854 & 0.00087 & -0.00952 & 0.00074 & -5.0135 & 0.0023 \\ 
54346.874033 & 35.53317 & 0.00072 & 7.47649 & 0.00158 & -0.00815 & 0.00134 & -4.9950 & 0.0070 \\ 
54347.810356 & 35.53424 & 0.00049 & 7.46904 & 0.00097 & -0.00759 & 0.00082 & -5.0139 & 0.0034 \\ 
54348.839344 & 35.52828 & 0.00050 & 7.46837 & 0.00099 & -0.00781 & 0.00084 & -5.0080 & 0.0036 \\ 
54349.824140 & 35.52340 & 0.00045 & 7.47036 & 0.00082 & -0.00605 & 0.00070 & -5.0072 & 0.0021 \\ 
54387.753382 & 35.54142 & 0.00057 & 7.46720 & 0.00115 & -0.00602 & 0.00098 & -5.0220 & 0.0032 \\ 
54392.709864 & 35.53606 & 0.00052 & 7.46935 & 0.00101 & -0.00638 & 0.00086 & -5.0227 & 0.0027 \\ 
54672.930833 & 35.52832 & 0.00053 & 7.47076 & 0.00111 & -0.00403 & 0.00094 & -5.0038 & 0.0030 \\ 
54674.936375 & 35.54180 & 0.00063 & 7.47600 & 0.00137 & -0.00680 & 0.00116 & -5.0044 & 0.0038 \\ 
54677.874182 & 35.53856 & 0.00047 & 7.47193 & 0.00094 & -0.00556 & 0.00080 & -4.9867 & 0.0023 \\ 
54681.883583 & 35.54058 & 0.00054 & 7.46681 & 0.00113 & -0.00706 & 0.00096 & -4.9866 & 0.0028 \\ 
54699.905951 & 35.52998 & 0.00052 & 7.47000 & 0.00106 & -0.00733 & 0.00090 & -5.0019 & 0.0026 \\ 
54700.879837 & 35.52411 & 0.00049 & 7.47149 & 0.00097 & -0.00581 & 0.00082 & -4.9976 & 0.0024 \\ 
54702.876507 & 35.52509 & 0.00056 & 7.47453 & 0.00118 & -0.00778 & 0.00100 & -5.0067 & 0.0031 \\ 
54705.862866 & 35.52832 & 0.00070 & 7.47295 & 0.00155 & -0.00673 & 0.00132 & -4.9921 & 0.0044 \\ 
54708.857586 & 35.52909 & 0.00048 & 7.47135 & 0.00097 & -0.00530 & 0.00082 & -4.9928 & 0.0023 \\ 
54710.817762 & 35.53887 & 0.00058 & 7.47319 & 0.00125 & -0.00447 & 0.00106 & -4.9916 & 0.0033 \\ 
54720.728040 & 35.52945 & 0.00049 & 7.47071 & 0.00099 & -0.00591 & 0.00084 & -5.0001 & 0.0024 \\ 
54721.812209 & 35.53551 & 0.00081 & 7.46498 & 0.00181 & -0.00670 & 0.00154 & -5.0050 & 0.0056 \\ 
54722.811958 & 35.53488 & 0.00048 & 7.47013 & 0.00097 & -0.00799 & 0.00082 & -4.9998 & 0.0023 \\ 
54730.790969 & 35.53325 & 0.00050 & 7.47161 & 0.00104 & -0.00690 & 0.00088 & -5.0047 & 0.0027 \\ 
54731.731648 & 35.53420 & 0.00043 & 7.46762 & 0.00082 & -0.00608 & 0.00070 & -4.9974 & 0.0020 \\ 
54732.735023 & 35.53884 & 0.00055 & 7.46672 & 0.00115 & -0.00623 & 0.00098 & -4.9992 & 0.0029 \\ 
54733.743640 & 35.53872 & 0.00043 & 7.46921 & 0.00082 & -0.00509 & 0.00070 & -4.9951 & 0.0019 \\ 
54734.759629 & 35.53230 & 0.00054 & 7.46849 & 0.00113 & -0.00613 & 0.00096 & -4.9997 & 0.0028 \\ 
54736.683562 & 35.52815 & 0.00057 & 7.47328 & 0.00120 & -0.00656 & 0.00102 & -5.0024 & 0.0032 \\ 
54737.736214 & 35.53397 & 0.00073 & 7.47191 & 0.00162 & -0.00589 & 0.00138 & -4.9969 & 0.0047 \\ 
54743.677098 & 35.54015 & 0.00052 & 7.47390 & 0.00108 & -0.00517 & 0.00092 & -4.9993 & 0.0027 \\ 
54745.663551 & 35.53661 & 0.00047 & 7.47070 & 0.00094 & -0.00448 & 0.00080 & -5.0042 & 0.0023 \\ 
54747.822128 & 35.52527 & 0.00047 & 7.46858 & 0.00094 & -0.00778 & 0.00080 & -4.9985 & 0.0023 \\ 
54748.786221 & 35.52669 & 0.00041 & 7.46656 & 0.00078 & -0.00511 & 0.00066 & -4.9992 & 0.0019 \\ 
54750.803894 & 35.53364 & 0.00052 & 7.46808 & 0.00108 & -0.00727 & 0.00092 & -4.9964 & 0.0027 \\ 
54758.682427 & 35.52975 & 0.00043 & 7.46884 & 0.00085 & -0.00534 & 0.00072 & -5.0056 & 0.0023 \\ 
54759.674222 & 35.52850 & 0.00042 & 7.46828 & 0.00080 & -0.00786 & 0.00068 & -5.0008 & 0.0020 \\ 
54760.686923 & 35.53114 & 0.00041 & 7.46624 & 0.00078 & -0.00590 & 0.00066 & -5.0017 & 0.0019 \\ 
54761.677044 & 35.53778 & 0.00043 & 7.46567 & 0.00082 & -0.00630 & 0.00070 & -5.0132 & 0.0022 \\ 
54763.663210 & 35.53430 & 0.00048 & 7.46798 & 0.00097 & -0.00800 & 0.00082 & -5.0116 & 0.0027 \\ 
54764.605129 & 35.53084 & 0.00087 & 7.47410 & 0.00198 & -0.00824 & 0.00168 & -4.9786 & 0.0071 \\ 
54765.638628 & 35.53012 & 0.00069 & 7.46935 & 0.00153 & -0.00946 & 0.00130 & -4.9948 & 0.0049 \\ 
54766.631754 & 35.53086 & 0.00069 & 7.46599 & 0.00153 & -0.00433 & 0.00130 & -4.9901 & 0.0049 \\ 
54767.620499 & 35.53819 & 0.00049 & 7.47105 & 0.00101 & -0.00623 & 0.00086 & -5.0054 & 0.0028 \\ 
54774.614543 & 35.54531 & 0.00045 & 7.47178 & 0.00089 & -0.00691 & 0.00076 & -4.9899 & 0.0021 \\ 
54775.656999 & 35.54138 & 0.00054 & 7.47336 & 0.00113 & -0.00571 & 0.00096 & -4.9855 & 0.0028 \\ 
54776.565884 & 35.53901 & 0.00054 & 7.47442 & 0.00113 & -0.00622 & 0.00096 & -4.9897 & 0.0028 \\ 
54777.586673 & 35.53941 & 0.00057 & 7.47079 & 0.00120 & -0.00695 & 0.00102 & -4.9872 & 0.0030 \\ 
54778.583910 & 35.54286 & 0.00043 & 7.47242 & 0.00082 & -0.00615 & 0.00070 & -4.9890 & 0.0019 \\ 
54779.561541 & 35.54268 & 0.00050 & 7.47296 & 0.00104 & -0.00683 & 0.00088 & -4.9926 & 0.0025 \\ 
54780.578359 & 35.53982 & 0.00047 & 7.47084 & 0.00094 & -0.00591 & 0.00080 & -4.9912 & 0.0022 \\ 
54799.586422 & 35.52535 & 0.00054 & 7.47137 & 0.00113 & -0.00551 & 0.00096 & -5.0000 & 0.0029 \\ 
54800.616245 & 35.52437 & 0.00044 & 7.46770 & 0.00085 & -0.00902 & 0.00072 & -4.9903 & 0.0020 \\ 
54801.621649 & 35.52932 & 0.00047 & 7.47042 & 0.00092 & -0.00427 & 0.00078 & -4.9960 & 0.0022 \\ 
54802.638602 & 35.53103 & 0.00051 & 7.47385 & 0.00106 & -0.00727 & 0.00090 & -5.0068 & 0.0031 \\ 
54803.641708 & 35.52828 & 0.00041 & 7.46591 & 0.00078 & -0.00583 & 0.00066 & -4.9962 & 0.0020 \\ 
54804.664408 & 35.52373 & 0.00049 & 7.46941 & 0.00099 & -0.00622 & 0.00084 & -5.0004 & 0.0025 \\ 
54805.597303 & 35.52220 & 0.00049 & 7.46922 & 0.00099 & -0.00556 & 0.00084 & -4.9989 & 0.0024 \\ 
54806.581063 & 35.52765 & 0.00059 & 7.46920 & 0.00127 & -0.00895 & 0.00108 & -5.0034 & 0.0038 \\ 
54848.523947 & 35.53092 & 0.00050 & 7.46837 & 0.00101 & -0.00581 & 0.00086 & -4.9993 & 0.0025 \\ 
54851.533166 & 35.52169 & 0.00044 & 7.46785 & 0.00085 & -0.00642 & 0.00072 & -4.9979 & 0.0020 \\ 
54853.542410 & 35.52842 & 0.00048 & 7.46739 & 0.00094 & -0.00456 & 0.00080 & -5.0000 & 0.0023 \\ 
55020.896436 & 35.54196 & 0.00044 & 7.46884 & 0.00087 & -0.00573 & 0.00074 & -4.9911 & 0.0020 \\ 
55021.904529 & 35.54027 & 0.00056 & 7.46627 & 0.00094 & -0.00383 & 0.00080 & -4.9850 & 0.0022 \\ 
55022.895816 & 35.54093 & 0.00057 & 7.47520 & 0.00122 & -0.00784 & 0.00104 & -4.9874 & 0.0029 \\ 
55024.901467 & 35.53184 & 0.00050 & 7.47442 & 0.00104 & -0.00552 & 0.00088 & -4.9924 & 0.0025 \\ 
55025.767586 & 35.53675 & 0.00089 & 7.47401 & 0.00203 & -0.00957 & 0.00172 & -4.9999 & 0.0068 \\ 
55037.933241 & 35.53957 & 0.00045 & 7.46853 & 0.00089 & -0.00720 & 0.00076 & -4.9987 & 0.0021 \\ 
55038.867190 & 35.53824 & 0.00044 & 7.47193 & 0.00087 & -0.00558 & 0.00074 & -4.9991 & 0.0021 \\ 
55039.919361 & 35.53814 & 0.00196 & 7.46885 & 0.00459 & -0.00664 & 0.00390 & -4.9725 & 0.0211 \\ 
55040.848622 & 35.53190 & 0.00063 & 7.47484 & 0.00139 & -0.00665 & 0.00118 & -5.0013 & 0.0038 \\ 
55041.805672 & 35.52934 & 0.00059 & 7.47448 & 0.00127 & -0.00914 & 0.00108 & -4.9981 & 0.0034 \\ 
55042.876081 & 35.53238 & 0.00042 & 7.46870 & 0.00082 & -0.00665 & 0.00070 & -4.9864 & 0.0019 \\ 
55044.885876 & 35.53455 & 0.00065 & 7.46926 & 0.00141 & -0.00550 & 0.00120 & -5.0119 & 0.0039 \\ 
55045.847936 & 35.52777 & 0.00054 & 7.47078 & 0.00113 & -0.00632 & 0.00096 & -4.9952 & 0.0028 \\ 
55067.829962 & 35.53283 & 0.00050 & 7.46967 & 0.00104 & -0.00583 & 0.00088 & -4.9912 & 0.0025 \\ 
55068.857573 & 35.53175 & 0.00059 & 7.46815 & 0.00104 & -0.00744 & 0.00088 & -4.9923 & 0.0026 \\ 
55069.809858 & 35.52925 & 0.00050 & 7.46719 & 0.00104 & -0.00676 & 0.00088 & -4.9903 & 0.0025 \\ 
55070.808393 & 35.53095 & 0.00063 & 7.47245 & 0.00139 & -0.00712 & 0.00118 & -4.9961 & 0.0038 \\ 
55071.806534 & 35.53285 & 0.00045 & 7.46862 & 0.00092 & -0.00712 & 0.00078 & -4.9956 & 0.0023 \\ 
55072.870486 & 35.53447 & 0.00044 & 7.46680 & 0.00089 & -0.00722 & 0.00076 & -4.9980 & 0.0024 \\ 
55074.843681 & 35.52982 & 0.00048 & 7.46561 & 0.00101 & -0.00398 & 0.00086 & -4.9955 & 0.0028 \\ 
55076.815394 & 35.52740 & 0.00077 & 7.46954 & 0.00172 & -0.00895 & 0.00146 & -4.9937 & 0.0059 \\ 
55095.813607 & 35.53214 & 0.00053 & 7.46398 & 0.00113 & -0.00612 & 0.00096 & -5.0037 & 0.0030 \\ 
55096.775849 & 35.52873 & 0.00052 & 7.46821 & 0.00108 & -0.00639 & 0.00092 & -4.9965 & 0.0027 \\ 
55097.862579 & 35.52432 & 0.00065 & 7.47114 & 0.00122 & -0.00791 & 0.00104 & -5.0068 & 0.0033 \\ 
55099.817834 & 35.52230 & 0.00047 & 7.46958 & 0.00094 & -0.00540 & 0.00080 & -4.9967 & 0.0023 \\ 
55101.809197 & 35.53485 & 0.00093 & 7.47209 & 0.00212 & -0.00525 & 0.00180 & -5.0196 & 0.0072 \\ 
55103.753359 & 35.52780 & 0.00070 & 7.47297 & 0.00155 & -0.00681 & 0.00132 & -4.9993 & 0.0044 \\ 
55106.712448 & 35.52879 & 0.00054 & 7.47053 & 0.00113 & -0.00590 & 0.00096 & -4.9983 & 0.0029 \\ 
55108.743117 & 35.52738 & 0.00056 & 7.46726 & 0.00118 & -0.00839 & 0.00100 & -4.9929 & 0.0030 \\ 
55110.698889 & 35.51991 & 0.00054 & 7.46586 & 0.00085 & -0.00613 & 0.00072 & -4.9889 & 0.0019 \\ 
55112.713465 & 35.52889 & 0.00060 & 7.46375 & 0.00130 & -0.00863 & 0.00110 & -5.0055 & 0.0036 \\ 
55113.697280 & 35.53144 & 0.00053 & 7.46735 & 0.00111 & -0.00845 & 0.00094 & -4.9932 & 0.0028 \\ 
55115.751504 & 35.52800 & 0.00068 & 7.46998 & 0.00151 & -0.00836 & 0.00128 & -5.0076 & 0.0041 \\ 
55121.790157 & 35.53599 & 0.00057 & 7.47652 & 0.00122 & -0.00850 & 0.00104 & -5.0148 & 0.0033 \\ 
55122.741306 & 35.53456 & 0.00054 & 7.47215 & 0.00113 & -0.00543 & 0.00096 & -5.0127 & 0.0029 \\ 
55123.724617 & 35.53441 & 0.00050 & 7.47332 & 0.00104 & -0.00654 & 0.00088 & -5.0017 & 0.0025 \\ 
55125.727881 & 35.53788 & 0.00050 & 7.46726 & 0.00104 & -0.00587 & 0.00088 & -5.0034 & 0.0026 \\ 
55128.677184 & 35.53364 & 0.00051 & 7.46875 & 0.00106 & -0.00650 & 0.00090 & -5.0029 & 0.0027 \\ 
55132.672170 & 35.53575 & 0.00044 & 7.46646 & 0.00087 & -0.00630 & 0.00074 & -5.0030 & 0.0021 \\ 
55133.726967 & 35.53641 & 0.00054 & 7.46395 & 0.00113 & -0.00808 & 0.00096 & -5.0145 & 0.0030 \\ 
55135.624472 & 35.54342 & 0.00045 & 7.46779 & 0.00089 & -0.00853 & 0.00076 & -5.0143 & 0.0022 \\ 
55136.690239 & 35.54648 & 0.00065 & 7.46744 & 0.00120 & -0.00482 & 0.00102 & -5.0123 & 0.0033 \\ 
55139.585732 & 35.53264 & 0.00060 & 7.47179 & 0.00130 & -0.00494 & 0.00110 & -5.0066 & 0.0035 \\ 
55142.620244 & 35.53528 & 0.00065 & 7.47202 & 0.00141 & -0.00813 & 0.00120 & -5.0062 & 0.0039 \\ 
55143.631659 & 35.52923 & 0.00067 & 7.46833 & 0.00146 & -0.00596 & 0.00124 & -5.0010 & 0.0040 \\ 
55151.694827 & 35.52843 & 0.00048 & 7.46810 & 0.00099 & -0.00656 & 0.00084 & -5.0073 & 0.0025 \\ 
55153.650165 & 35.53639 & 0.00055 & 7.46894 & 0.00115 & -0.00734 & 0.00098 & -5.0069 & 0.0029 \\ 
55156.630744 & 35.52411 & 0.00106 & 7.46449 & 0.00243 & -0.00498 & 0.00206 & -5.0209 & 0.0087 \\ 
55160.621921 & 35.53008 & 0.00052 & 7.46712 & 0.00108 & -0.00558 & 0.00092 & -5.0111 & 0.0028 \\ 
55161.619721 & 35.52798 & 0.00074 & 7.46891 & 0.00146 & -0.00530 & 0.00124 & -5.0035 & 0.0041 \\ 
55349.912174 & 35.53404 & 0.00062 & 7.47053 & 0.00137 & -0.00604 & 0.00116 & -4.9978 & 0.0040 \\ 
55350.949040 & 35.52715 & 0.00070 & 7.46860 & 0.00155 & -0.00416 & 0.00132 & -5.0109 & 0.0048 \\ 
55351.941713 & 35.52050 & 0.00052 & 7.47369 & 0.00111 & -0.00959 & 0.00094 & -5.0109 & 0.0031 \\ 
55352.935047 & 35.52156 & 0.00052 & 7.47300 & 0.00108 & -0.00805 & 0.00092 & -4.9973 & 0.0030 \\ 
55353.955486 & 35.52173 & 0.00057 & 7.46834 & 0.00125 & -0.00543 & 0.00106 & -4.9997 & 0.0040 \\ 
55354.949932 & 35.52597 & 0.00061 & 7.46902 & 0.00132 & -0.00548 & 0.00112 & -4.9994 & 0.0036 \\ 
55355.919340 & 35.52818 & 0.00068 & 7.47083 & 0.00151 & -0.00600 & 0.00128 & -4.9987 & 0.0045 \\ 
55357.946100 & 35.52250 & 0.00079 & 7.46987 & 0.00177 & -0.00963 & 0.00150 & -5.0072 & 0.0057 \\ 
55358.928483 & 35.52082 & 0.00069 & 7.47357 & 0.00153 & -0.00878 & 0.00130 & -5.0043 & 0.0047 \\ 
55359.943460 & 35.52614 & 0.00065 & 7.47068 & 0.00144 & -0.00856 & 0.00122 & -5.0108 & 0.0045 \\ 
55372.929844 & 35.53892 & 0.00054 & 7.46661 & 0.00115 & -0.00524 & 0.00098 & -4.9982 & 0.0037 \\ 
55373.937747 & 35.53347 & 0.00080 & 7.46892 & 0.00181 & -0.00715 & 0.00154 & -5.0029 & 0.0070 \\ 
55375.902072 & 35.53158 & 0.00066 & 7.47566 & 0.00148 & -0.00535 & 0.00126 & -5.0238 & 0.0058 \\ 
55376.895057 & 35.53517 & 0.00060 & 7.46819 & 0.00130 & -0.00835 & 0.00110 & -5.0147 & 0.0048 \\ 
\end{longtable}
}

\end{document}